\documentclass[12pt]{article}
\usepackage{epsfig}
\usepackage{amsmath}
\usepackage{hhline}
\usepackage{amssymb}
\usepackage{times}
\usepackage{cite}
\usepackage{lscape}

\newlength{\dinwidth}
\newlength{\dinmargin}
\begin{document}
\newcommand{\pom}{{I\!\!P}}
\newcommand{\reg}{{I\!\!R}}
\def\gsim{\,\lower.25ex\hbox{$\scriptstyle\sim$}\kern-1.30ex%
\raise 0.55ex\hbox{$\scriptstyle >$}\,}
\def\lsim{\,\lower.25ex\hbox{$\scriptstyle\sim$}\kern-1.30ex%
\raise 0.55ex\hbox{$\scriptstyle <$}\,}
\newcommand{\trm}{m_{\perp}}
\newcommand{\trp}{p_{\perp}}
\newcommand{\trmm}{m_{\perp}^2}
\newcommand{\trpp}{p_{\perp}^2}
\newcommand{\alp}{\alpha_s}
\newcommand{\alps}{\alpha_s}
\newcommand{\sqrts}{$\sqrt{s}$}
\newcommand{\LO}{$O(\alpha_s^0)$}
\newcommand{\Oa}{$O(\alpha_s)$}
\newcommand{\Oaa}{$O(\alpha_s^2)$}
\newcommand{\PT}{p_{\perp}}
\newcommand{\JPSI}{J/\psi}
\newcommand{\PO}{I\!\!P}
\newcommand{\xbj}{x}
\newcommand{\xpom}{x_{\PO}}
\newcommand{\dgr}{^\circ}
\newcommand{\gev}{\,\mbox{Ge}}
\newcommand{\GeV}{\rm GeV}
\newcommand{\xp}{x_p}
\newcommand{\xpi}{x_\pi}
\newcommand{\xg}{x_\gamma}
\newcommand{\xgj}{x_\gamma^{jet}}
\newcommand{\xpj}{x_p^{jet}}
\newcommand{\xpij}{x_\pi^{jet}}
\renewcommand{\deg}{^\circ}
\newcommand{\qsq}{\ensuremath{Q^2} }
\newcommand{\gevsq}{\ensuremath{\mathrm{GeV}^2} }
\newcommand{\et}{\ensuremath{E_t^*} }
\newcommand{\rap}{\ensuremath{\eta^*} }
\newcommand{\gp}{\ensuremath{\gamma^*}p }
\newcommand{\dsiget}{\ensuremath{{\rm d}\sigma_{ep}/{\rm d}E_t^*} }
\newcommand{\dsigrap}{\ensuremath{{\rm d}\sigma_{ep}/{\rm d}\eta^*} }
\newcommand {\gapprox}
   {\, \raisebox{-0.7ex}{$\stackrel {\textstyle>}{\sim} \,$}}
\newcommand {\lapprox}
   {\, \raisebox{-0.7ex}{$\stackrel {\textstyle<}{\sim} \,$}}

\def\Journal#1#2#3#4{{#1} {\bf #2}, #4 (#3)}
\def\NCA{\em Nuovo Cimento}
\def\NIM{\em Nucl. Instrum. Methods}
\def\NIMA{{\em Nucl. Instrum. Methods} {\bf A}}
\def\NPB{{\em Nucl. Phys.}   {\bf B}}
\def\PLB{{\em Phys. Lett.}   {\bf B}}
\def\PRL{\em Phys. Rev. Lett.}
\def\PRD{{\em Phys. Rev.}    {\bf D}}
\def\PR{{\em Phys. Rev.}    }
\def\PRP{{\em Phys. Rep.}    }
\def\ZPC{{\em Z. Phys.}      {\bf C}}
\def\ZP{{\em Z. Phys.}      }
\def\EJC{{\em Eur. Phys. J.} {\bf C}}
\def\EJA{{\em Eur. Phys. J.} {\bf A}}
\def\CPC{\em Comp. Phys. Commun.}
\def\SJNP{{\em Sov. J. Nucl. Phys.}}
\def\SPJETP{{\em Sov. Phys. JETP}}
\def\JETPL{{\em JETP Lett.}}


\begin{titlepage}


\noindent
  \begin{flushleft}
  DESY 10-095 \hfill ISSN 0418-9833\\
  June 2010
\end{flushleft}


\vspace*{2cm}

\begin{center}
\begin{Large}

{\boldmath \bf
    Measurement of the Diffractive Deep-Inelastic Scattering Cross Section 
    with a Leading Proton at HERA
}

\vspace{2cm}

H1 Collaboration

\end{Large}
\end{center}

\vspace{2cm}

\begin{abstract}
\noindent
 The cross section for the diffractive deep-inelastic scattering process
 $ep \to e X p$ is measured,
 with the leading final state proton detected in the H1 Forward Proton
 Spectrometer. The data sample covers the range $\xpom<0.1$ in fractional proton longitudinal momentum loss,  $0.1 < |t| < 0.7~{\rm 
 GeV}^2$ in squared four-momentum transfer at the proton vertex and $4 < Q^2 < 700~{\rm
 GeV}^2$ in  photon virtuality.
 The cross section is measured four-fold 
 differentially in $t, \xpom, Q^2$ and $\beta=x/\xpom$,  where $x$ is the Bjorken scaling variable.
 The $t$ and $\xpom$ dependences are interpreted in terms 
 of an effective pomeron trajectory and a sub-leading exchange. The data
 are compared to perturbative QCD predictions at next-to-leading order based on
 diffractive parton distribution functions previously
 extracted from complementary measurements of inclusive diffractive deep-inelastic scattering. 
 The ratio of the diffractive to  the inclusive $ep$ cross section is studied as a function of 
 $Q^2, \beta$ and $\xpom$. 
\end{abstract}

\vspace{1cm}

\begin{center}
  Submitted to \EJC
\end{center}

\end{titlepage}

\begin{flushleft}
F.D.~Aaron$^{5,49}$,           
C.~Alexa$^{5}$,                
V.~Andreev$^{25}$,             
S.~Backovic$^{30}$,            
A.~Baghdasaryan$^{38}$,        
E.~Barrelet$^{29}$,            
W.~Bartel$^{11}$,              
K.~Begzsuren$^{35}$,           
A.~Belousov$^{25}$,            
J.C.~Bizot$^{27}$
% ...

\bigskip{\it
% ...
$ ^{5}$ National Institute for Physics and Nuclear Engineering (NIPNE) ,
          Bucharest, Romania$^{ m}$ \\
% ...
$ ^{11}$ DESY, Hamburg, Germany \\
% ...
$ ^{25}$ Lebedev Physical Institute, Moscow, Russia$^{ e}$ \\
% ...
$ ^{27}$ LAL, Universit\'e Paris-Sud, CNRS/IN2P3, Orsay, France \\
% ...
$ ^{29}$ LPNHE, Universit\'e Pierre et Marie Curie Paris 6,
          Universit\'e Denis Diderot Paris 7, CNRS/IN2P3, Paris, France \\
 $ ^{30}$ Faculty of Science, University of Montenegro,
          Podgorica, Montenegro$^{ e}$ \\
% ...
$ ^{35}$ Institute of Physics and Technology of the Mongolian
          Academy of Sciences, Ulaanbaatar, Mongolia \\
% ...
$ ^{38}$ Yerevan Physics Institute, Yerevan, Armenia \\
% ...
$ ^{49}$ Also at Faculty of Physics, University of Bucharest,
          Bucharest, Romania \\
% ...
$ ^e$ Supported by the Deutsche Forschungsgemeinschaft \\
% ...
$ ^m$ Supported by the Romanian National Authority for Scientific Research
      under the contract PN 09370101 \\
% ...

}
\end{flushleft}

\newpage

\section{Introduction}

Diffractive processes such as $ep \rightarrow eXp$ have been
studied  extensively in deep-inelastic electron\footnote{In this paper ``electron'' is used to denote both electron and positron.}-proton
scattering (DIS) at the HERA
collider~\cite{H1Diff,H1Diff2,H1Diff94,ZEUSDiff,ZEUSDiff2,ZEUSDiff3,ZEUSDiff4,
H1FPS,H1LRG,H1DiJets,H1FPS1,H1FPS2,ZEUSLPS,ZEUSLPS2,ZEUSMX,ZEUSMX2,ZEUSDPDF},
and provide an essential input for the understanding of
quantum chromodynamics (QCD) at high parton densities.
The photon virtuality $Q^2$ supplies a hard scale, which allows the application of
perturbative QCD. Diffractive
DIS events can be viewed as resulting from processes in which
the photon probes a net colour singlet
combination of exchanged partons.
A hard scattering QCD collinear factorisation theorem \cite{Collins}
allows the definition of `diffractive parton distribution functions' (DPDFs)
for a given scattered proton
four-momentum. 
The dependence of diffractive DIS on $x$, the Bjorken scaling variable , and $Q^2$
can thus be treated in a manner similar to inclusive DIS, e.g. through
the application of the
DGLAP parton evolution equations~\cite{DGLAP,DGLAP2,DGLAP3,DGLAP4}.

\begin{figure}[h] \unitlength 1mm
 \begin{center}
 \begin{picture}(100,60)
  \put(25,-6){\epsfig{file=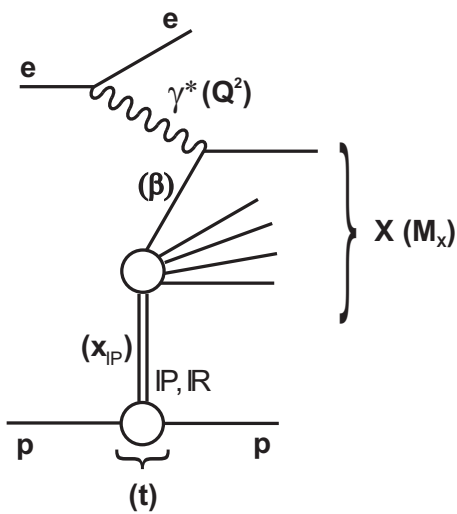 ,width=0.375\textwidth}}
 \end{picture}
 \end{center}
 \caption{Schematic illustration of the diffractive
DIS process $ep \rightarrow eXp$ and the kinematic variables used for
its description in a model in which the pomeron ($\pom$) and a sub-leading
($\reg$) trajectory are exchanged.}
\label{reggefac}
\end{figure}

Within Regge phenomenology,
diffractive cross sections are described
by the exchange of a pomeron ($\pom$)
trajectory, as illustrated in figure~\ref{reggefac}. In previous measurements 
at HERA~\cite{H1LRG,H1DiJets}
diffractive DIS cross sections 
are interpreted in a combined framework, which applies
the QCD factorisation theorem to the $x$ and $Q^2$ dependence and
uses a Regge inspired approach to express the dependence on $\xpom$, 
the fraction of the incident proton longitudinal
momentum carried by the colour singlet exchange.
In this framework the data at low $\xpom$ are well described and
DPDFs and a pomeron trajectory intercept
are extracted. In order to describe the data at
larger $\xpom$, it is necessary to include
a sub-leading exchange trajectory ($\reg$), with an intercept which is
consistent with the approximately degenerate trajectories associated
with the exchange of $\rho$, $\omega$, $a_2$ and $f_2$ mesons.

In many previous analyses diffractive DIS events
are selected on the basis
of the presence of
a large rapidity gap (LRG) between the leading
proton and the remainder of the hadronic final state $X$~\cite{H1Diff94,H1LRG}. The main advantage of the LRG method is high 
acceptance
for diffractive processes.
A complementary way to study diffractive processes is by direct
measurement of the outgoing proton. This is achieved in H1 using the 
Forward Proton Spectrometer (FPS)~\cite{H1FPS,H1FPS1,FPS}. Although the FPS detector has low acceptance, 
the FPS method of studying
diffraction has several advantages.
In contrast to the LRG case, the
squared four-momentum transfer $t$ at the proton vertex
can be reconstructed.
The FPS method selects
events in which the proton scatters elastically, whereas
the LRG method does not distinguish between the case where the scattered proton remains intact 
or where it dissociates into a system of low mass $M_Y$.
The FPS method also allows measurements up to
higher values of $\xpom$
than is possible with the LRG method, extending into regions
where the sub-leading trajectory is the dominant exchange.  
The FPS and LRG methods provide means to investigate whether the hard scattering process characterised
by the variables $\beta = x / \xpom$ and $Q^2$ depends also on the variables $\xpom$, $t$ and $M_Y$ associated with the 
proton vertex. According to the proton vertex factorisation hypothesis, the cross section can be written as the product 
of two factors, one characterising the hard interaction depending on $\beta$ and $Q^2$, the other characterising 
the proton vertex depending on $\xpom$ and $t$.

In this paper, a measurement of the cross section for the diffractive
DIS process $ep \rightarrow e X p$
is presented, using H1 FPS data with statistics increased by a factor $20$ compared to the previous
analysis~\cite{H1FPS}. In addition the kinematic range of the FPS measurement is extended to higher $Q^2$. The high statistics 
of the 
present 
data make the FPS results competitive in precision with the results of the LRG method.   
Reduced diffractive  cross sections, $\sigma_r^{D(4)}(\beta,Q^2,\xpom,t)$
and $\sigma_r^{D(3)}(\beta,Q^2,\xpom)$, 
are measured.
These measurements are used to
to extract the parameters of the pomeron trajectory and
to quantify the sub-leading exchange contribution. The proton vertex factorisation hypothesis is tested.  
The cross section dependence on the hard scattering variables, $\beta$ and $Q^2$, is further studied.
The ratio of diffractive to inclusive $ep$ cross sections is measured as a function of
$Q^2, \beta$ and $\xpom$. The data are compared with similar measurements of the ZEUS experiment~\cite{ZEUSLPS,ZEUSLPS2}.
The data are also compared
directly with the LRG measurement \cite{H1LRG} in order
to test the compatibility between 
the two measurement techniques and to
quantify the proton dissociation contribution in the LRG data.

\section{Experimental Technique}

The data used in this analysis correspond to an integrated
luminosity of $156.6\,\rm pb^{-1}$ and
were collected with the
H1 detector in $e^-p$ interactions (luminosity of $77.2\, \rm pb^{-1}$) and $e^+p$ interactions (luminosity of 
$79.4\, \rm pb^{-1}$) during the HERA II running period from 2005 to 2007. During this period the HERA collider was operated
at electron and proton beam energies of $E_e=27.6~\GeV$ and $E_p=920~\GeV$,
respectively, corresponding to an $ep$ centre of mass energy of
$\sqrt{s} = 319 \ {\rm GeV}$.

\subsection{H1 detector}
\label{detector}

A detailed description of the H1 detector can be found elsewhere
\cite{h1detector,h1detector2}. Here, the components most
relevant for the present measurement are described briefly.
A right handed coordinate system is employed with the origin at the position of the
nominal interaction point that has its $z$-axis pointing in the proton beam, or forward,
direction and $x(y)$ pointing in the horizontal (vertical) direction.
Transverse momenta are measured with respect to the beam axis.

The Central Tracking Detector (CTD), with  a polar angle coverage of
$20\deg<\theta<160\deg$, is used
to reconstruct the interaction vertex and
to measure the momentum of
charged particles from the curvature of their trajectories
in the $1.16\,{\rm T}$ field provided by a superconducting solenoid.

Scattered electrons with polar angles
in the range $154\deg<\theta_e^\prime <176\deg$ are measured in
a lead\,/\,scintillating-fibre calorimeter, the SpaCal \cite{SPACAL}.
The energy resolution is
$\sigma(E)/E\approx 7\%/\sqrt{E[\GeV]}\oplus 1\%$
and the energy scale uncertainty is $1\%$.
A Backward Proportional Chamber (BPC) in front of the SpaCal
is used to measure the electron polar angle with a precision
of $1$~mrad. Hadrons are measured in the Spacal with an energy scale precision
of $7\%$.

The finely segmented Liquid Argon (LAr) sampling calorimeter \cite{LAR,LAR2}
surrounds the tracking system and covers the range in polar angle
$4\deg<\theta<154\deg$. The LAr calorimeter is used to reconstruct the scattered electron 
in DIS processes at high $Q^2$.
The LAr calorimeter consists of an electromagnetic section with
lead as absorber, and
a hadronic section with steel as absorber.
Its total depth varies with $\theta$ between $4.5$ and $8$ interaction lengths.
Its energy resolution,
 determined in test beam  measurements, 
is $\sigma(E)/E\approx
11\%/\sqrt{E[\GeV]}\oplus 1\%$ for electrons and
$\sigma(E)/E\approx 50\%/\sqrt{E[\GeV]}\oplus 2\%$ for hadrons.
The absolute electromagnetic energy scale is known to $1\%$ precision.

The hadronic final state is reconstructed using an
energy flow algorithm which
combines charged particles measured in the CTD with  information from
the SpaCal and LAr calorimeters~\cite{hadroo2}.
The absolute hadronic energy scale is known with a
precision of $4\%$ for the measurements presented here.

The luminosity is determined  with a precision of $3.7\%$  by
detecting photons from the Bethe-Heitler process
$ep\rightarrow ep\gamma$ in a calorimeter 
located at $z=-103$~m.

The energy and scattering angle of the leading proton are obtained
from track measurements in the FPS \cite{FPS,H1FPS1}.
Protons scattered through small angles
are deflected by the proton beam-line magnets into a system of
detectors placed within the proton beam pipe inside two movable
stations, known as Roman Pots. Each Roman Pot station contains
four planes of five scintillating fibres, which together measure two
orthogonal coordinates in the $(x,y)$ plane. The fibre coordinate planes are
sandwiched between planes of 
scintillator tiles used for the trigger.
The stations
approach the beam horizontally from outside
the proton ring and
are positioned at
$z = 61 \ {\rm m}$ and $z = 80 \ {\rm m}$.
The detectors are sensitive to scattered protons which
lose less than $10\%$ of their energy in the $ep$
interaction and which are scattered through angles below $1 \ {\rm mrad}$.

The leading proton energy resolution is approximately $5$~GeV, independent of energy within the
measured range. The absolute energy scale uncertainty is $1$~GeV.
The effective resolution in the reconstruction of the transverse 
momentum components of the scattered proton with respect to the incident proton  
is determined to be
$\sim 50 \ {\rm MeV}$ for $p_x$
and $\sim 150 \ {\rm MeV}$ for $p_y$, dominated by the intrinsic transverse momentum
spread of the proton beam at the interaction point.
The corresponding $t$-resolution varies over the measured range from
$0.06 \ {\rm GeV^2}$ at $|t| = 0.1 \ {\rm GeV^2}$
to $0.17 \ {\rm GeV^2}$ at $|t| = 0.7 \ {\rm GeV^2}$.
 The calibration of the FPS is performed  using a sample of elastic $e p \rightarrow e 
 \rho^0 p$ photoproduction events with $\rho^0 \rightarrow \pi^+ \pi^-$ decays 
 by comparing the variables reconstructed in the CTD with the values measured in the FPS.
The scale uncertainties in the
transverse momentum measurements are $10 \ {\rm MeV}$ for
$p_x$ and $30 \ {\rm MeV}$ for $p_y$. 
Further details of the analysis of the FPS resolution and scale uncertainties can be found in~\cite{H1FPS}.   
For a leading proton which passes through
both FPS stations, the average overall track reconstruction efficiency is
$48\%$.

\subsection{Event selection and kinematic reconstruction}
\label{recsec}

The events used in this analysis are triggered on the basis of a coincidence
between the FPS trigger scintillator tile signals
and an electromagnetic cluster signal in the SpaCal or LAr calorimeter.
The trigger efficiency is around $99\%$ on average.

Several selection criteria are applied to the data in order to select the DIS event sample and to suppress
beam related backgrounds,
 photoproduction processes
and events in which
the incoming electron loses
significant energy through QED
radiation. The DIS selection criteria are summarised below.
\begin{itemize}
\item The reconstructed $z$ coordinate  of the event vertex is required to
lie within $35 \ {\rm cm}(\sim 3\sigma)$ of the mean position. At
least one track originating from the interaction vertex and
reconstructed in the CTD is required to have a transverse momentum
above $0.1$~GeV.
\item The 
the energy $E_e^\prime$ and the polar angle $\theta_e^\prime$ of the scattered electron
are determined from the
SpaCal (LAr) cluster, linked to a reconstructed charged particle
track in the BPC (CTD), and
the interaction vertex reconstructed in the CTD.
The electron candidate is required to satisfy either 
$154\deg<\theta^{\prime}_e<176\deg$ and $E_e^\prime > 8 \ {\rm GeV}$ in the Spacal calorimeter or
$\theta^{\prime}_e<154\deg$ and $E_e^\prime > 10 \ {\rm GeV}$ in the LAr calorimeter.
\item The quantity $E - p_z$,
calculated from the energies and longitudinal momenta of
all reconstructed particles including the electron, is required to lie
between  $35~\GeV$ and $70~\GeV$. For
neutral current DIS events
this quantity is expected to be twice the
electron beam energy neglecting detector effects
and QED radiation.
\end{itemize}
\noindent
The following requirements are applied
to the leading proton measured in the FPS.
\begin{itemize}
\item The measurement is restricted to the region where the FPS acceptance is
high by requiring the horizontal
and vertical projections of the transverse momentum to 
lie in the ranges $-0.63 < p_x < -0.27~\GeV$ and
$|p_y| < 0.8~\GeV$, respectively, and
the energy of the leading proton $E_p^\prime$ to be greater than $90\%$ of 
the proton beam energy $E_p$,
where $E_p$ is the energy of the proton beam.
\item The quantity $E + p_z$, summed
over all reconstructed
particles including the leading proton, is required to be
below $1900~\GeV$. For neutral current DIS events this
quantity is expected to be twice the
proton beam energy neglecting detector effects. This requirement is applied to suppress
cases where a DIS event reconstructed in the central detector coincides with background
in the FPS, for example due to an off-momentum beam proton (beam halo).
\end{itemize}

\noindent
The inclusive DIS variables $Q^2$, $x$ and the inelasticity $y$ are reconstructed by combining information  
from the scattered electron and
the hadronic final state using the method introduced in \cite{H1Diff94}:
\begin{eqnarray}
  y = y_e^2+y_d(1-y_d) \ \ \ ; \ \ \
Q^2=\frac{4E_e^2 (1-y)}{\tan ^2(\theta_e^\prime/2)} \ \ \ ; \ \ \
x=\frac{Q^2}{sy} \ .
\label{eq:y}
\end{eqnarray}
Here, $y_e$ and $y_d$ denote the values of $y$
obtained from the scattered electron only (`electron method') and
from the angles
of the electron and the hadronic final state (`double angle
method'), respectively~\cite{dameth}.

Variables specific to diffractive DIS are defined as
\begin{eqnarray}
 \xpom =\frac{q \cdot (P-P^\prime)}{q \cdot P} \ \ \ ; \ \ \
 \beta =\frac{Q^2}{2q \cdot (P-P^\prime)} \ ,
\label{eq:xpomdef}
\end{eqnarray}
with $q$, $P$ and $P^\prime$ denoting the four-vectors of the exchanged
virtual photon
and the incoming and outgoing proton, respectively. The variable
$\beta$ can be interpreted as the fraction
of the longitudinal momentum of the colourless exchange which is carried
by the struck quark.
The variable $\xpom$ is reconstructed directly from the energy of the leading proton, such that
\begin{eqnarray}
 \xpom = 1 - E'_p/E_p \ .
\label{eq:xpomfps}
\end{eqnarray}
Two methods are used to reconstruct $\beta$ in order to obtain the optimal resolution across the measured $\xpom$ range.
It is reconstructed as $\beta = x/\xpom$ in the range $\xpom \geq 0.012$. For $\xpom < 0.012$ 
the hadronic final state is used for the reconstruction according to:
\begin{eqnarray}
 \beta = \frac{Q^2}{Q^2+M_X^2} \ \ \ . \ \ \
\label{eq:betadef}
\end{eqnarray}
The mass $M_X$ of the hadronic system $X$ is 
obtained from
\begin{eqnarray}
 M_X^2=(E^2-p_x^2-p_y^2-p_z^2)_{\rm had} \cdot \frac{y}{y_h} \ ,
\label{eq:mxdef}
\end{eqnarray}
where the subscript `${\rm had}$' represents a sum over all hadronic final
state particles excluding the leading proton
and $y_h$ is the value of $y$ reconstructed using only the hadronic
final state \cite{hadmeth}. Including the factor $y/y_h$
leads to cancellations of several measurement inaccuracies.

The squared four-momentum transfer $t = (P-P^\prime)^2$ is
reconstructed using the transverse momentum $p_t$ of
the leading proton measured with the FPS and the value of $\xpom$
as described above, such that
\begin{eqnarray}
 t = t_{\rm min} - \frac{p_t^2}{1-\xpom} \ \ \ ; \ \ \
t_{\rm min} = - \frac{\xpom^2 m_p^2}{1-\xpom} \ ,
\label{eq:tdef}
\end{eqnarray}
\noindent where $|t_{\rm min}|$ is the minimum kinematically accessible value of $|t|$
and $m_p$ is the proton mass.

In this measurement, the
reconstructed $|t|$ is required to lie in the range
$0.1 < |t| < 0.7~\GeV^2$ and $\xpom$ in the range $\xpom < 0.1$.
The measurement is restricted to a `medium' $Q^2$
region of $4 < Q^2 < 110~\GeV^2$, $0.03 < y < 0.7$ and a `high' $Q^2$ region of $120 < Q^2 < 700~\GeV^2$, 
$0.03 < y < 0.8$ for data with electron candidates reconstructed in the Spacal and LAr calorimeters, respectively.
The final data sample contains about $68\,200$ events at medium $Q^2$ and about $400$ events at high $Q^2$.

\section{Monte Carlo Simulation and Corrections to the Data}
\label{mc}

Monte Carlo simulations are used to correct the data for the effects of
detector acceptances, inefficiencies, migrations
between measurement intervals due to finite resolutions and
QED radiation. The reaction $ep \rightarrow eXp$
is simulated with the RAPGAP program~\cite{RAPGAP} 
using the H1 2006 DPDF Fit B set~\cite{H1LRG}.
The H1 2006 DPDF Fit A and H1 2006 DPDF Fit B parameterisations give a consistent description of diffractive inclusive DIS 
processes~\cite{H1LRG}, but the H1 2006 DPDF Fit B predictions are in better agreement with the diffractive 
di-jet production cross sections measured in DIS~\cite{H1DiJets}. 
Contributions from both leading ($\pom$) and sub-leading ($\reg$) exchanges are considered.
Hadronisation is simulated using the Lund string model~\cite{Lund} implemented within the
PYTHIA program~\cite{Pythia}. An additional $\rho$-meson contribution 
relevant for the low $M_X$ domain is simulated
using the DIFFVM generator~\cite{DIFFVM}.

The background from photoproduction processes, where
the electron is scattered into the backward beam pipe and a particle from the
hadronic final state fakes the electron signature, 
is estimated using the PHOJET Monte Carlo model~\cite{PHOJET}. This background
is negligible except at the highest $y$ values and
is $3\%$ at most.
The proton dissociation background, where the leading proton originates
from the decay of a higher mass state, is estimated using an implementation
in RAPGAP of the proton dissociation model originally developed for the
DIFFVM Monte Carlo generator. This background
is negligible except at the highest $\xpom$ values, where it reaches $2\%$.

The response of the H1 detector is simulated in detail using the GEANT3 program~\cite{Geant3} and the events
are passed through the same analysis chain as is used for the data.

\begin{figure}[p] \unitlength 1mm
  \begin{center}
\vspace{-0.5cm}

  \begin{picture}(160,190)
     \put(15,0){\epsfig{file=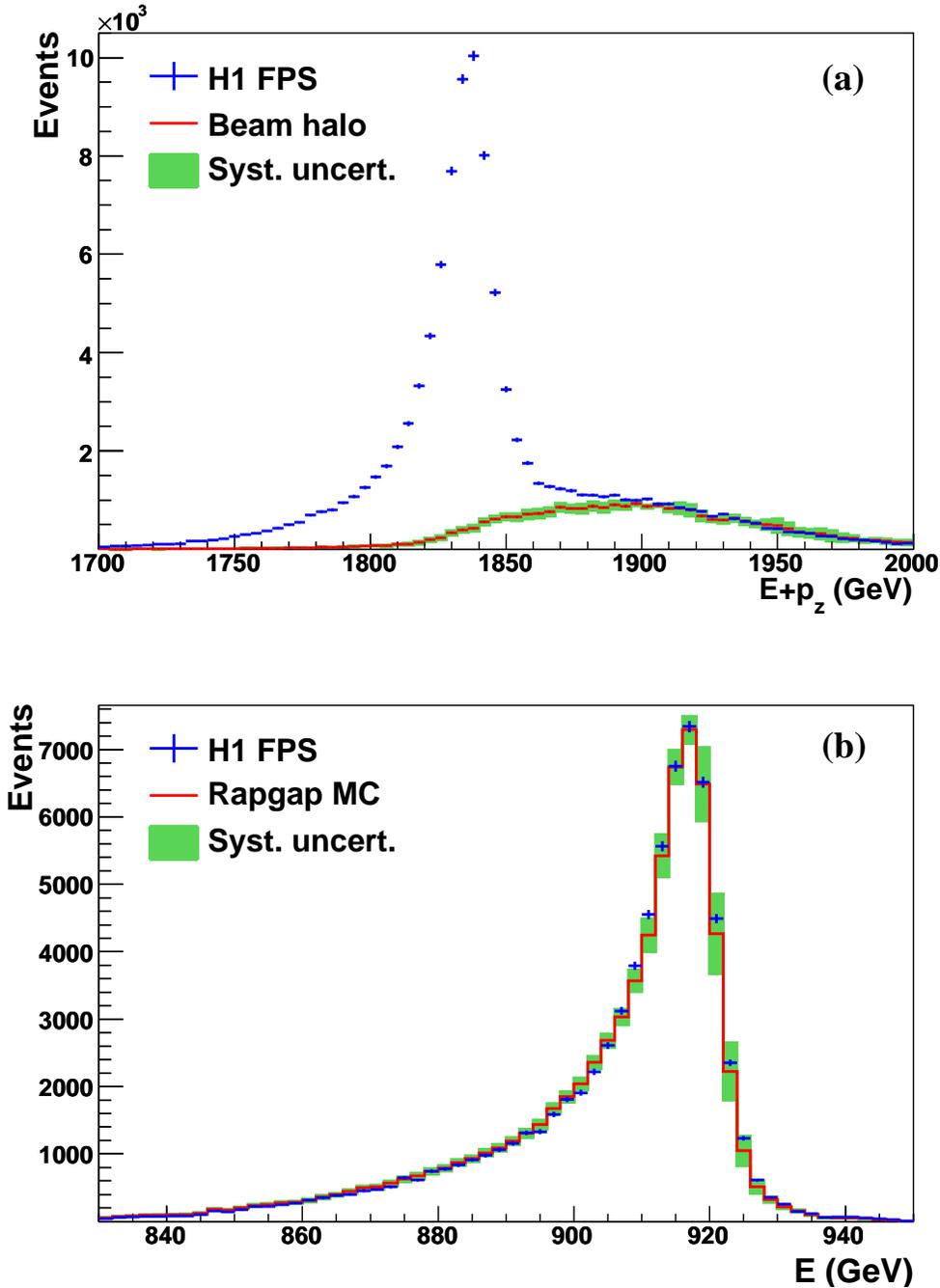 ,width=0.8\linewidth}}
     \put(127,165){\bf{\large{(a)}}}
     \put(127,73){\bf{\large{(b)}}}
  \end{picture}
 \end{center}
\caption{(a) The distribution of $E+p_z$ for FPS DIS events (histogram with error bars) and for random 
  coincidences of DIS events reconstructed in the H1 central detector with beam-halo protons giving a signal
  in the FPS (histogram with shaded bands). The systematic uncertainties on the beam-halo background are  presented as shaded bands 
  around the beam-halo histogram. 
  (b) The distribution of the leading proton energy reconstructed in the FPS (histogram with error bars). The beam-halo background 
  is subtracted from the data. The RAPGAP  Monte 
  Carlo simulation is shown as a histogram with shaded bands indicating the experimental systematic uncertainties.}
\label{fig:epzplot}
\end{figure}

\begin{figure}[p] \unitlength 1mm
  \begin{center}
  \begin{picture}(160,190)
     \put(5,0){\epsfig{file=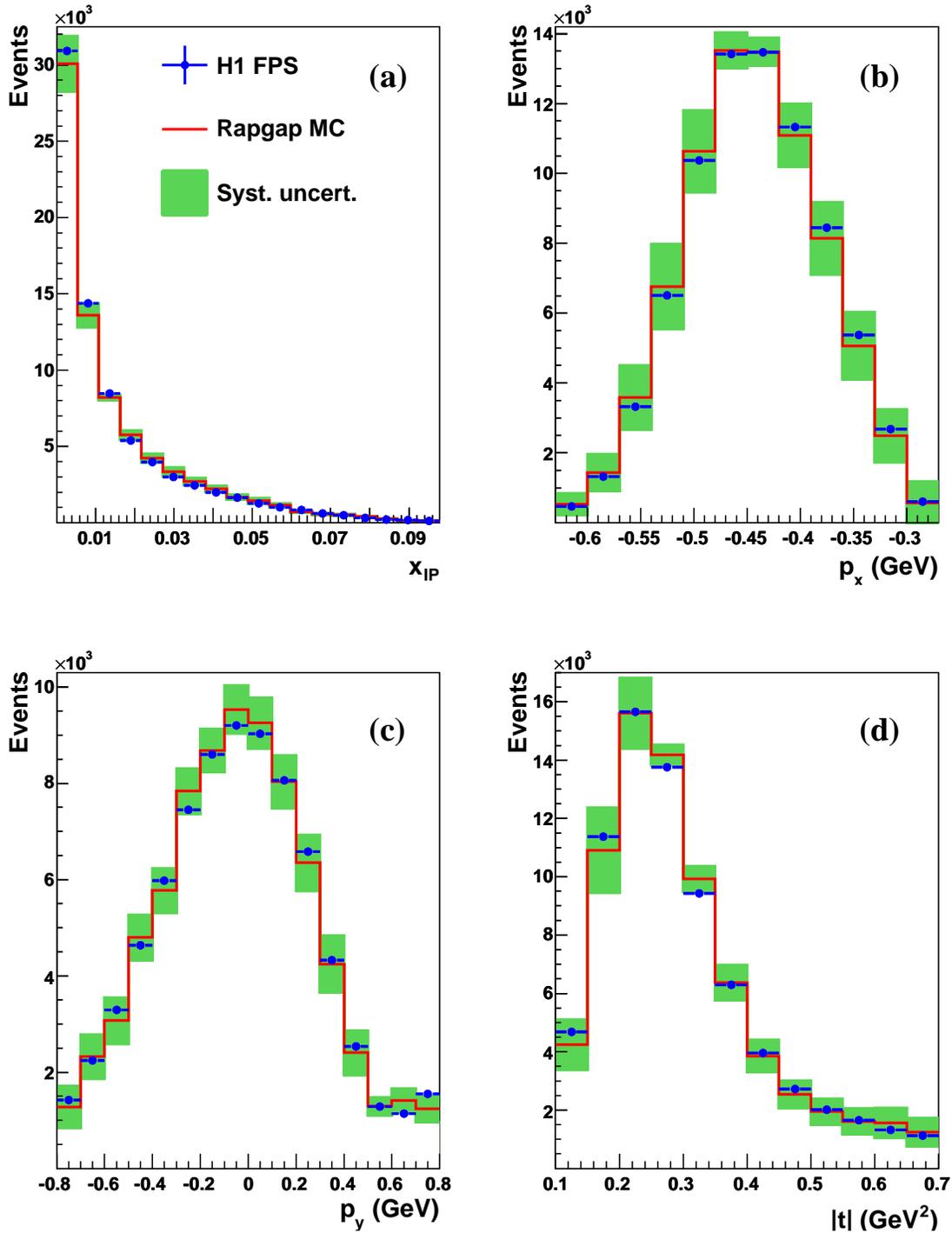 ,width=0.9\linewidth}}
      \put(60,175){\bf{\large{(a)}}}
      \put(135,175){\bf{\large{(b)}}}
      \put(60,75){\bf{\large{(c)}}}
      \put(135,75){\bf{\large{(d)}}}
  \end{picture}
 \end{center}
\caption{
  The distributions of the variables (a) $\xpom$, (b) $p_x$, (c) $p_y$ and (d) $|t|$ reconstructed using the FPS 
  (histogram with error bars). The beam-halo background is subtracted from the data. The RAPGAP Monte Carlo simulation is shown as a 
  histogram. The experimental systematic uncertainties are presented as shaded
  bands around the Monte Carlo histogram.}
\label{fig:fpsplots}
\end{figure}

Background mainly arises from random
coincidences of DIS events resulting in activity in the central detector
with off-momentum beam protons originating from interactions of beam protons with the residual gas in the beam-pipe or 
with the beam collimators (beam-halo background) giving a signal
in the FPS. This contribution is estimated statistically by
combining the quantity $E+p_z$ for all reconstructed particles in the central 
detector in DIS events (without the requirement of a track in the
FPS) with the quantity $E+p_z$ for beam-halo protons from randomly triggered events.
The resulting background distribution is normalised to the FPS DIS data distribution
in the range $E+p_z>1900~\GeV$ where beam-halo background dominates. 
After the selection cut of $E+p_z<1900~\GeV$ the background amounts to $13\%$ on average.
The $E+p_z$ spectra for leading proton and beam-halo 
DIS events are shown in figure~\ref{fig:epzplot}a. 
The background is determined using the reconstructed $E+p_z$ distribution as a function of the variables $x$, $Q^2$ 
and $t$.
A comparison of the FPS data after background subtraction and the RAPGAP simulation is presented in 
figure~\ref{fig:epzplot}b for the energy of the leading 
proton $E'_p$ and in figure~\ref{fig:fpsplots} for the variables $\xpom$, $p_x$, $p_y$ and $|t|$. The beam-halo 
background is subtracted from the data. The Monte 
Carlo simulation reproduces the data within the experimental systematic uncertainties (section~\ref{systs}).   

Cross sections are obtained at the Born level by applying corrections for QED
radiative effects to the measured values. These corrrections amount to about
$10\%$ and are obtained using the HERACLES~\cite{HERACLES} program within the
RAPGAP event generator.
The measured cross sections are quoted at the bin centres in 
$Q^2, \beta, \xpom$ and $|t|$. Bin centre corrections are applied
for the influence of the
finite bin sizes using pomeron and sub-leading exchange parameterisations in the framework of the H1 2006 DPDF Fit B~\cite{H1LRG} for the $Q^2$, $\beta$ and 
$\xpom$ dependences
and the measured $t$ dependences at each ($Q^2$, $\beta$, $\xpom$) value (section~\ref{f2d4}).

\section{Systematic Uncertainties on the Measured Cross Sections}
\label{systs}

Systematic uncertainties are considered
from the following sources.

\begin{itemize}
\item The uncertainties in the leading proton
energy and in the horizontal and vertical projections of the proton
transverse momentum are $1$~GeV, $10$~MeV and $30$~MeV,
respectively (section~\ref{detector}).
The corresponding average uncertainties on the
$\sigma_r^{D(3)}$ and $\sigma_r^{D(4)}$  measurements are
$2.5\%$, $4.8\%$ and $1.8\%$. 
The dominant uncertainty originates from the FPS acceptance variation as a
function of the leading proton transverse momentum in the horizontal
projection.

\item The electromagnetic energy scale uncertainty implies an error
of $1\%$  on the $E'_e$ measurement,
which leads to an average systematic error of $1.2\%$ on the $\sigma_r^D$
measurements.
Possible biases in the $\theta_e^\prime$ measurement in the SpaCal (LAr) calorimeter
at the level of $\pm 1\,{\rm mrad}$ ($\pm 3\,{\rm mrad}$) lead to an average systematic error of
$2.5\%$.

\item The systematic uncertainties arising from
the hadronic final state
reconstruction are determined by varying the
hadronic energy scale of the LAr
calorimeter by $\pm 4\%$ and that of the Spacal calorimeter by $\pm 7\%$.
These sources lead to an uncertainty
in the $\sigma_r^D$ measurements of around $1\%$.

\item The model dependence of the acceptance and
migration corrections is estimated by varying the shapes of the
distributions in the kinematic variables $\xpom$, $\beta$, $Q^2$ and $t$
in the RAPGAP simulation within the constraints imposed on those distributions to describe the
present
data. The $\xpom$ distribution is reweighted by $(1/ \xpom )^{\pm 0.05}$,
the $\beta$ distribution by $\beta^{\pm 0.05}$ and
$(1- \beta )^{\mp 0.05}$, the $Q^2$ distribution by 
$\log(Q^2)^{\pm 0.2}$. These sources result in an uncertainty in the $\sigma_r^D$ measurements of $1\%$. Reweighting 
the
$t$ distribution by $e^{\pm t}$ results in an uncertainty of $1.4\%$ for the measured range of $0.1 < |t|  < 0.7~\GeV^2$.

\item The model dependence of the bin centre corrections for the
 reduced cross section is estimated 
 by comparing the results obtained in the framework of the 
 H1 2006 DPDF Fit~B and H1 2006 DPDF Fit~A parameterisations~\cite{H1LRG} for the kinematic variables $\beta$ and $Q^2$.
 The $\xpom$ parameterisation is reweighted by $(1/ \xpom )^{\pm 0.05}$.
 The average uncertainty for the reduced cross section is around $1\%$.
 Reweighting the $t$ distribution by $e^{\pm t}$ results in bin centre correction uncertainties of $1.6\%$ for 
 the $\sigma_r^{D(4)}$ measurements.

\item The uncertainties related to the subtraction of background are
at most $2\%$ for proton dissociation, $3\%$ for photoproduction
and $3\%$ for the proton beam-halo contribution (section~\ref{mc}).

\item The systematic error related to the reconstruction of the event vertex is on average $1\%$, as evaluated by comparing 
the
reconstruction efficiencies for the data and Monte Carlo simulation.

\item A normalisation uncertainty of $1\%$ is
attributed to the trigger efficiencies (section~\ref{recsec}), evaluated using event samples obtained with
independent triggers.

\item The uncertainty in the FPS track reconstruction efficiency
results in the normalisation uncertainty of $2\%$.

\item A further normalisation uncertainty of $3.7\%$ arises from
the luminosity measurement.

\item The extrapolation in $t$ from the measured FPS
range of $0.1 < |t|  < 0.7~\GeV^2$ to the region
$|t_{\rm min}| < |t| < 1~{\GeV}^2$ covered by the LRG data \cite{H1LRG}
results in an additional normalisation error of
$4\%$ for the $\sigma_r^{D(3)}$ data (section~\ref{f2d3sec}).

\end{itemize}

\noindent
The systematic errors shown in the figures 
are obtained by adding in quadrature
all contributions except the normalisation uncertainty,
leading to an average uncertainty of $8\%$ for the data.
The overall
normalisation uncertainties are of $4.3\%$ and $6\%$ for the $\sigma_r^{D(4)}$ and $\sigma_r^{D(3)}$ 
measurements, respectively.

\section{Results}

\subsection{The reduced cross section {\boldmath $\sigma_r^{D(4)}$}}
\label{f2d4}

The dependence of diffractive DIS on $\beta,Q^2,\xpom$ and $t$
is studied in terms of the reduced diffractive cross section
$\sigma_r^{D(4)}$. This observable is related to the measured
differential cross section by
\begin{eqnarray}
 \frac{{\rm d}^4 \sigma^{ep \rightarrow eXp}}{{\rm d}\beta {\rm d}Q^2 {\rm d}\xpom {\rm d}t} =
  \frac{4 \pi \alpha^2}{\beta Q^4} \cdot
  \left( 1 - y + \frac{y^2}{2} \right)
  \cdot \sigma_r^{D(4)}(\beta,Q^2,\xpom,t) \ .
 \label{eq:sigrddef}
\end{eqnarray}
The reduced cross section depends on the diffractive structure functions $F_2^{D(4)}$ and $F_L^{D(4)}$
according to 
\begin{eqnarray}
  \sigma_r^{D(4)} = F_2^{D(4)} - \frac{y^2}{1+(1-y)^2} \ F_L^{D(4)} \ .
 \label{eq:sigrf2fl}
\end{eqnarray}
To a good approximation the reduced cross section is equal to the diffractive structure
function $F_2^{D(4)} (\beta,Q^2,\xpom,t)$ in
the region of relatively low $y$ values covered by the current analysis. Results for
$\sigma_r^{D(4)}$ are obtained in three $t$ ranges, $0.1 \leq |t| <
0.3~{\rm \GeV^2}, 0.3 \leq |t| < 0.5~{\rm \GeV^2}$ and $0.5 \leq |t| < 0.7~{\rm
\GeV^2}$, and are interpolated to the values $|t|=0.2,~ 0.4,~
0.6~{\rm \GeV^2}$ using the measured $t$ dependence at each $\xpom$,
$\beta$ and $Q^2$ value. Only the high statistics medium $Q^2$ data are used to 
evaluate the four-dimensional distributions $\sigma_r^{D(4)}$. 

\begin{figure}[p] \unitlength 1mm
 \begin{center}
 \begin{picture}(160,190)
    \put(0,0){\epsfig{file=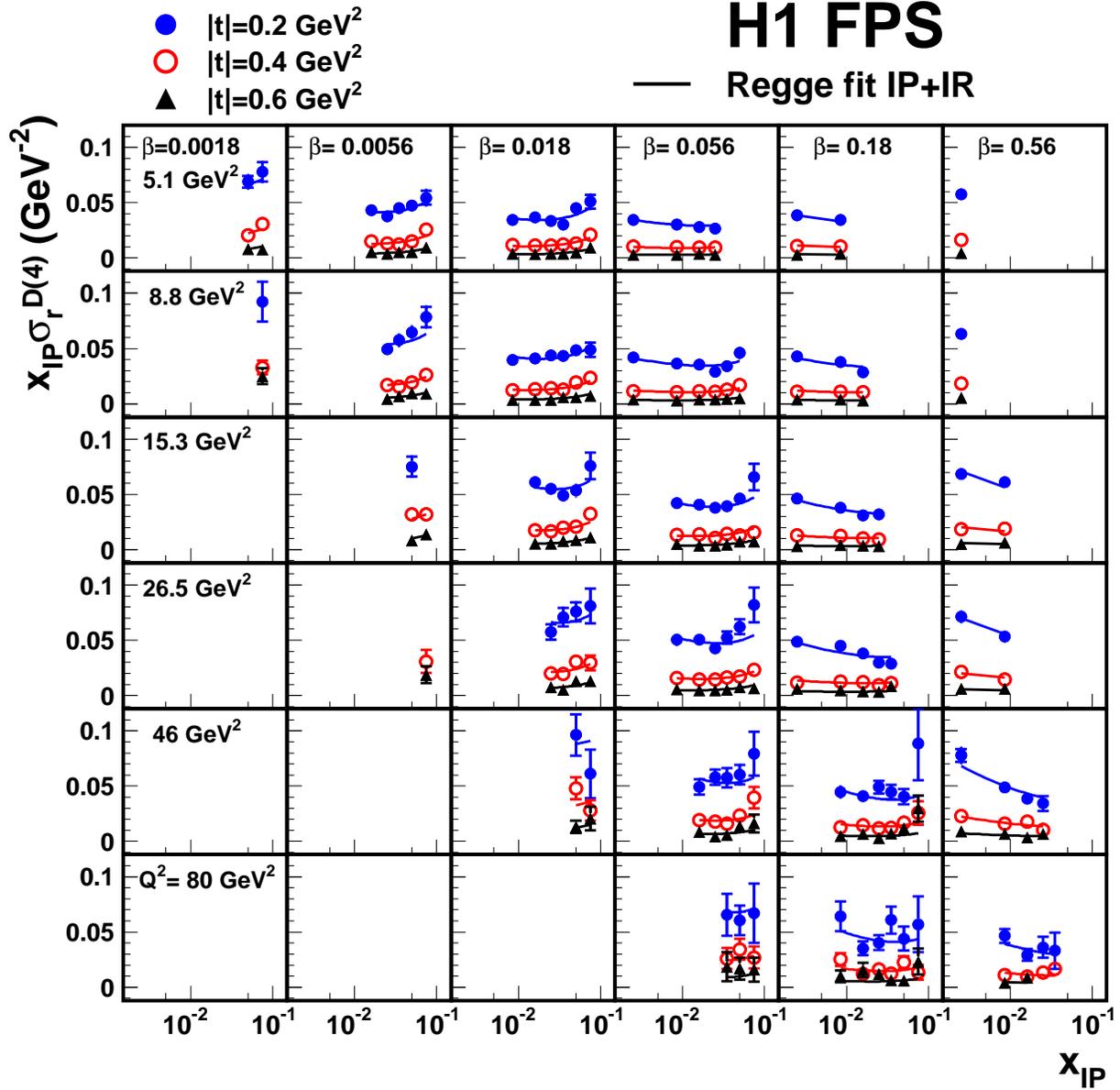 ,width=\linewidth}}
 \end{picture}
 \end{center}
 \caption{The  reduced diffractive cross section
 $\xpom \, \sigma_r^{D(4)}(\beta,Q^2,\xpom,t)$
 shown as a function of $\xpom$ for
 different values of $t$, $\beta$ and $Q^2$.
 The error bars indicate the statistical and systematic
 errors added in quadrature. The overall
 normalisation uncertainty of $4.3\%$ is not shown. The solid curves represent the results of the phenomenological Regge fit
 to the data, including both the pomeron ($\pom$) and a sub-leading ($\reg$) exchange.}
\label{fig:f2d4}
\end{figure}

The reduced cross section $\xpom \, \sigma_r^{D(4)}$ is presented in table~\ref{table:f2d4}.
Figure~\ref{fig:f2d4} shows $\xpom \, \sigma_r^{D(4)}$ as a function
of $\xpom$ for different $|t|$, $\beta$ and $Q^2$ values. At medium and large
$\beta$ values, $\xpom \, \sigma_r^{D(4)}$ falls or is flat as a
function of $\xpom$. Qualitatively this behaviour is consistent with
a dominant contribution of the pomeron exchange described in the Regge framework by a linear trajectory $\alpha_\pom(t) 
= \alpha_\pom(0) + \alpha_\pom^\prime t$ with an intercept 
$\alpha_{\pom}(0)
\gtrsim 1$~\cite{softpom2}. At low $\beta$ values $\xpom \sigma_r^{D(4)}$ rises with $\xpom$ at
the highest $\xpom$, which can be interpreted as 
 a contribution from a sub-leading exchange ($\reg$) with an intercept
$\alpha_\reg(0) < 1$. This observation is consistent with the previous H1 FPS 
analysis~\cite{H1FPS}.

\subsection{Cross section dependence on {\boldmath $\xpom$} and
{\boldmath $t$} and extraction of the pomeron trajectory}
\label{f2d4sec}

The structure function $F_2^{D(4)}$ is obtained by correcting $\sigma_r^{D(4)}$ for the 
small
$F_L^{D(4)}$ contribution using the prediction of H1 2006 DPDF Fit B given in
\cite{H1LRG}. To describe the $\xpom$ and $t$ dependence quantitatively, the structure function $F_2^{D(4)}$
is parameterised by the form
\begin{eqnarray}
 F_2^{D(4)} = f_{\pom}(\xpom,t) F_{\pom}(\beta,Q^2) + n_{\reg} \cdot f_{\reg}(\xpom,t) F_{\reg}(\beta,Q^2)  \ ,
\label{eq:regfit}
\end{eqnarray}
\noindent which assumes proton
vertex factorisation of the $\xpom$ and $t$ dependences from those
on $\beta$ and $Q^2$ for both the pomeron and any sub-leading
exchange, with no interference between the two contributions. The $\xpom$ and $t$ dependences are parameterised using flux 
factors $f_{\pom}$ and $f_{\reg}$ motivated by Regge phenomenology,
\begin{eqnarray}
 f_{\pom}(\xpom,t) = A_{\pom} \cdot \frac{e^{B_{\pom} \, t}}{(\xpom)^{2\alpha_{\pom}(t)-1}} \ \ ; \ \
 f_{\reg}(\xpom,t) = A_{\reg} \, . \, \frac{e^{B_{\reg} \, t}}{(\xpom)^{2\alpha_{\reg}(t)-1}} \ ,
\label{eq:fluxfac}
\end{eqnarray}
\noindent assuming that the sub-leading exchanges have a linear
trajectory, $\alpha_\reg(t) = \alpha_\reg(0) + \alpha_\reg^\prime t$,
as is also assumed for the pomeron. Following the convention of \cite{H1Diff94}, the values of $A_{\pom}$ and $A_{\reg}$ are
chosen such that $\xpom \cdot \int_{t_{cut}}^{t_{\rm min}}
f_{\pom,\reg}(\xpom,t) \, {\rm d} t = 1$
 at $\xpom = 0.003$ with $t_{cut}=-1~\GeV^2$.
Fitting the form of equation~\ref{eq:regfit} to the experimental  $F_2^{D(4)}$ data, the free parameters in the fit are the 
intercept and slope of the
pomeron trajectory, $\alpha_{\pom}(t) = \alpha_{\pom}(0) +
\alpha_{\pom}^\prime t$, the exponential $t$-slope parameter $B_{\pom}$ for $\xpom \rightarrow 1$,
the pomeron structure function
 $F_{\pom}(\beta,Q^2)$ 
at each of the ($\beta, Q^2$) values considered, and the 
single parameter $n_{\reg}$ describing the normalisation of the
sub-leading exchange contribution. As in \cite{H1Diff94,H1FPS,H1LRG}, the
structure function $F_{\reg}(\beta,Q^2)$ for the sub-leading
exchange in each $\beta$ and $Q^2$ bin are taken from a
parameterisation of the pion structure function~\cite{Owens}.

The behaviour of $F_2^{D(4)}$ at large $\xpom$ and low $\beta$ is
sensitive to the parameters of the sub-leading exchange $\alpha_\reg (0)$, $\alpha_\reg^\prime$
and $B_\reg$. They are taken to be the same as in the previous fits to the H1
$F_2^D$ data \cite{H1FPS,H1LRG} in order to compare the normalisation
parameters for the the sub-leading exchange contribution between the measurements. 
The intercept $\alpha_\reg (0) = 0.50$ of the sub-leading
exchange is taken from~\cite{H1Diff94}. The parameters
$\alpha_\reg^\prime = 0.3~{\rm GeV^{-2}}$ and $B_\reg
= 1.6~{\rm GeV^{-2}}$ are obtained from a
parameterisation of the previously published H1 FPS
data~\cite{H1FPS}. 
The model dependent 
uncertainty is determined by repeating the fit with the fixed parameters made free one after another in the fit.
The fitted parameters of the sub-leading exchange are consistent with the values given above. 
The influence of neglecting the $F_L^{D(4)}$ contribution to $\sigma_r^{D(4)}$ 
is also included in the model dependent uncertainty.
The experimental systematic uncertainties on the free parameters are
evaluated by repeating the fit after shifting the data points
according to each individual uncertainty listed in
section~\ref{systs}. 

The fit to equation~\ref{eq:regfit} provides
a good description of the $\xpom$ and $t$ dependences of the data, with a
minimum $\chi^2 = 273$ for $289$ degrees of freedom, combining statistical 
and uncorrelated systematic errors.
The data hence support the proton vertex factorisation hypothesis for both the pomeron and the sub-leading contribution 
as given by the fit.

\renewcommand{\arraystretch}{1.35}
\begin{table}[h]
\centering
\begin{tabular}{|l|r@{$\,$}l|}
\hline
Parameter & \multicolumn{2}{c|}{Value} \\
\hline
$\alpha_\pom(0)$  & $1.10$~ & $\pm~~0.02~{\rm (exp.)} \pm~~0.03~{\rm (model)}$ \\
$\alpha_\pom'$    & $0.04$~ & $\pm~~ 0.02~{\rm (exp.)}_{\,-\,~0.06}^{\,+\,~0.08}~{\rm (model)~GeV}^{-2}$ \\
$B_\pom$          & $5.73$~ & $\pm~~ 0.25~{\rm (exp.)}_{\,-\,~0.90}^{\,+\,~0.80}~{\rm (model)~GeV}^{-2}$ \\
$n_\reg$          & $[0.87$~ & $\pm~~ 0.10~{\rm (exp.)}_{\,-\,~0.40}^{\,+\,~0.60}~{\rm (model)}] \cdot 10^{-3}$ \\
\hline
\end{tabular}
\caption{The central values of the Regge model parameters extracted from a fit to $F_2^{D(4)}$
and their experimental and model uncertainties. The experimental uncertainty is defined as the quadratic sum
of the statistical and systematic uncertainties. The model
uncertainty is determined by varying the fixed parameters in the fit as explained in the text.}
\label{table:fit_results}
\end{table}
\renewcommand{\arraystretch}{1.}

The results for the free parameters of the fit are summarised in table~\ref{table:fit_results}. 
The experimental uncertainty of the fit parameters is defined as the quadratic sum
of the statistical and systematic uncertainties. The overall normalisation uncertainty of 
$\sigma_r^{D(4)}$ contributes only to the experimental uncertainty of the 
sub-leading exchange normalisation parameter $n_\reg$. 
The result for $\alpha_\pom (0)$ is compatible with that 
obtained
from H1 data previously measured using the LRG and FPS
methods~\cite{H1LRG,H1FPS} and with the ZEUS
measurements~\cite{ZEUSLPS2,ZEUSMX}. It is also consistent with
the pomeron intercept describing soft hadronic scattering,
$\alpha_{\pom}(0) \simeq 1.08$ \cite{softpom,softpom2,softpom3}.

In a Regge approach with a single linear exchanged
pomeron trajectory, $\alpha_{\pom}(t) = \alpha_{\pom}(0) + \alpha_\pom^\prime t$, the
exponential $t$-slope parameter $B$ of the diffractive cross section is expected to decrease 
logarithmically with increasing $\xpom$ according to
\begin{eqnarray}
B = B_\pom - 2\alpha_{\pom}^{\prime}\ln \xpom \ ,
\label{eq:tslope}
\end{eqnarray}
\noindent an effect which is often referred to as `shrinkage' of the
diffractive peak. The degree of shrinkage depends on the slope of
the pomeron trajectory, $\alpha_{\pom}^{\prime}$. 
The present FPS data favour a small value of $\alpha_\pom^\prime$, as expected in perturbative models of the
pomeron~\cite{bfkl,bfkl2}. This result is incompatible with the value of $\alpha_\pom^\prime \simeq 0.25~\GeV^{-2}$ obtained 
from soft hadron-hadron scattering at high energies~\cite{softpom,softpom2,softpom3}. Vector meson measurements at HERA
have also resulted in smaller values of $\alpha_{\pom}^{\prime}$,
whether a hard scale is present \cite{zeus:jpsi,h1:jpsi,h1:rhodis} or not \cite{zeus:rho}.
The present FPS results for $\alpha_\pom^\prime$
and $B_\pom$ are compatible with those obtained previously from the H1 / ZEUS data using the FPS / LPS 
detectors~\cite{H1FPS,ZEUSLPS2}.
Although the value of $B_\pom$ measured in the H1 experiment is
lower than that from the ZEUS data ($7.1\pm 0.7~{\rm (stat.)}\pm^{1.4}_{0.7}~{\rm (syst.)~GeV}^{-2}$), the 
results are consistent within uncertainties.

The result for the sub-leading exchange normalisation parameter is   
slightly smaller but agrees within experimental uncertainties with $n_{\reg} = [1.0 \pm 0.2~{\rm (exp.)}] \cdot 10^{-3}$ 
extracted from the previously published H1 FPS data~\cite{H1FPS} and  
$n_{\reg} = [1.4 \pm 0.4~{\rm (exp.)}] \cdot 10^{-3}$ obtained from the H1 2006 DPDF Fit B to the H1 LRG data \cite{H1LRG}.
The FPS and LRG measurements give consistent results as expected for a dominantly
isosinglet sub-leading trajectory ($\omega$ or $f_2$, rather than $\rho$ or $a_2$ exchanges). 
The sub-leading exchange is important at low $\beta$ and high $\xpom$,
contributing up to $50\%$ of the cross section at the highest bin centre value of $\xpom = 0.075$.

\subsection{Test of proton vertex factorisation}
\label{Q2dep}

\begin{figure}[p] \unitlength 1mm
 \begin{center}
 \begin{picture}(160,190)
    \put(-3,95){\epsfig{file=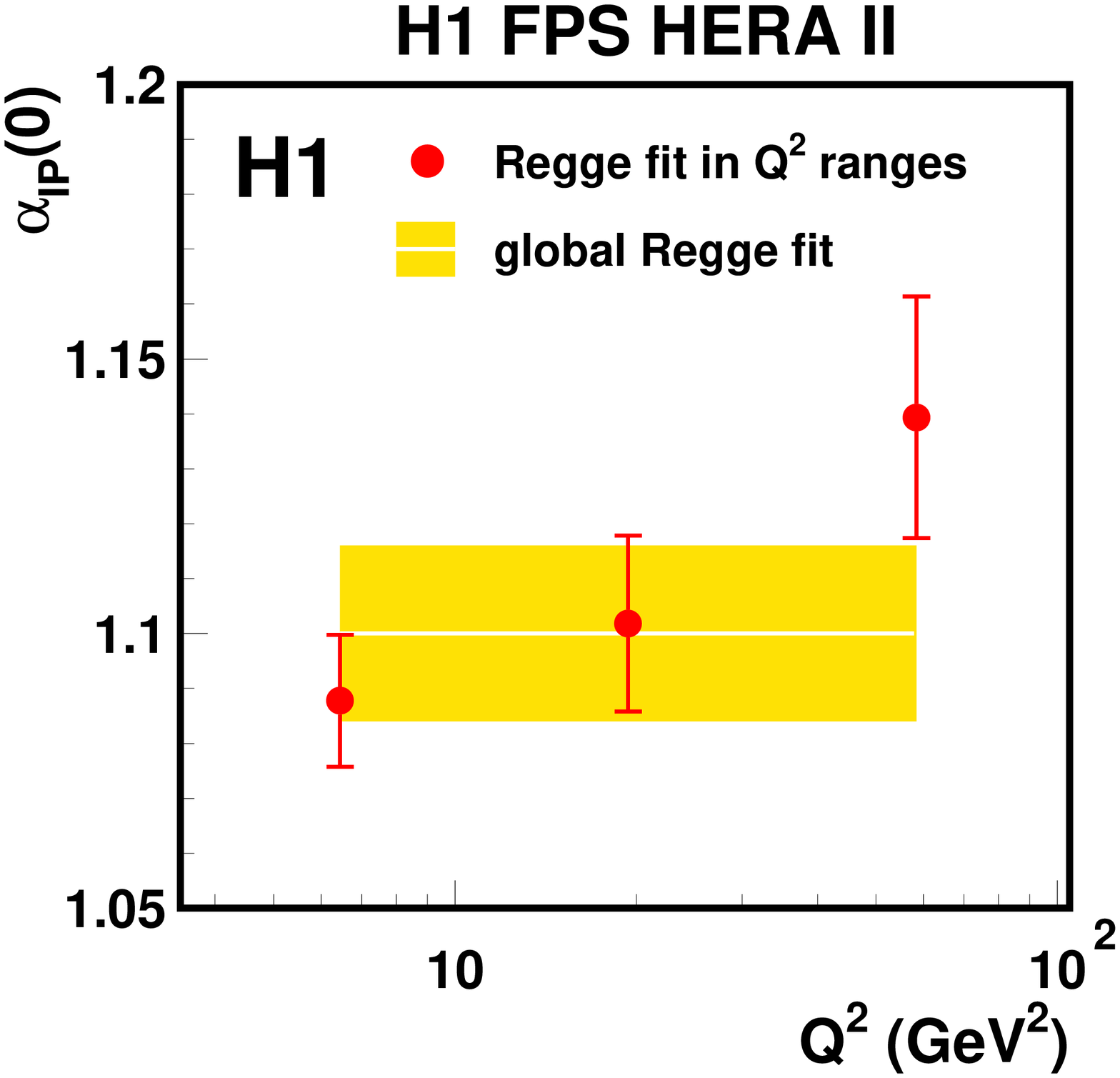,width=0.5\linewidth}}
    \put(80,95){\epsfig{file=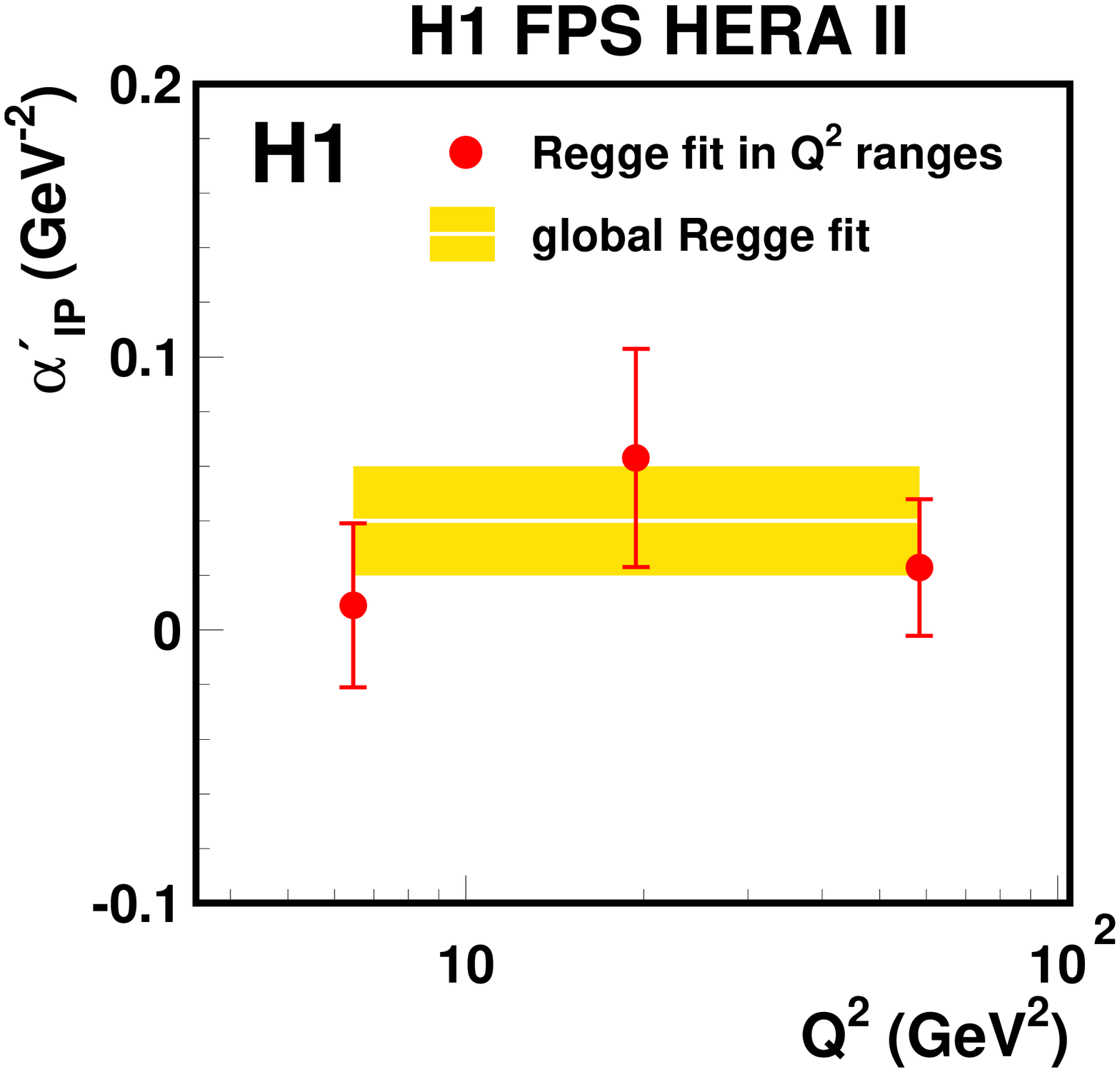,width=0.5\linewidth}}
    \put(40,-3){\epsfig{file=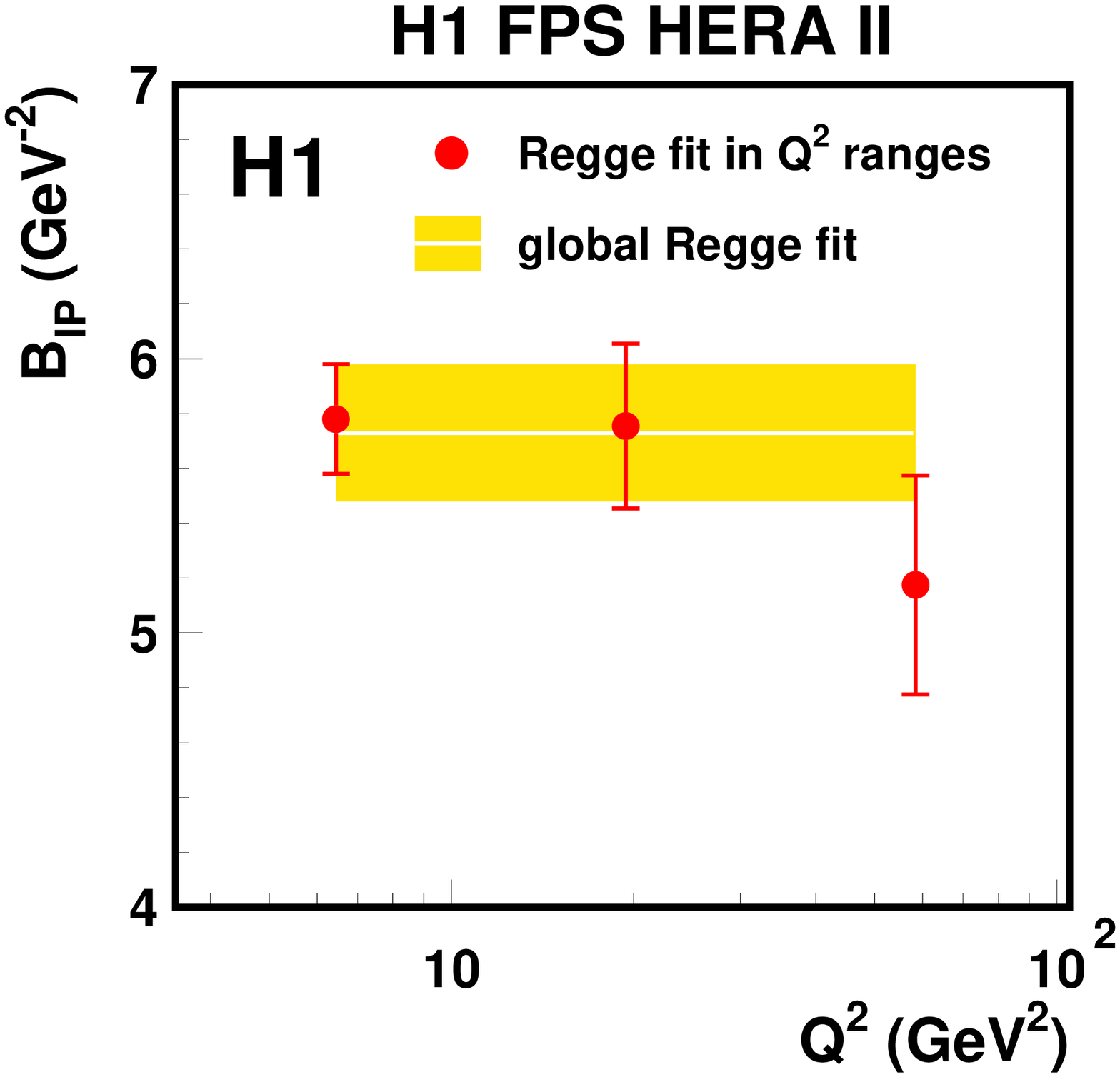,width=0.5\linewidth}}
    \put(13,150){\bf{\large{(a)}}}  
    \put(98,150){\bf{\large{(b)}}}
    \put(58,52){\bf{\large{(c)}}}
 \end{picture}
 \end{center}
 \caption{Results for (a) $\alpha_\pom(0)$, (b)
 $\alpha_\pom'$ and (c) $B_\pom$ obtained from a modified version of the Regge fit performed in three different 
 ranges of $Q^2$. The 
 error bars correspond to the experimental uncertainties which are the statistical and 
 uncorrelated systematic uncertainties added in quadrature. The 
 white lines and shaded bands show the result and experimental uncertainty
 from the standard fit over the whole $Q^2$ range.}
\label{fig:fitq2}
\end{figure}

To test in more detail a possible breakdown of
proton vertex factorisation the dependence of $\alpha_\pom(0)$, $\alpha_\pom^\prime$ and
$B_\pom$ on $Q^2$ is studied by repeating the fit described above in three different ranges of $Q^2$. 
The results of the fits, shown in figure~\ref{fig:fitq2} and
table~\ref{table:fitq2}, indicate no strong dependence on $Q^2$.
The experimental uncertainty is defined as the 
quadratic sum of the statistical and uncorrelated systematic uncertainties. 
In the fit procedure, the normalisation
factor $n_{\reg}$ for the sub-leading exchange contribution is fixed to the central value, presented in
table~\ref{table:fit_results}, as it is found to be insensitive to $Q^2$. 
 
\renewcommand{\arraystretch}{1.35}
\begin{table}[h]
\centering
\begin{tabular}{|l|l|l|l|}
\hline
\multicolumn{1}{|c|}{$Q^2$ range of Fit ($\rm GeV^{2}$)} &
\multicolumn{1}{c|}{$\alpha_\pom(0)$} &
\multicolumn{1}{c|}{$\alpha_\pom^\prime$ ($\rm GeV^{-2}$)} &
\multicolumn{1}{c|}{$B_\pom$ ($\rm GeV^{-2}$)} \\
\hline
$4 < Q^2 < 12$               & $1.088 \pm 0.012~{\rm (exp.)}$
                             & $0.009 \pm 0.031~{\rm (exp.)}$
                             & $5.78  \pm 0.20~~{\rm (exp.)}$ \\
$12 < Q^2 < 36$              & $1.102 \pm 0.016~{\rm (exp.)}$
                             & $0.063 \pm 0.041~{\rm (exp.)}$
                             & $5.75  \pm 0.30~~{\rm (exp.)}$ \\
$36 < Q^2 < 110$             & $1.139 \pm 0.022~{\rm (exp.)}$
                             & $0.023 \pm 0.026~{\rm (exp.)}$
                             & $5.17  \pm 0.40~~{\rm (exp.)}$ \\
\hline
\end{tabular}
\caption{The central values of $\alpha_\pom'$ and $B_\pom$ and their experimental uncertainties extracted from fits to 
$F_2^{D(4)}$ performed in three different ranges of $Q^2$.} 
\label{table:fitq2}
\end{table}
\renewcommand{\arraystretch}{1.}

In order to quantify a possible breakdown of proton vertex factorization, the data are fitted using parameterisations of the form $A + D \cdot \ln 
(Q^2/1~\GeV^2)$ for $\alpha_\pom(0)$ and $B_{\pom}$. 
The logarithmic derivatives of $\alpha_{\pom}(0)$ and $B_{\pom}$ are found to be 
$D(\alpha_{\pom}(0))= 0.018 \pm 0.013~{\rm(exp.)}$ 
and $D(B_{\pom})= -0.20 \pm 0.14~{\rm (exp.)~GeV}^{-2}$, respectively. 
Given the experimental uncertainties, the values of the logarithmic derivatives are within 
$1.5\sigma$ from zero and hence do not contradict to proton vertex factorisation.
 
\begin{figure}[p] \unitlength 1mm
 \begin{center}
 \begin{picture}(160,190)
    \put(0,0){\epsfig{file=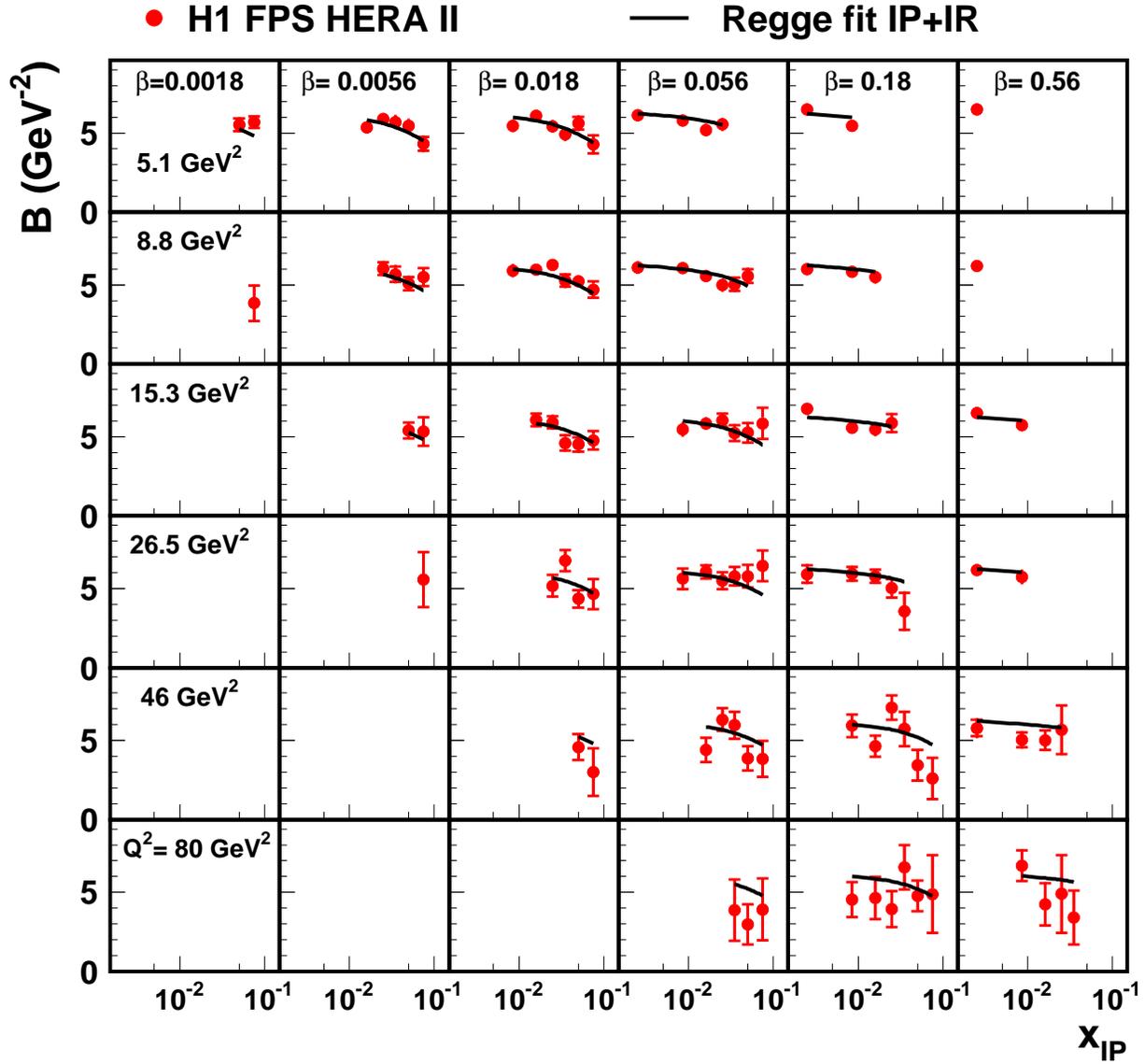 ,width=\linewidth}}
 \end{picture}
 \end{center}
 \caption{Results for the slope parameter $B$ obtained from a fit of the form
${\rm d}\sigma/{\rm d}t \propto e^{Bt}$
shown as a function of $\xpom$ for
different values of $\beta$ and $Q^2$.
The error bars indicate the statistical and systematic
errors added in quadrature. The solid curves represent the results of the phenomenological Regge fit to $F_2^{D(4)}$
including both the pomeron ($\pom$) and a sub-leading ($\reg$) exchange.}
\label{fig:bslope}
\end{figure}

Assuming an exponential $t$-dependence of the cross section, ${\rm d}\sigma/{\rm d}t \propto e^{Bt}$, the slope
para\-meter $B$ is measured as function of $\xpom$ at fixed values of $Q^2$ and $\beta$. The results are presented in 
figure~\ref{fig:bslope}.
The results 
for $B$ are compared with 
a parameterisation of the
$t$-dependence from the fit to $F_2^{D(4)}$ (section \ref{f2d4sec}) as shown in figure~\ref{fig:bslope}. 
The fit results are shown as curves of the form:
\begin{eqnarray*}
B(\xpom,\beta,Q^2)=[1-w_\reg(\xpom,\beta,Q^2)]
[B_\pom-2\alpha_{\pom}^{\prime}\ln \xpom]+w_\reg(\xpom,\beta,Q^2)
[B_\reg-2\alpha_{\reg}^{\prime}\ln \xpom] \ ,
\end{eqnarray*}
\noindent $w_\reg (\xpom,\beta,Q^2)$ being the fraction of
$F_2^{D(4)}$ which is due to the sub-leading exchange 
A good description of the data over the full $\xpom$, $Q^2$ and $\beta$ range is obtained. 
At low $\xpom$, the data are compatible with a constant slope parameter, $B \simeq 6 \ {\rm 
GeV^{-2}}$. No significant $Q^2$ or $\beta$
dependence of the slope parameter $B$ is observed for data points with $\xpom \leq 0.025$. The 
sub-leading exchange contribution integrated over this kinematic range is $5\%$. A parameterisation of the data in this $\xpom$ 
range with a constant 
slope parameter $B$ gives $\chi^2 = 89$ for $75$ data points, where the errors include the combined statistical and 
uncorrelated systematic uncertainties. 
Within uncertainties,
the $t$ dependence of the cross section in the pomeron dominated low $\xpom$
region can therefore be factorised from the
$Q^2$ and $\beta$ dependences.

\begin{figure}[p] \unitlength 1mm
 \begin{center}
\vspace{-0.5cm}

 \begin{picture}(160,190)
    \put(0,0){\epsfig{file=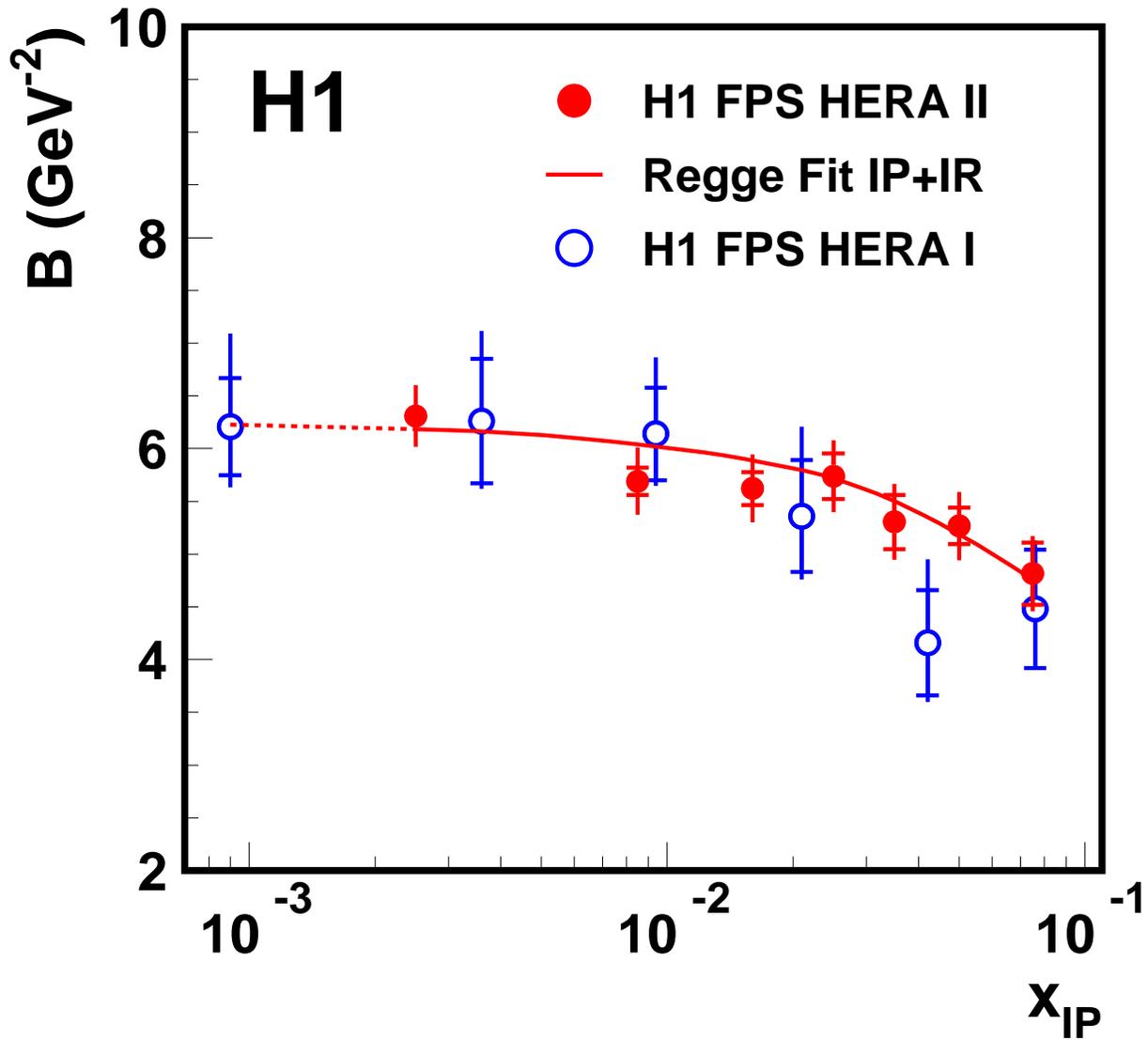 ,width=\linewidth}}
 \end{picture}
 \end{center}
 \caption{The slope parameter $B$ obtained from a fit of the form
${\rm d}\sigma/{\rm d}t \propto e^{Bt}$ shown as a function of $\xpom$. The data are averaged over $Q^2$ and $\beta$.
The inner error bars represent the statistical errors.
The outer error bars indicate the statistical and systematic
errors added in quadrature. The solid curve represents the results of the phenomenological Regge fit
to the data, including both the pomeron ($\pom$) and a sub-leading ($\reg$) exchange. The dashed curve represents the 
prediction beyond the $\xpom$ range used in the fit. The previously published H1 FPS results~\cite{H1FPS} are also shown (open circles).}
\label{fig:bslopexp}
\end{figure}

Since no significant $Q^2$ or $\beta$ dependence is observed, the slope parameter $B$ is obtained by averaging over the 
$Q^2$ and $\beta$ 
and compared with the result of a parameterisation of the
\mbox{$t$-dependence} from the fit to $F_2^{D(4)}$.
The result is shown as a function of $\xpom$ in figure~\ref{fig:bslopexp}.
 The previously published H1 FPS results~\cite{H1FPS} are also shown.
A weak decrease of the $B$ parameter value from $6 \ {\rm GeV^{-2}}$ to less than $5 \ {\rm GeV^{-2}}$ is observed
towards larger values of $\xpom \gapprox 0.05$,
where the contribution
from the sub-leading exchange is
significant. 
This reduction of the slope parameter indicates that the
size of the interaction region is reduced for $\reg$ exchange, as compared to $\pom$ exchange.

\subsection{The reduced cross section {\boldmath $\sigma_r^{D(3)}$} and comparison with other measurements}
\label{f2d3}

The reduced cross section $\sigma_r^{D(3)} (\beta,Q^2,\xpom)$, defined as the integral of
$\sigma_r^{D(4)} (\beta,Q^2,\xpom, t)$ over $t$ in the range  
$|t_{\rm min}| < |t| < 1 \ {\rm GeV^2}$, is obtained by extrapolating 
the FPS data from the measured
range $0.1 < |t| < 0.7~\GeV^2$
using the $t$-dependence
at each ($\xpom, \beta, Q^2$) value (section~\ref{Q2dep}).
This extrapolation factor, which amounts to a value of $1.8$ with an uncertainty of $4\%$, depends only weakly
on $\xpom$.

\begin{figure}[p] \unitlength 1mm
 \begin{center}
 \begin{picture}(160,190)
    \put(0,0){\epsfig{file=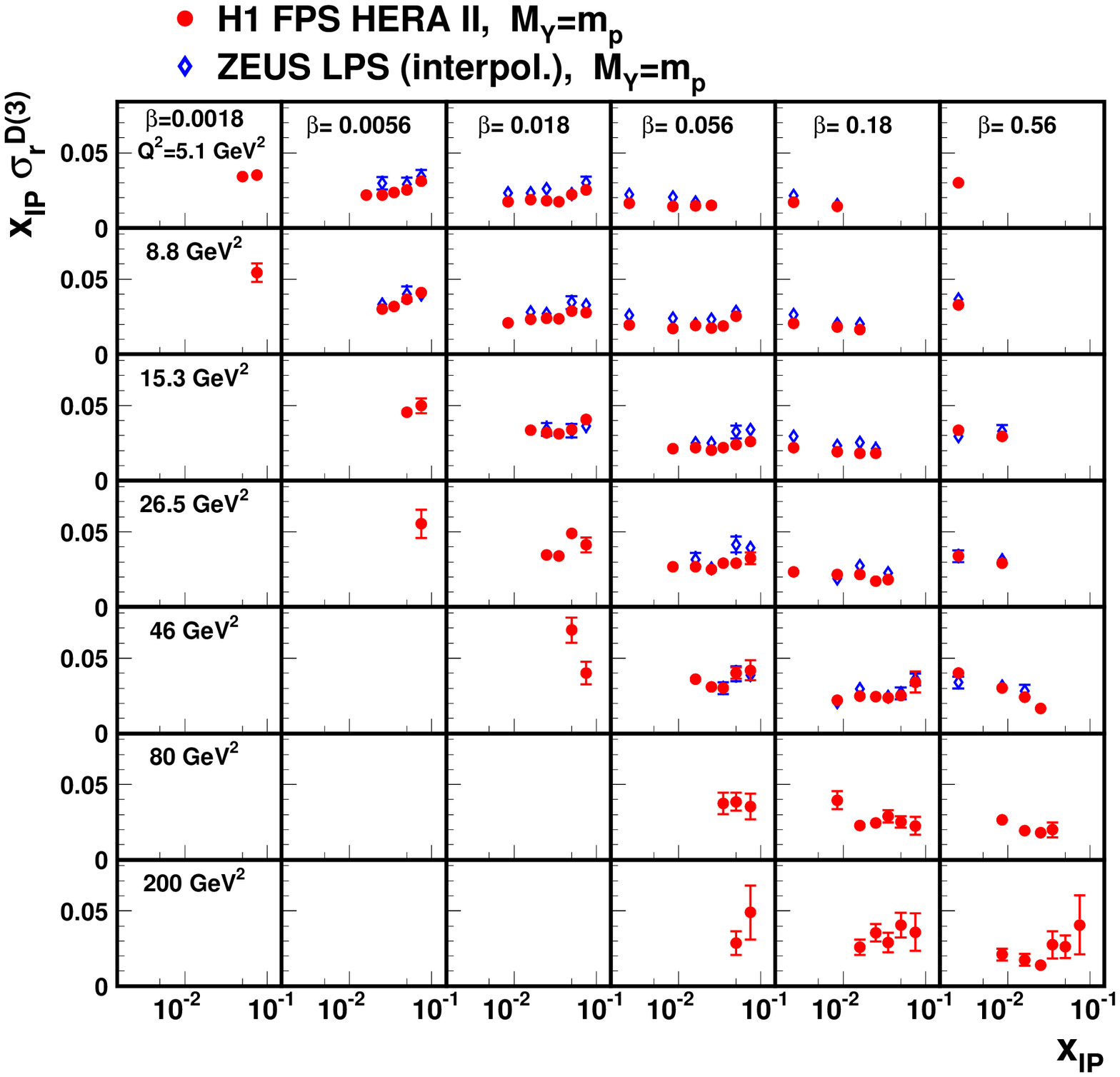 ,width=\linewidth}}
 \end{picture}
 \end{center}
 \caption{The  reduced diffractive  cross section
$\xpom \, \sigma_r^{D(3)}(\beta,Q^2,\xpom)$
for $|t| < 1 \ {\rm GeV^2}$,
shown as a function of $\xpom$ for
different values of $\beta$ and $Q^2$.
The error bars indicate the statistical and systematic
errors added in quadrature. The H1 FPS data are compared
with the ZEUS LPS results \cite{ZEUSLPS2} interpolated to the FPS  $\beta,Q^2, 
\xpom$ values. The overall
normalisation uncertainty of $6\%$ on the H1 FPS data and the normalisation 
uncertainty of $^{+11}_{-\ \,7}\%$ on the ZEUS LPS 
data are
not shown.}
\label{fig:f2d3xp_zeuslps}
\end{figure}

In figure~\ref{fig:f2d3xp_zeuslps} the H1 FPS measurements of $\xpom \, \sigma_r^{D(3)}$ are presented. They are compared with 
those of the
ZEUS collaboration, measured using their Leading Proton Spectrometer (LPS)~\cite{ZEUSLPS2}. The ZEUS data points are interpolated to the
$\beta, Q^2$ and $\xpom$ values of this measurement using a parameterisation of the ZEUS DPDF SJ fit~\cite{ZEUSDPDF}.
The ratio of the H1 FPS to ZEUS LPS data averaged over the measured
kinematic range is
$~0.85\pm{\rm 0.01(stat.)}\pm{\rm 0.03(syst.)}^{+0.09}_{-0.12}{\rm(norm.)}$,
which is consistent with unity taking into account the
normalisation uncertainties of $6\%$ and $^{+11}_{-\ \,7}\%$ for the H1 FPS and ZEUS LPS data, respectively.
Within the errors, there is no strong $\xpom$, $\beta$ or $Q^2$ dependence of the ratio. 
The FPS data extend the kinematic range of the cross section measurement to higher $Q^2$ and low $\beta$.

\begin{figure}[p] \unitlength 1mm
 \begin{center}
 \begin{picture}(160,190)
    \put(0,0){\epsfig{file=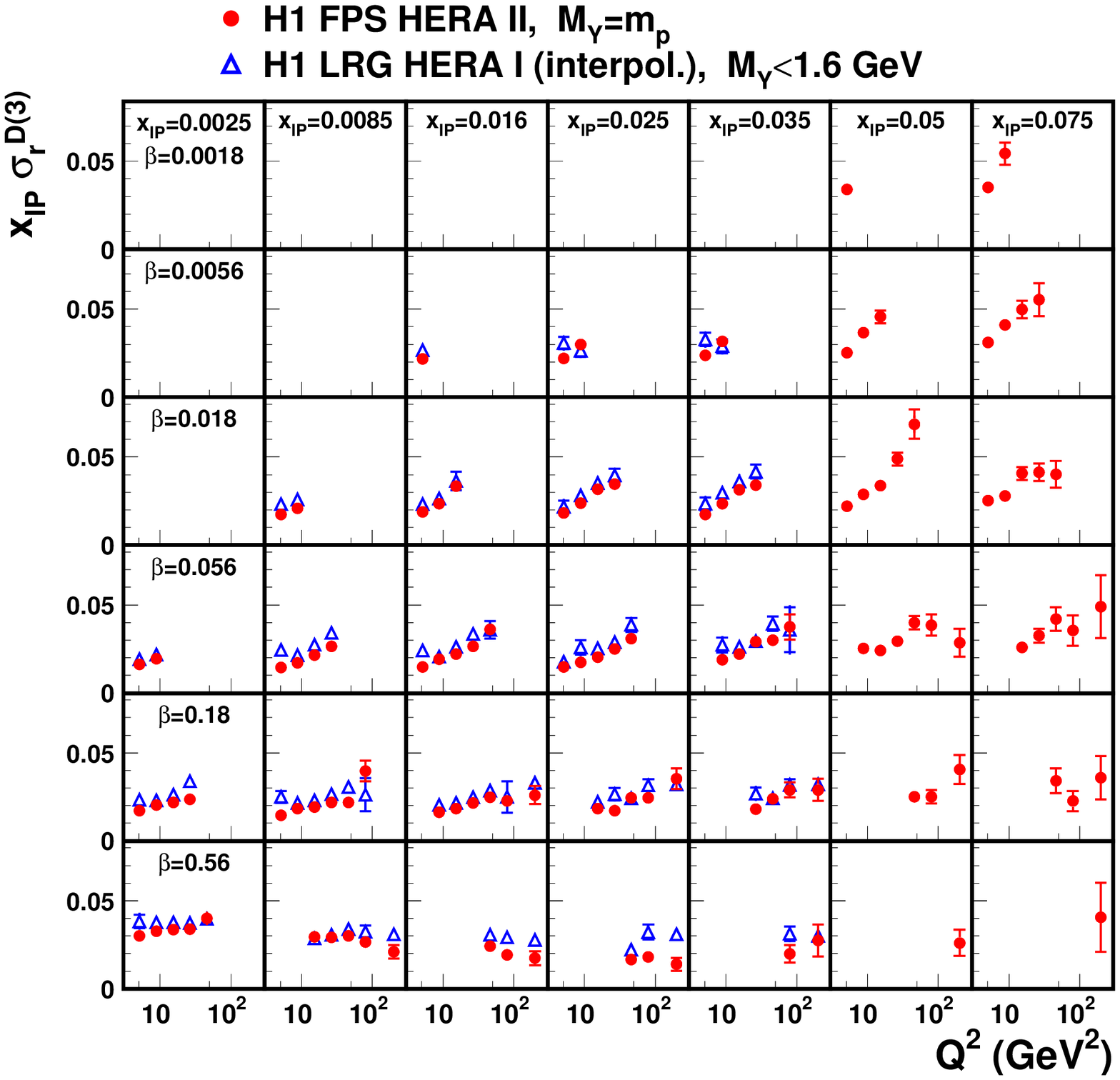 ,width=\linewidth}}
 \end{picture}
 \end{center}
 \caption{The  reduced diffractive  cross section
$\xpom \, \sigma_r^{D(3)}(\beta,Q^2,\xpom)$
for $|t| < 1 \ {\rm GeV^2}$,
shown as a function of $Q^2$ for
different values of $\xpom$ and $\beta$.
The error bars indicate the statistical and systematic
errors added in quadrature.
The H1 FPS data are compared
with the H1 LRG results interpolated to the FPS $\beta,Q^2, \xpom$ values~\cite{H1LRG}. The overall
normalisation uncertainty of $6\%$ on the FPS data and the normalisation uncertainty of $6.2\%$ on the LRG data are not 
shown.}
\label{fig:f2d3q2_h1lrg}
\end{figure}

\begin{figure}[p] \unitlength 1mm
 \begin{center}
\vspace{-0.5cm}

 \begin{picture}(160,190)
    \put(0,0){\epsfig{file=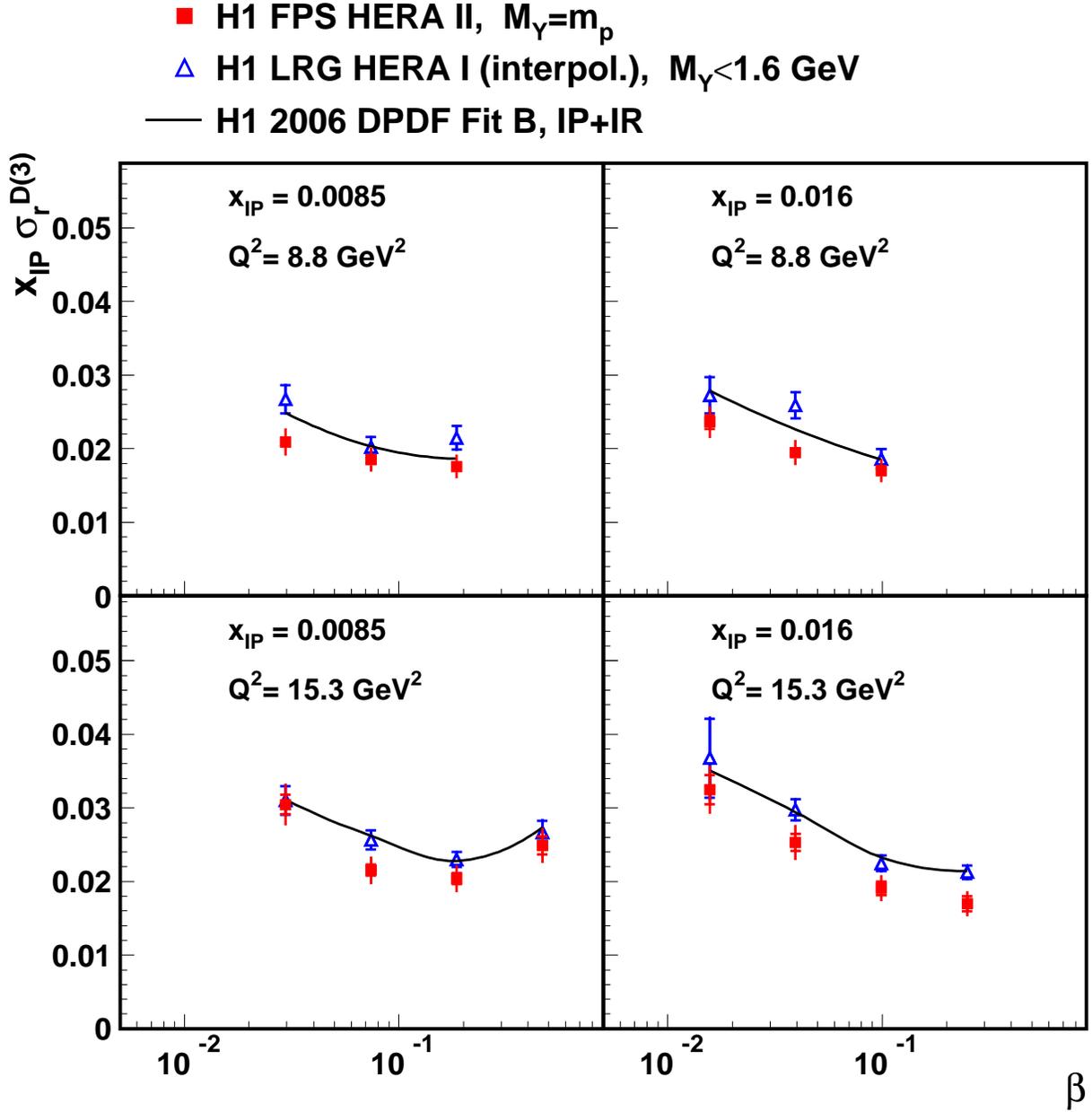 ,width=\linewidth}}
 \end{picture}
 \end{center}
 \caption{The  reduced diffractive  cross section
 $\xpom \, \sigma_r^{D(3)}(\beta,Q^2,\xpom)$
 for $|t| < 1 \ {\rm GeV^2}$,
 shown as a function of $\beta$ for
 selected values of $\xpom$ and $Q^2$.
 The inner error bars represent the statistical errors.
 The outer error bars indicate the statistical and systematic
 errors added in quadrature. The H1 FPS data are compared
 with the H1 LRG results~\cite{H1LRG} interpolated to the FPS  $\beta,Q^2, \xpom$ values.
 The solid curves represent H1 2006 DPDF Fit B to the LRG data. The overall 
normalisation uncertainty of $6\%$ on the FPS data and the normalisation uncertainty of $6.2\%$ on the LRG data are not
shown.}
 \label{fig:f2d3beta_sel}
\end{figure}

The reduced cross section $\sigma_r^{D(3)}$ can be compared with H1 measurements
obtained using the LRG technique~\cite{H1LRG} 
after taking into account the slightly different cross section
definitions in the two cases. 
The cross section $ep \rightarrow eXY$ measured with the
LRG data is defined to
include proton dissociation to any system $Y$ with a mass in
the range $M_Y<1.6~\GeV$, whereas $Y$ is defined to be a proton in the
cross section measured with the FPS.
The results on $\xpom \,\sigma_r^{D(3)}$ measured using the FPS and LRG methods are shown in 
figure~\ref{fig:f2d3q2_h1lrg} as a function $Q^2$ in bins of 
$\beta$ and $\xpom$,
 and in figure~\ref{fig:f2d3beta_sel} as a function of $\beta$ in selected bins of $Q^2$ and $\xpom$.
The kinematic range of the measurements is extended to higher $\xpom$. The experimental 
uncertainties of the two measurements are defined as the quadratic sum
of the statistical and uncorrelated systematic uncertainties.  As can be seen in figure~\ref{fig:f2d3beta_sel}, 
the present FPS measurement has a precision comparable to the 
measurement~\cite{H1LRG} obtained using the LRG method. 
The LRG results are interpolated to the $Q^2$, $\beta$ and $\xpom$ bin centre values
of the FPS data using the parameterisation H1 2006 DPDF Fit~B~\cite{H1LRG}.

Since the two data sets are statistically independent and
the dominant sources of systematic errors are different, correlations between
the uncertainties of the FPS and LRG data are negligible.
The ratio of the two measurements is formed for each ($Q^2$, $\beta$, $\xpom$) point
in the range $\xpom < 0.04$, where LRG data are available.
The dependence of this ratio on each kinematic variable is studied by taking
statistically weighted averages over the other two variables.

\begin{figure}[p] \unitlength 1mm
 \begin{center}
\vspace{-0.5cm}

 \begin{picture}(160,190)
    \put(-3,95){\epsfig{file=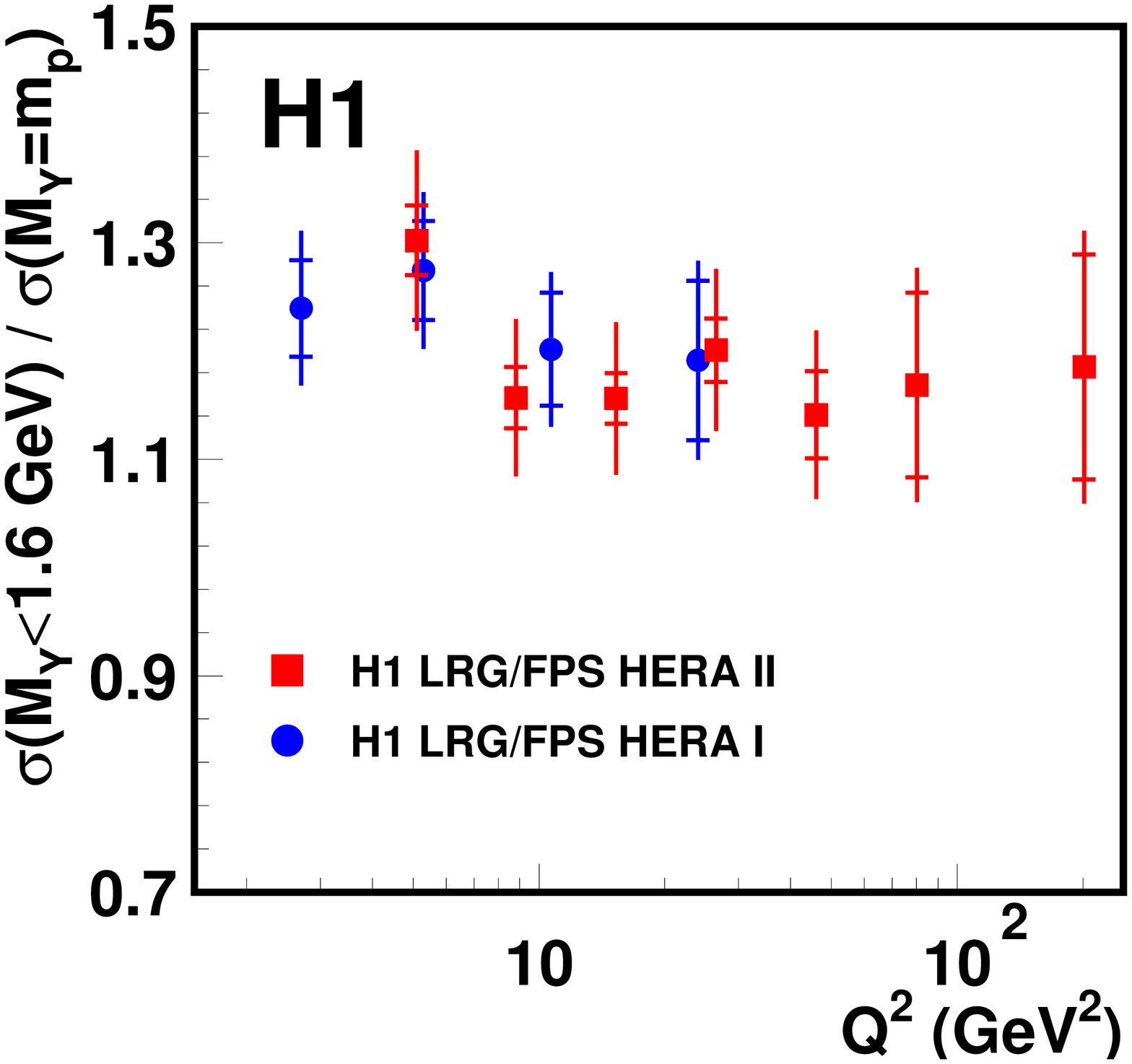,width=0.5\linewidth}}
    \put(80,95){\epsfig{file=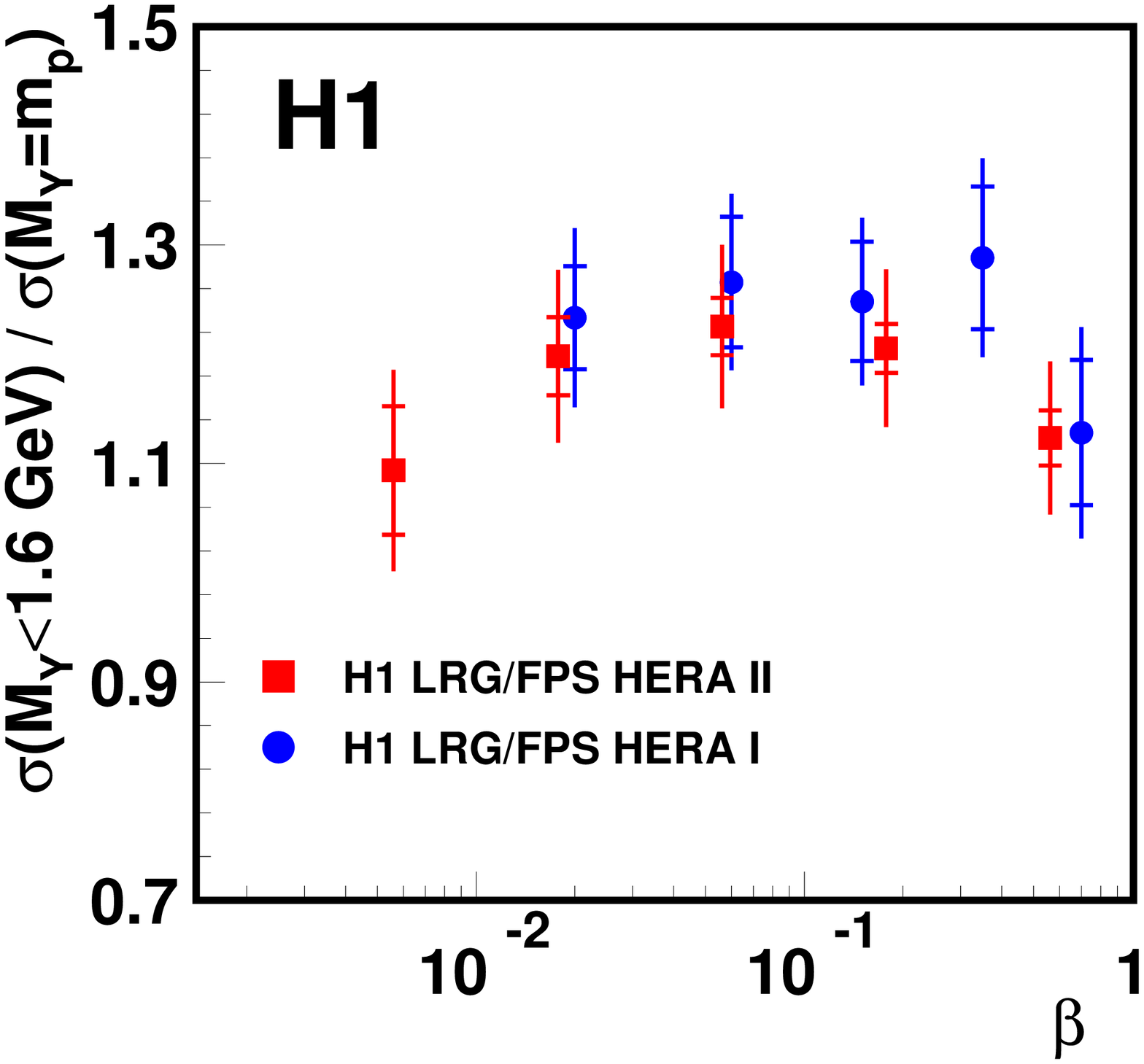,width=0.5\linewidth}}
    \put(40,0){\epsfig{file=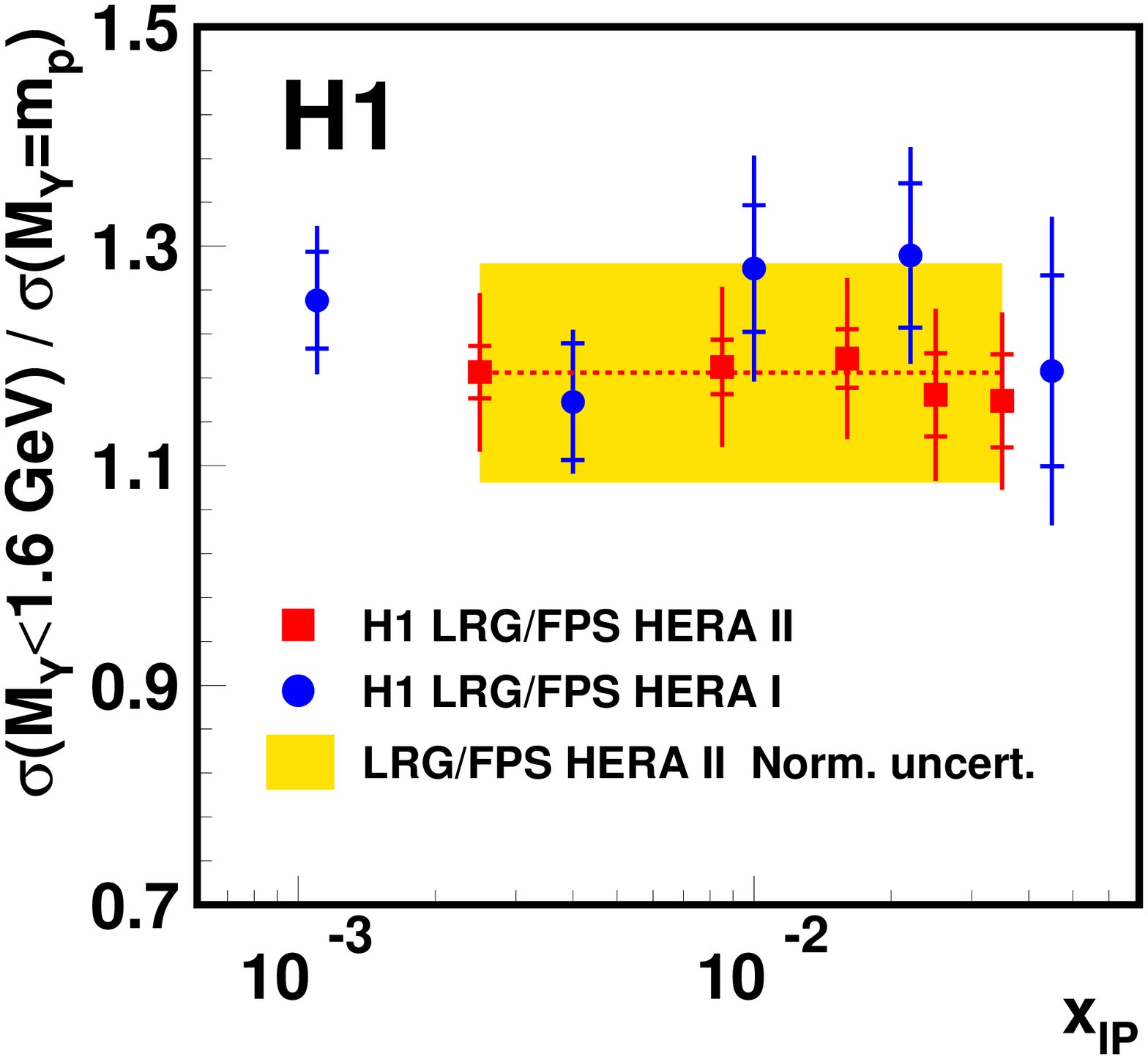,width=0.5\linewidth}}
    \put(65,160){\bf{\large{(a)}}}
    \put(148,160){\bf{\large{(b)}}}
    \put(108,65){\bf{\large{(c)}}}
 \end{picture}
 \end{center}
 \caption{The ratio of the reduced diffractive cross section  $\sigma_r^{D(3)}$ for $M_Y < 1.6 \ {\rm GeV}$ and
$|t| < 1 \ {\rm GeV^2}$ obtained using the H1 LRG data~\cite{H1LRG} to that for $M_Y = m_p$ and $|t| < 1 \ {\rm GeV^2}$,
obtained from the present and previously published FPS data~\cite{H1FPS}.
The results are shown
as a function of (a) $Q^2$, (b) $\beta$ and (c) $\xpom$, after
averaging over the other variables.
The inner error bars represent the statistical errors.
The outer error bars indicate the statistical and uncorrelated systematic errors added in
quadrature. The combined normalisation uncertainty of $8.5\%$ is shown 
as a band in figure (c).
The dashed line in figure (c) represents the result of a fit to the data in the region shown assuming no
dependence on $\xpom$.}
\label{fig:LRGratio}
\end{figure}

The ratio of the LRG to the FPS cross section is shown in
figure~\ref{fig:LRGratio} as a function of $Q^2$, $\beta$ and $\xpom$.
Within the uncorrelated uncertainties of typically $6\%$ per data point,
there is no significant dependence on $\beta$, $Q^2$ or $\xpom$.
The ratio of overall normalisations, LRG / FPS, is 
$1.18 \pm 0.01 {\rm (stat.)} \pm 0.06 {\rm (uncor.syst.)} \pm 0.10 {\rm (norm.)}$, 
the dominant uncertainties arising from the normalisations of the FPS and LRG data.
This result is in agreement within uncertainties with the value of $1.23 \pm 0.03 {\rm (stat.)} \pm 0.16 {\rm (syst.)}$
obtained from the previously published H1 LRG and FPS data~\cite{H1FPS}.  
Combining the result of~\cite{H1FPS} with the present measurement leads to a more precise value 
of the cross section ratio: 
\begin{eqnarray}
\frac{\sigma (M_Y < 1.6 \ {\rm GeV})}{\sigma (M_Y = m_p)} =
1.20 \ \pm \ 0.11 \ {\rm (exp.)} \ . \ 
\label{eq:myratio}
\end{eqnarray}
where the experimental uncertainty is a combination of the statistical, uncorrelated systematic and 
normalisation 
uncertainties of the measurements. 
The result is consistent with the prediction
of $1.15 _{-0.08}^{+0.15}$ from the DIFFVM generator, where the total
proton elastic and proton dissociation cross sections are taken to
be equal for the central value and their ratio is varied in the range $1:2$ to $2:1$ for the uncertainties \cite{H1LRG,DIFFVM}.

The good agreement, after accounting for proton dissociation,
between the LRG
and the FPS data confirms that the two measurement methods lead
to compatible results, despite
their very different systematics.
The lack of any kinematic dependence of the ratio of the two cross
sections shows, within the uncertainties, that proton dissociation
with $M_Y < 1.6 \ {\rm GeV}$ can be treated in a similar way to the
elastic proton case. 
The result confirms that contributions from proton dissociation in the
LRG measurement do not significantly alter the measured $\beta$,
$Q^2$ or $\xpom$ dependences and hence cannot have a large influence
on the diffractive parton densities extracted from the LRG data up to the normalisation difference.

\subsection{Cross section dependence on
{\boldmath $Q^2$} and {\boldmath $\beta$}}
\label{f2d3sec}

\begin{figure}[p] \unitlength 1mm
 \begin{center}
 \begin{picture}(160,190)
    \put(0,0){\epsfig{file=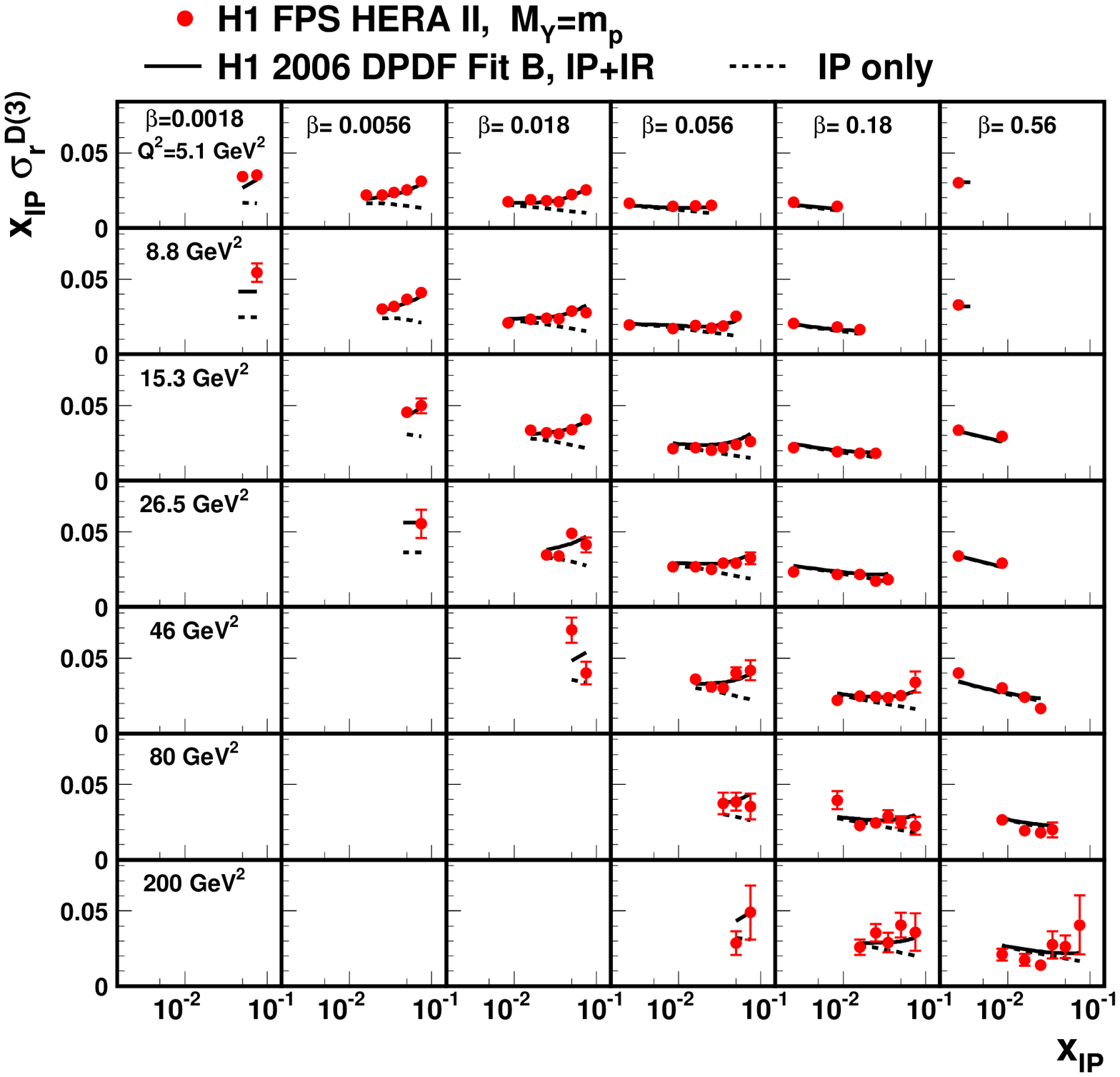 ,width=\linewidth}}
 \end{picture}
 \end{center}
 \caption{The  reduced diffractive cross section
$\xpom \, \sigma_r^{D(3)}(\beta,Q^2,\xpom)$
for $|t| < 1 \ {\rm GeV^2}$,
shown as a function of $\xpom$ for
different values of $\beta$ and $Q^2$.
The error bars indicate the statistical and systematic
errors added in quadrature. The overall
normalisation uncertainty of $6\%$ is not shown.
The solid curves represent the results of the H1 2006 DPDF
Fit B to the LRG data \cite{H1LRG} reduced by a global factor $1.20$ to correct for the contributions of proton 
dissociation 
processes. The dashed curves indicate the contribution of pomeron exchange in 
this model.}
\label{fig:f2d3xp_fit}
\end{figure}

\begin{figure}[p] \unitlength 1mm
 \begin{center}
 \begin{picture}(160,190)
    \put(0,0){\epsfig{file=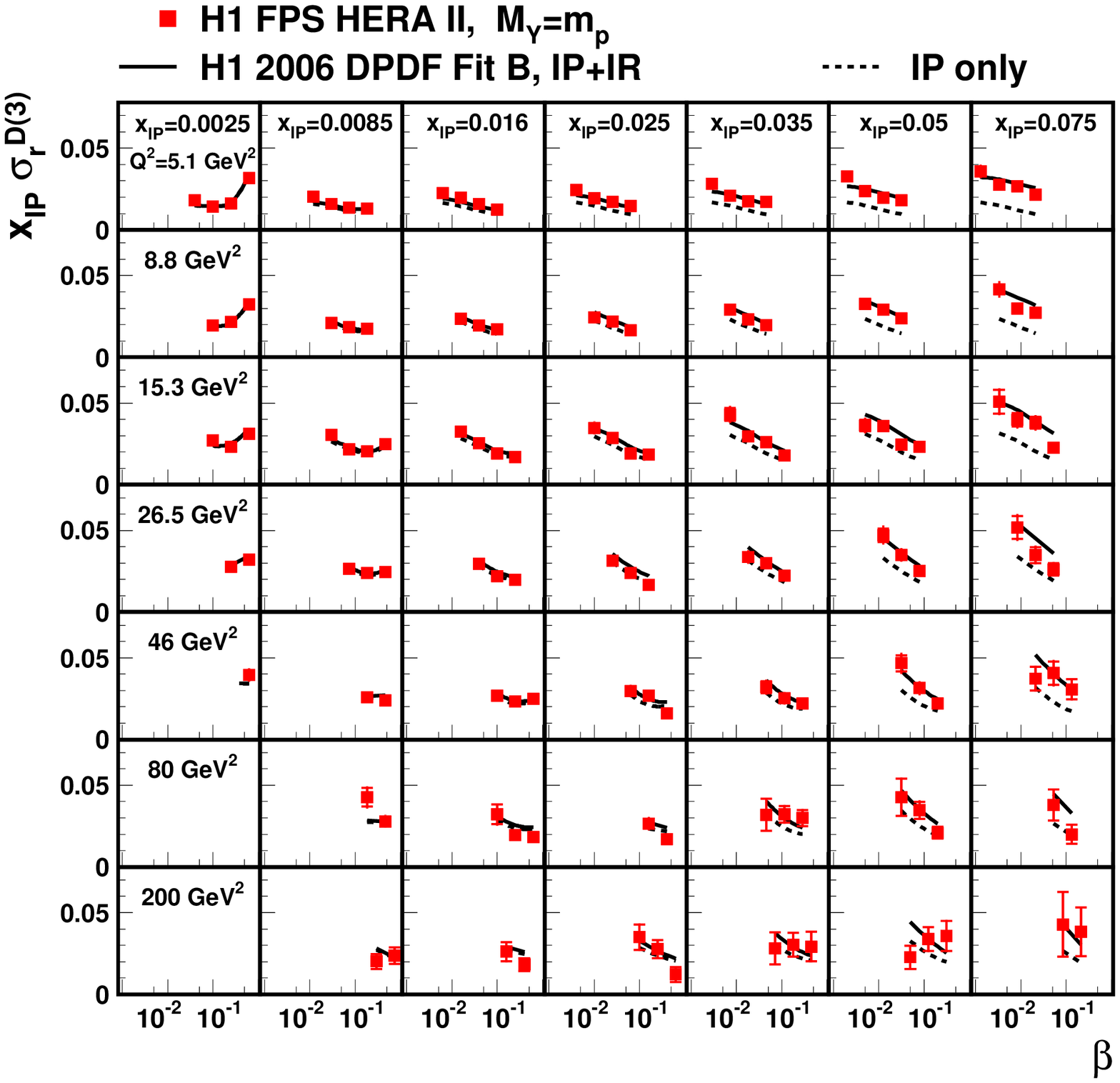 ,width=\linewidth}}
 \end{picture}
 \end{center}
 \caption{The  reduced diffractive  cross section
$\xpom \, \sigma_r^{D(3)}(\beta,Q^2,\xpom)$
for $|t| < 1 \ {\rm GeV^2}$,
shown as a function of $\beta$ for
different values of $\xpom$ and $Q^2$.
The error bars indicate the statistical and systematic
errors added in quadrature. The overall
normalisation uncertainty of $6\%$ is not shown.
The solid curves represent the results of the H1 2006 DPDF
Fit B to the LRG data \cite{H1LRG} reduced by a global factor $1.20$ to correct for the contributions of proton 
dissociation
processes. 
The dashed curves indicate the contribution of pomeron exchange in
this model.}
\label{fig:f2d3beta_fit}
\end{figure}

\begin{figure}[p] \unitlength 1mm
 \begin{center}
 \begin{picture}(160,190)
    \put(0,0){\epsfig{file=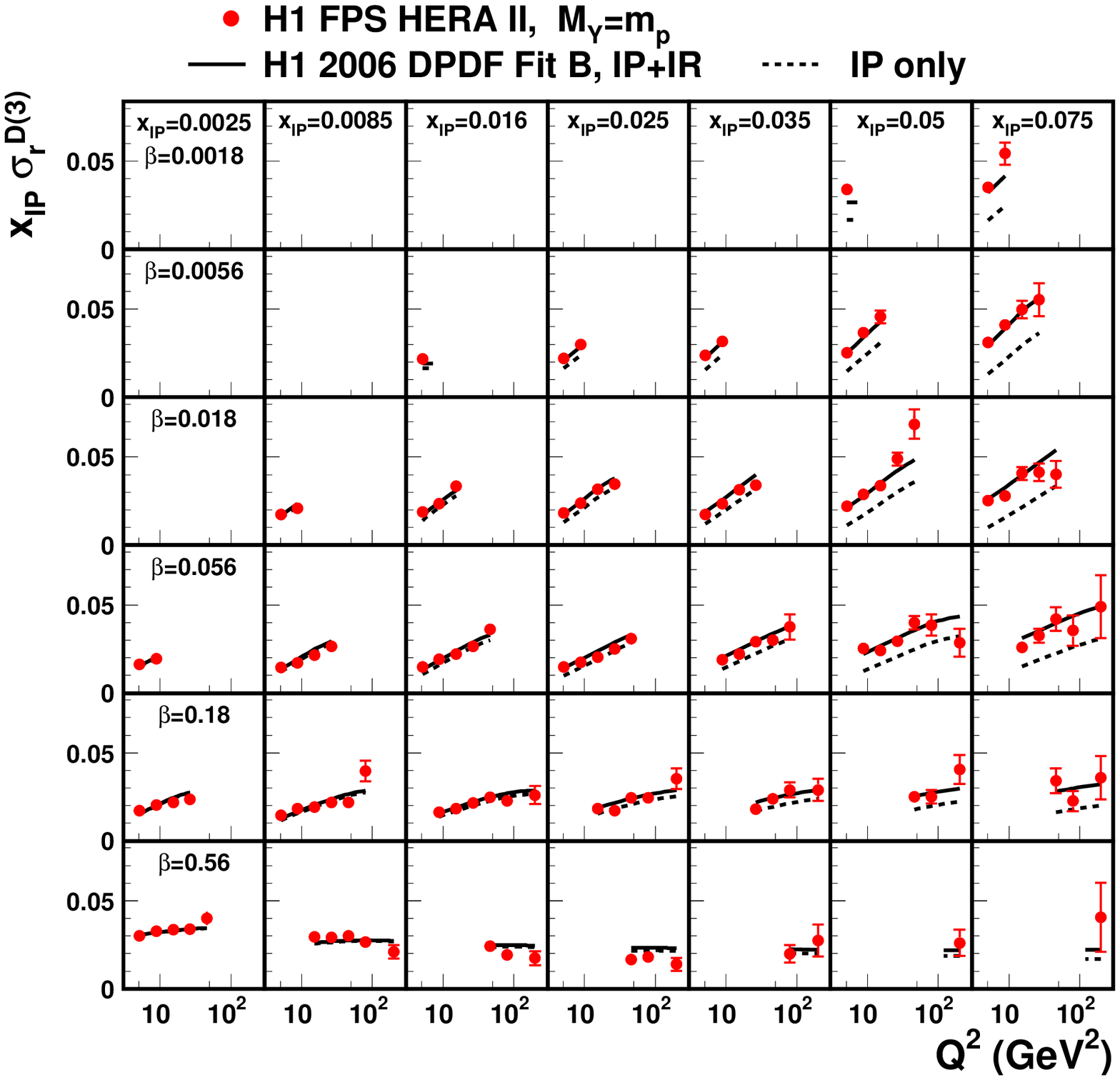 ,width=\linewidth}}
 \end{picture}
 \end{center}
 \caption{The  reduced diffractive  cross section
$\xpom \, \sigma_r^{D(3)}(\beta,Q^2,\xpom)$
for $|t| < 1 \ {\rm GeV^2}$,
shown as a function of $Q^2$ for
different values of $\xpom$ and $\beta$.
The error bars indicate the statistical and systematic
errors added in quadrature. The overall
normalisation uncertainty of $6\%$ is not shown.
The solid curves represent the results of the H1 2006 DPDF
Fit B to the LRG data \cite{H1LRG} reduced by a global factor $1.20$ to correct for the contributions of proton 
dissociation
processes. The dashed curves indicate the contribution of pomeron exchange in 
this model.}
\label{fig:f2d3q2_fit}
\end{figure}

The measured reduced cross sections 
$\xpom \, \sigma_r^{D(3)}$ are
presented in figures~\ref{fig:f2d3xp_fit}-\ref{fig:f2d3q2_fit} and in table~\ref{table:f2d3} as a function of $\xpom$, $\beta$ and $Q^2$.
The FPS data are compared with QCD predictions at next-to-leading order derived from the
H1 2006 DPDF Fit B to the H1 LRG cross sections~\cite{H1LRG}, 
which include both the pomeron and a sub-leading exchange. The 
normalisation of the
H1 2006 DPDF Fit B predictions is reduced by a global factor $1.20$ to correct for the
contributions of proton dissociation processes to the LRG cross sections, as evaluated in 
section~\ref{f2d3}. 

As can be seen in figure~\ref{fig:f2d3xp_fit}, the rise of the data at large $\xpom$ is in 
agreement with a significant contribution from a sub-leading exchange.
The reduced cross section $\sigma_r^{D(3)}$ shown in figure~\ref{fig:f2d3beta_fit} decreases with $\beta$ over most of the 
kinematic range. However, it clearly 
rises as $\beta \rightarrow 1$ at low $Q^2$ and $\xpom$. Within the framework of DPDFs, this can be explained in terms of 
diffractive quark densities peaking at high fractional momentum at low $Q^2$ \cite{H1LRG,H1Diff94}. 

\begin{figure}[p] \unitlength 1mm
 \begin{center}
 \begin{picture}(160,190)
    \put(0,0){\epsfig{file=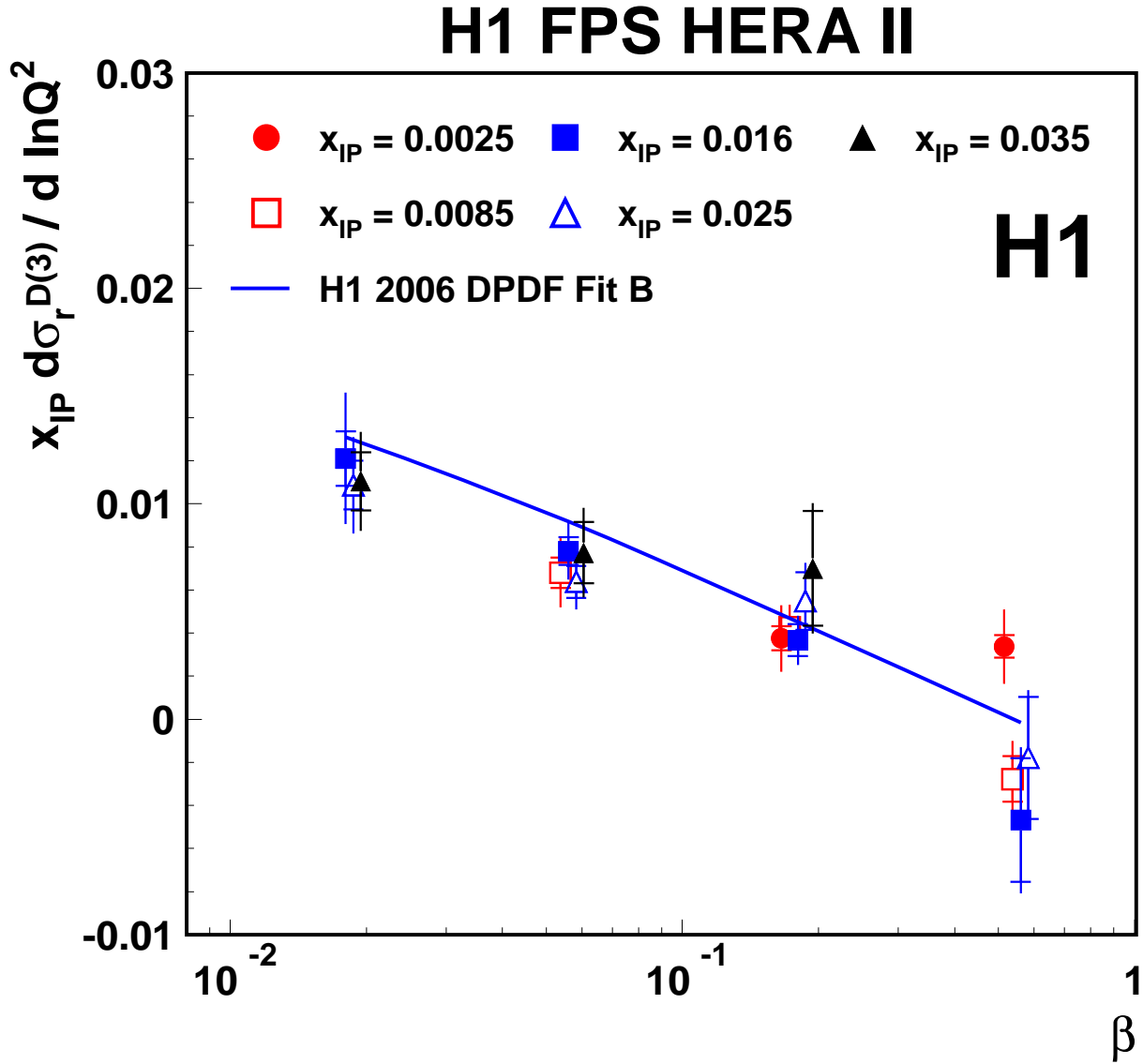 ,width=\linewidth}}
 \end{picture}
 \end{center}
 \caption{The logarithmic $Q^2$ derivative of the reduced diffractive  cross section
 $\xpom \, \sigma_r^{D(3)}(\beta,Q^2,\xpom)$ at different fixed values of $\xpom$ and $\beta$.
 The inner error bars represent the statistical errors.
 The outer error bars indicate the statistical and systematic
 errors added in quadrature. The solid curve represents the results of the H1 2006 DPDF
 Fit B~\cite{H1LRG} at $\xpom =0.016$ reduced by a global factor $1.20$ to correct for the contributions of proton dissociation
 processes.} 
\label{fig:sigmafitq2}
\end{figure}

The figure~\ref{fig:f2d3q2_fit} shows the $Q^2$ dependence of $\sigma_r^{D(3)}$ at fixed $\xpom$ and
$\beta$.
Positive scaling
violations ($\partial \, \sigma_r^{D(3)} / \partial \, {\rm ln}~Q^2>0$)
are observed throughout the kinematic range, except at
the highest $\beta = 0.56$.
This observation is consistent with previous H1
measurements using the LRG
method~\cite{H1Diff94,H1LRG} and
implies a large gluonic component in the DPDFs.
As can be seen from QCD predictions,
the positive scaling violations
may be attributed to the pomeron contribution
even at the highest $\xpom$ values, where the sub-leading
exchange is largest.
The $Q^2$ dependence is quantified by fitting the data at fixed $x_{\pom}$ and $\beta$ to the form
\begin{eqnarray}
x_{\pom}\sigma^{D(3)}_r ( \beta, x_{\pom}, Q^2) = a_D(\beta, x_{\pom}) + b_D(\beta, x_{\pom}) \ln (Q^2/1~\GeV^2)
\label{eq:q2slope}
\end{eqnarray}
\noindent such that $b_D(\beta, x_{\pom})$
is the derivative of the reduced
cross section with respect to $\ln Q^2$ multiplied by $\xpom$. This form is fitted to data points 
 for which the $\xpom$ bin centre values satisfy $\xpom \leq 0.035$ and for
  which the $\beta$ bin contains at least 3 data points. The
sub-leading exchange contribution at $\xpom = 0.035$ is below $15\%$.
The resulting logarithmic derivatives are shown 
in figure \ref{fig:sigmafitq2}.  Although the logarithmic derivatives at
different $x_{\pom}$ values cover different $Q^2$ regions, they are similar when viewed as a
function of $\beta$. This confirms the applicability of the proton vertex factorisation framework to
the description of the current data. The FPS results  are consistent with predictions derived from the
H1 2006 DPDF Fit B to the H1 LRG data also shown in figure \ref{fig:sigmafitq2}.

\subsection{Comparison of the diffractive and inclusive DIS cross sections}
\label{inclcomp}

\begin{figure}[p] \unitlength 1mm
 \begin{center}
 \begin{picture}(160,190)
    \put(0,0){\epsfig{file=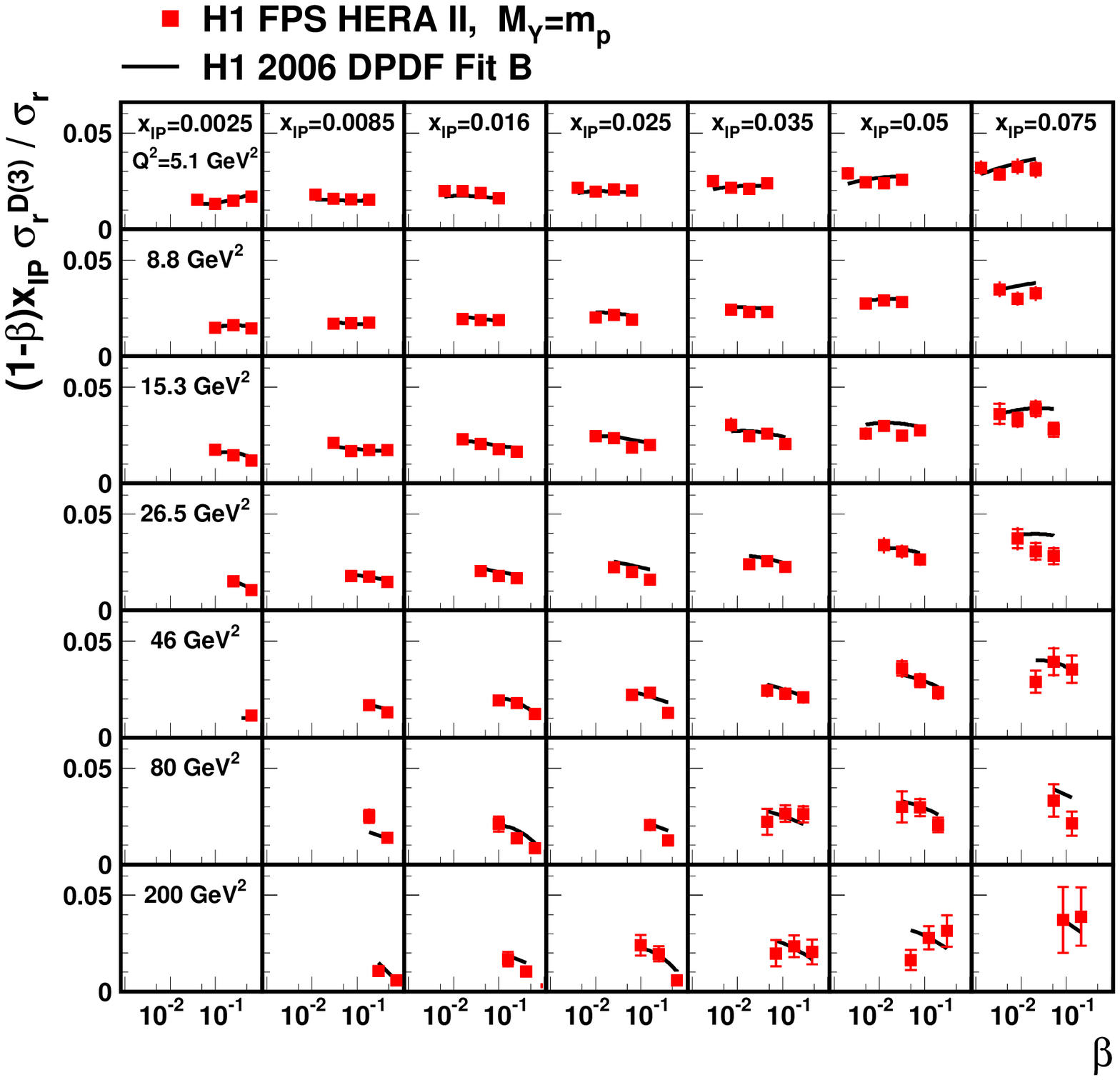 ,width=\linewidth}}
 \end{picture}
 \end{center}
 \caption{The ratio of the reduced diffractive  cross section 
 $\sigma_r^{D(3)}(\beta,Q^2,\xpom)$
 to the reduced inclusive cross section $\sigma_r(x=\beta \xpom,Q^2)$ 
 obtained using the parameterisation H1PDF 2009, 
 multiplied by $(1-\beta)\xpom$,  
 shown as a function of $\beta$ for
 different values of $\xpom$ and $Q^2$.
 The error bars indicate the statistical and systematic
 errors added in quadrature. The solid curves represent predictions obtained using the H1 2006 DPDF
 Fit B for the diffractive cross sections and H1PDF 2009 set for the inclusive cross sections. The 
 results for the ratio derived from the PDF predictions are reduced by a global factor $1.20$ to correct
 for the contribution of proton dissociation processes.}
\label{fig:ratioincl}
\end{figure}

By analogy with hadronic scattering, the diffractive and the total
cross sections in DIS can be related via the generalisation of the optical theorem to virtual photon 
scattering~\cite{opttheorem}.
Many models of low $x$ DIS~\cite{lowx,lowx2,lowx3,lowx4,lowx5,lowx6} assume links between these quantities.
Comparing the $Q^2$ and $x$ dynamics of the diffractive with the inclusive cross
section is therefore a powerful means of testing models and of
comparing the properties of the DPDFs with their inclusive counterparts. A detailed comparison of
the diffractive and inclusive cross section is performed in~\cite{H1LRG}. 
Following~\cite{H1LRG}, the evolution of the reduced diffractive cross section with $Q^2$ is 
compared with that of the
reduced inclusive DIS cross section $\sigma_r$  by forming the quantity
$(1-\beta)\frac{\xpom\sigma_r^{D(3)}(\xpom,\beta,Q^2)}
{\sigma_r(x=\beta \xpom,Q^2)}$
at fixed $Q^2, \beta$ and $\xpom$, using a parameterisation of the $\sigma_r$ data from~\cite{f2incl}.
 This quantity is equivalent to the ratio of
diffractive to inclusive DIS cross section $M_X^2 \frac{{\rm d}\sigma^{D(3)}(M_X,W,Q^2)}{{\rm d}M_X^2} / \sigma_{incl}^{\gamma*p 
\rightarrow X}(W,Q^2)$ studied
in~\cite{ZEUSMX,ZEUSMX2,ZEUSLPS} as a function of the $\gamma^*p$ centre of mass energy $W$ and $Q^2$ in ranges of
$M_X$. The ratio  is shown in figure~\ref{fig:ratioincl} as a function of $\beta$ at fixed $\xpom$
and $Q^2$.

The ratio of the diffractive to the inclusive cross section is 
approximately constant as a function of $\beta$ at fixed $Q^2$ and $\xpom$ except at high $\beta$.
As can be seen in
figure~\ref{fig:ratioincl}, the decrease of the ratio towards high $\beta$ is reproduced by QCD predictions based on 
diffractive and inclusive proton PDFs~\cite{H1LRG,f2incl}. 
The ratio also rises towards larger  
values of $\xpom$ where the sub-leading exchange contribution to the diffractive cross section is not negligible.

\begin{figure}[p] \unitlength 1mm
 \begin{center}
 \begin{picture}(160,190)
    \put(0,0){\epsfig{file=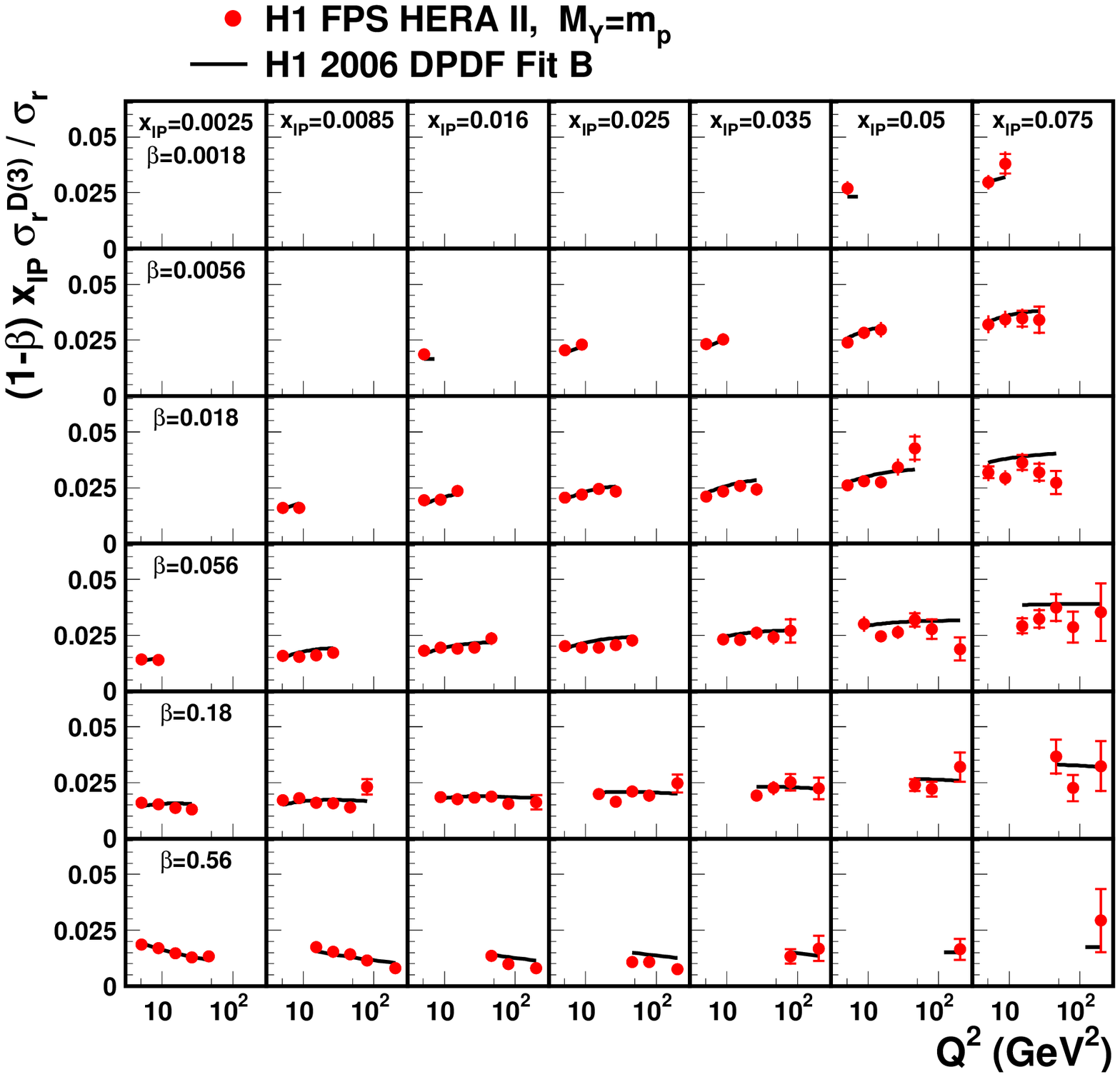 ,width=\linewidth}}
 \end{picture}
 \end{center}
 \caption{The ratio of the reduced diffractive  cross section 
 $\sigma_r^{D(3)}(\beta,Q^2,\xpom)$
 to the reduced inclusive cross section $\sigma_r(x=\beta \xpom,Q^2)$ 
 obtained using the parameterisation H1PDF 2009, 
 multiplied by $(1-\beta)\xpom$,  
 shown as a function of $Q^2$ for
 different values of $\xpom$ and $\beta$.
 The error bars indicate the statistical and systematic
 errors added in quadrature. 
 The solid curves represent predictions obtained using the H1 2006 DPDF
 Fit B for the diffractive cross sections and H1PDF 2009 set for the inclusive cross sections.
 The results for the ratio derived from the 
 PDF predictions are reduced by a global factor $1.20$ to correct 
 for the contribution of proton dissociation processes.}
\label{fig:ratioinclq2}
\end{figure}

\begin{figure}[p] \unitlength 1mm
 \begin{center}
\vspace*{-0.5cm}

 \begin{picture}(160,190)
    \put(0,0){\epsfig{file=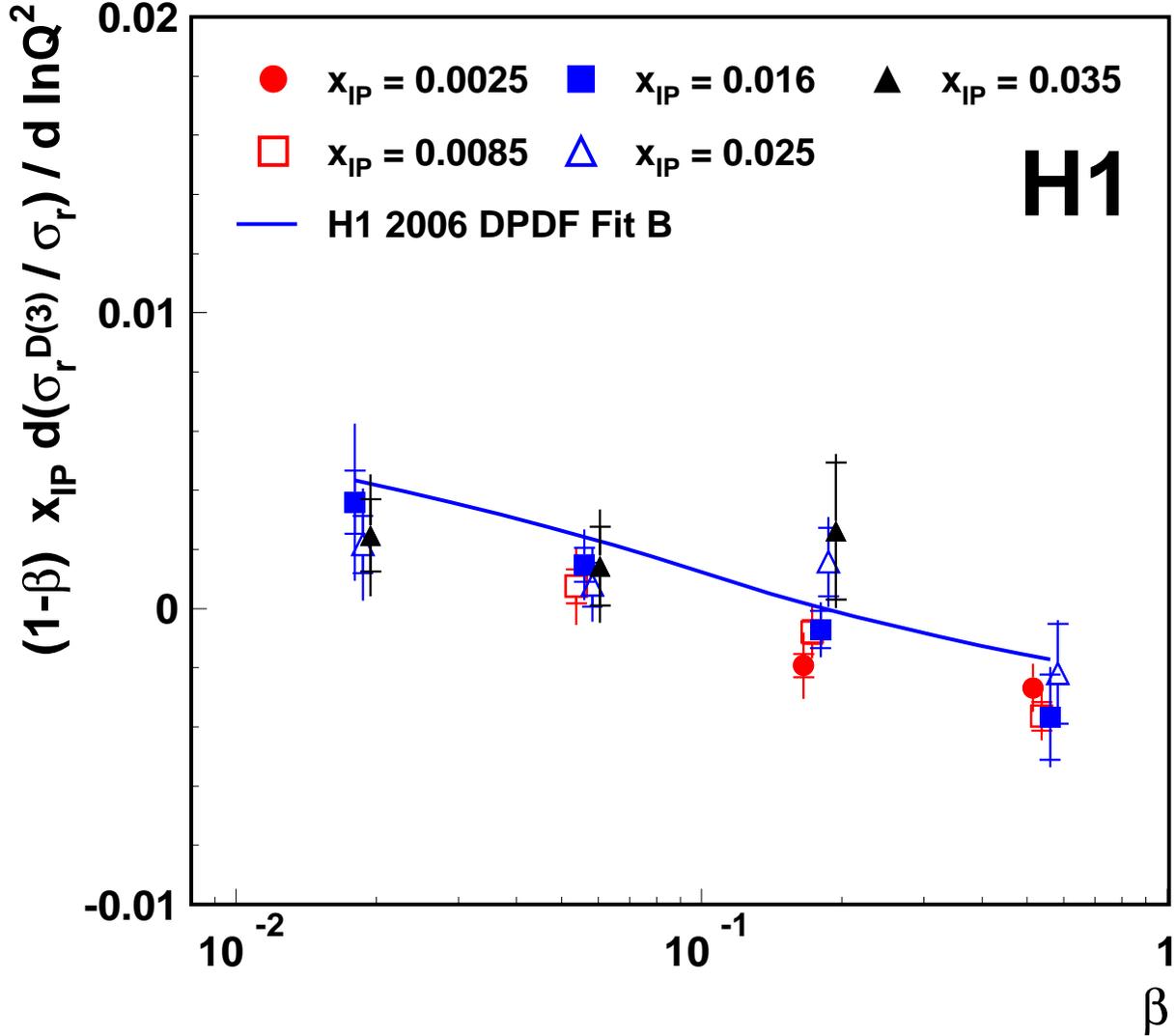 ,width=\linewidth}}
 \end{picture}
 \end{center}
\vspace*{-0.5cm}

 \caption{The logarithmic $Q^2$ derivative of the ratio of the reduced diffractive cross section
 $\sigma_r^{D(3)}(\beta,Q^2,\xpom)$ to the reduced inclusive  cross section 
 $\sigma_r(x=\beta \xpom,Q^2)$ 
 obtained using the parameterisation H1PDF 2009
 multiplied by $(1~-~\beta)~\xpom$, shown at different fixed values of $\xpom$ and $\beta$.
 The inner error bars represent the statistical errors.
 The outer error bars indicate the statistical and systematic
 errors added in quadrature.  
 The solid curves represent predictions at $\xpom =0.016$ obtained using the H1 2006 DPDF
 Fit B for the diffractive cross sections and H1PDF 2009 set for the inclusive cross sections.
 The results for the ratio derived from the 
 PDF  predictions are reduced by a global factor $1.20$ to correct
 for the contribution of proton dissociation processes.} 
\label{fig:ratiofitq2}
\end{figure}

The ratio, shown in figure~\ref{fig:ratioinclq2} as a function of $Q^2$ at fixed $\xpom$ and $\beta$,
depends only weakly on $Q^2$ for most $\beta$ and $\xpom$ values. 
In order to compare the $Q^2$ dependences of the
diffractive and the inclusive cross sections quantitatively, the derivative $b_r$ of the ratio with respect to $\ln Q^2$
is extracted through fits of the form $a_r(\beta,\xpom)+b_r(\beta,\xpom)\cdot\ln(Q^2/1~\GeV^2)$.
To reduce the influence of the sub-leading exchange contribution, only data points with bin centres at $\xpom \leq 
0.035$ are included in the analysis of $b_r$. 
The resulting values of $b_r$ are shown in figure~\ref{fig:ratiofitq2}. 
They are consistent with zero within $3\sigma$. 
At fixed $\beta$, there is no significant dependence of the logarithmic derivative on
$\xpom$. 
Whereas the reduced diffractive and inclusive cross sections are
closely related to their respective quark densities, the logarithmic
derivatives are approximately proportional to the relevant gluon
densities in regions where the $Q^2$ evolution is dominated by the
$g \to \bar{q}q$ splitting~\cite{Pritz}. The compatibility of
$b_r$ with zero thus implies that the ratio of
the quark to the gluon density is similar in the diffractive and
inclusive DIS when considered at the same low $x$ value.
As can be seen in
figure~\ref{fig:ratiofitq2}, QCD predictions based on proton PDFs extracted in diffractive and inclusive DIS~\cite{H1LRG,f2incl} reproduce a weak 
decrease of the logarithmic derivative towards larger $\beta$.

\section{Summary}

A cross section measurement is presented
for the diffractive deep-inelastic scattering
process $ep \rightarrow eXp$.
The results are obtained using high statistics data taken with the H1 detector at HERA.
In the process studied, the scattered
proton carries at least $90\%$
of the incoming proton momentum and is measured in the Forward Proton
Spectrometer (FPS). The data
 cover the range $\xpom <0.1$ in fractional proton longitudinal momentum
 loss, $0.1 < |t| < 0.7~{\rm GeV}^2$
 in squared four-momentum transfer at the proton vertex, $4 < Q^2 < 700~{\rm GeV}^2$
 in photon virtuality and
 $0.001 < \beta = x / \xpom < 1$. 
The measurement is performed in the range of the inelasticity variable 
$0.03 < y < 0.7$ for $4 < Q^2 < 110~\GeV^2$  and in the range $0.03 < y < 0.8$ for $120 < Q^2 < 700~\GeV^2$.
The new H1 FPS data are in good agreement with the previously published H1 FPS results  and are consistent within 
uncertainties with results of the ZEUS collaboration obtained with their Leading Proton Spectrometer.
The new measurements extend the kinematic range to higher $Q^2$ values.

The reduced diffractive cross section
$\sigma_r^{D(4)}(\beta,Q^2,\xpom,t)$ is measured. The
$\xpom$ and $t$ dependences are described using
a model which is motivated by Regge phenomenology, in which a leading
pomeron and a sub-leading exchange contribute. 
The effective pomeron intercept describing the data is
$\alpha_{\pom}(0) = 1.10~\pm~0.02~{\rm (exp.)}~\pm~0.03~{\rm (model)}$, 
which is compatible within uncertainties with
 the pomeron intercept measured in soft hadron-hadron scattering.
The slope of the pomeron trajectory $\alpha_{\pom}^{\prime}$ is consistent 
with zero and smaller than
 the value $\sim 0.25~{\rm GeV}^{-2} $ obtained from soft hadron-hadron scattering data. The $t$-dependence of the 
pomeron exchange is described by an exponential function with constant slope parameter 
$B_{\pom} = 5.73~\pm~0.25~{\rm (exp.)~}_{-\,0.90}^{+\,0.80}~{\rm (model)}~\GeV^{-2}$. 
The measured values of the slope of the pomeron trajectory and the 
$t$-dependences are 
characteristic of 
diffractive hard scattering processes.
The $Q^2$ dependence of the parameters $\alpha_{\pom}(0), \alpha_{\pom}^{\prime}$ and $B_{\pom}$ is studied. The logarithmic derivatives are consistent with zero within 
$1.5\sigma$ of the experimental uncertainties thereby supporting the proton vertex factorisation
hypothesis.

The data are also analysed in terms of the
reduced diffractive cross section $\sigma_r^{D(3)}$, obtained by integrating
$\sigma_r^{D(4)}$ over the range $|t_{\rm min}| < |t| < 1~{\GeV}^2$.
At fixed $\xpom$, a decrease of $\sigma_r^{D(3)}$ with $\beta$ is
observed over most of the kinematic range, except at the lowest values of $Q^2$ and $\xpom$.
The data display positive scaling violations
except at the highest $\beta$ value of $0.56$. The size of the measured scaling violations implies a large gluonic
component in the diffractive parton distributions, in agreement with previous observations. 

The FPS results  are compared with those obtained from an earlier
H1 measurement using events selected on the basis of a large rapidity
gap  rather than a leading proton. This LRG measurement 
includes proton dissociation to states $Y$ with masses $M_Y < 1.6 \ {\rm GeV}$.
The FPS data extend the kinematic range to higher $\xpom$ and thus constrain the 
sub-leading exchange contribution. 
The ratio of the LRG to the FPS cross section is
$1.20~ \pm \ 0.11~{\rm (exp.)}$. It is
 independent of $Q^2$, $\beta$ and $\xpom$ within uncertainties,
confirming that contributions from proton dissociation in the LRG measurement do not significantly alter the measured $Q^2$, 
$\beta$ or $\xpom$ dependences. 

The ratio of the diffractive to the inclusive $ep$ cross sections is measured as a function
of $Q^2, \beta$ and $\xpom$. At fixed $\xpom$ the ratio depends only weakly on $Q^2$ or
on $\beta$ except at the highest $\beta$. QCD predictions based on diffractive and inclusive proton PDFs reproduce the 
behaviour of the ratio. This result 
implies that the ratio of quark to
gluon distributions at low $x$ is similar in the diffractive and inclusive processes.

\section*{Acknowledgements}

We are grateful to the HERA machine group whose outstanding
efforts have made this experiment possible.
We thank the engineers and technicians for their work in constructing and
maintaining the H1 detector, our funding agencies for
financial support, the DESY technical staff for continual assistance
and the DESY directorate for support and for the
hospitality which they extend to the non-DESY members of the
collaboration.


\renewcommand{\arraystretch}{1.35}

\newpage
\begin{landscape}
\begin{table}[ht]
\centering
\begin{tiny}
  \begin{tabular}{| c| c| c| c| c| c| c| c| c| c| c| c| c| c| c| c| c| c| c| c| c|}
\hline
 $Q^2$ & $\beta$ & $\xpom$ & $-t$ & $\xpom\sigma_r^{D(4)}$ & $\delta_{stat}$ & $\delta_{sys}$ & $\delta_{tot}$ & 
 $\delta_{had}$ & $\delta_{ele}$ & $\delta_{\theta}$ & $\delta_{\beta}$ & $\delta_{\xpom}$ & $\delta_{t}$ & 
 $\delta_{E_p}$ & $\delta_{p_x}$ & $\delta_{p_y}$ & $\delta_{Q^2}$ & $\delta_{vtx}$ & $\delta_{bgn}$ & $\delta_{bcc}$ \\
 $[\GeV^2]$ & & & $[\GeV^2]$ & $[\GeV^{-2}]$ & $[\%]$ & $[\%]$ & $[\%]$ & $[\%]$ & $[\%]$ & $[\%]$ & $[\%]$ & $[\%]$ & $[\%]$ & $[\%]$ & $[\%]$ & 
 $[\%]$ & $[\%]$ & $[\%]$ & $[\%]$ & $[\%]$ \\
\hline
$  5.1 $&$ 0.0018 $&$ 0.0500 $&$  0.2 $&$ 0.0689 $&$   7.6 $&$   8.7 $&$  11.5 $&$  -0.1 $&$   2.7 $&$   1.8 $&$  -1.2 $&$   0.0 $&$  -3.9 $&$   0.8 $&$   5.2 $&$   0.0 $&$   1.6 $&$  -3.5 $&$  -1.5 $&$   1.9 $\\
$  5.1 $&$ 0.0018 $&$ 0.0500 $&$  0.4 $&$ 0.0204 $&$  10.2 $&$   8.7 $&$  13.4 $&$  -0.6 $&$   1.3 $&$   0.8 $&$  -1.0 $&$   0.0 $&$   0.8 $&$   0.2 $&$   6.6 $&$   4.4 $&$   1.0 $&$  -1.7 $&$  -0.8 $&$   1.9 $\\
$  5.1 $&$ 0.0018 $&$ 0.0500 $&$  0.6 $&$ 0.0080 $&$  15.1 $&$   8.7 $&$  17.4 $&$   0.0 $&$   4.2 $&$  -4.2 $&$  -0.4 $&$  -0.1 $&$   0.3 $&$   2.5 $&$  -0.4 $&$  -3.6 $&$   3.3 $&$  -2.1 $&$  -0.9 $&$   1.9 $\\
$  5.1 $&$ 0.0018 $&$ 0.0750 $&$  0.2 $&$ 0.0786 $&$  11.3 $&$   7.3 $&$  13.5 $&$  -0.2 $&$   0.7 $&$   2.1 $&$  -0.3 $&$   0.1 $&$  -1.8 $&$   0.5 $&$   6.2 $&$   0.8 $&$   1.1 $&$  -0.9 $&$  -1.6 $&$   1.6 $\\
$  5.1 $&$ 0.0018 $&$ 0.0750 $&$  0.4 $&$ 0.0297 $&$  11.4 $&$   7.3 $&$  13.6 $&$  -0.1 $&$   0.6 $&$   4.2 $&$  -0.7 $&$   0.2 $&$   1.4 $&$   1.2 $&$   1.7 $&$   0.8 $&$   2.0 $&$  -2.4 $&$  -3.9 $&$   1.6 $\\
$  5.1 $&$ 0.0018 $&$ 0.0750 $&$  0.6 $&$ 0.0078 $&$  17.7 $&$   7.3 $&$  19.2 $&$   0.0 $&$   0.8 $&$   4.4 $&$  -0.9 $&$   0.1 $&$   0.2 $&$   1.5 $&$   4.1 $&$   1.2 $&$   1.5 $&$  -1.2 $&$  -2.6 $&$   1.6 $\\
$  5.1 $&$ 0.0056 $&$ 0.0160 $&$  0.2 $&$ 0.0433 $&$   4.9 $&$   9.5 $&$  10.7 $&$  -0.3 $&$   1.7 $&$   3.2 $&$  -1.1 $&$  -0.1 $&$  -2.5 $&$   2.0 $&$   6.3 $&$   0.8 $&$   2.2 $&$  -3.6 $&$  -1.9 $&$   2.0 $\\
$  5.1 $&$ 0.0056 $&$ 0.0160 $&$  0.4 $&$ 0.0152 $&$   6.1 $&$   9.5 $&$  11.3 $&$  -0.9 $&$   0.0 $&$   4.2 $&$  -0.6 $&$  -0.1 $&$   1.7 $&$   2.4 $&$   6.3 $&$  -1.6 $&$   2.2 $&$  -3.1 $&$  -1.6 $&$   2.0 $\\
$  5.1 $&$ 0.0056 $&$ 0.0160 $&$  0.6 $&$ 0.0050 $&$  13.1 $&$   9.5 $&$  16.2 $&$   0.0 $&$   4.8 $&$  -5.8 $&$  -0.9 $&$   0.0 $&$   0.1 $&$   3.4 $&$   2.8 $&$   2.9 $&$   0.0 $&$  -1.0 $&$  -0.5 $&$   2.0 $\\
$  5.1 $&$ 0.0056 $&$ 0.0250 $&$  0.2 $&$ 0.0376 $&$   4.7 $&$   8.4 $&$   9.6 $&$   0.0 $&$   2.0 $&$   2.9 $&$  -0.5 $&$  -0.1 $&$  -2.5 $&$   1.5 $&$   6.2 $&$  -0.3 $&$   2.0 $&$  -1.9 $&$  -0.7 $&$   1.7 $\\
$  5.1 $&$ 0.0056 $&$ 0.0250 $&$  0.4 $&$ 0.0129 $&$   6.3 $&$   8.4 $&$  10.5 $&$   0.0 $&$  -0.8 $&$   3.9 $&$  -0.2 $&$   0.0 $&$   1.0 $&$   0.6 $&$   6.4 $&$   2.4 $&$   1.3 $&$  -1.1 $&$  -0.5 $&$   1.7 $\\
$  5.1 $&$ 0.0056 $&$ 0.0250 $&$  0.6 $&$ 0.0036 $&$  13.6 $&$   8.4 $&$  15.9 $&$   0.0 $&$   5.2 $&$   1.9 $&$  -1.5 $&$  -0.3 $&$  -0.8 $&$  -1.1 $&$   4.1 $&$   1.1 $&$   3.1 $&$  -1.7 $&$  -1.5 $&$   1.7 $\\
$  5.1 $&$ 0.0056 $&$ 0.0350 $&$  0.2 $&$ 0.0449 $&$   5.8 $&$   7.4 $&$   9.4 $&$  -0.3 $&$   0.8 $&$   3.2 $&$   0.0 $&$   0.1 $&$  -1.9 $&$   2.1 $&$   5.4 $&$  -0.1 $&$   1.3 $&$  -1.0 $&$  -0.3 $&$   1.6 $\\
$  5.1 $&$ 0.0056 $&$ 0.0350 $&$  0.4 $&$ 0.0123 $&$   8.1 $&$   7.4 $&$  10.9 $&$  -0.3 $&$   1.1 $&$   5.1 $&$   0.1 $&$   0.0 $&$   1.4 $&$  -0.1 $&$   2.5 $&$   3.8 $&$   0.9 $&$  -0.8 $&$  -0.3 $&$   1.6 $\\
$  5.1 $&$ 0.0056 $&$ 0.0350 $&$  0.6 $&$ 0.0051 $&$  12.9 $&$   7.4 $&$  14.8 $&$   0.0 $&$  -1.8 $&$  -4.2 $&$   0.4 $&$   0.1 $&$   1.2 $&$   2.3 $&$   4.2 $&$  -1.8 $&$   1.4 $&$  -0.7 $&$  -0.3 $&$   1.6 $\\
$  5.1 $&$ 0.0056 $&$ 0.0500 $&$  0.2 $&$ 0.0473 $&$   6.3 $&$   9.0 $&$  11.0 $&$   0.0 $&$   0.3 $&$   6.2 $&$  -0.4 $&$   0.2 $&$  -3.6 $&$   0.3 $&$   4.4 $&$   0.9 $&$   1.9 $&$  -1.3 $&$  -0.5 $&$   1.6 $\\
$  5.1 $&$ 0.0056 $&$ 0.0500 $&$  0.4 $&$ 0.0151 $&$   7.5 $&$   9.0 $&$  11.7 $&$   0.0 $&$  -1.5 $&$   7.9 $&$   0.0 $&$   0.1 $&$   0.2 $&$   1.1 $&$   2.6 $&$   1.1 $&$   1.8 $&$  -0.8 $&$  -0.4 $&$   1.6 $\\
$  5.1 $&$ 0.0056 $&$ 0.0500 $&$  0.6 $&$ 0.0054 $&$  12.9 $&$   9.0 $&$  15.7 $&$   0.0 $&$  -2.0 $&$   4.7 $&$  -0.4 $&$   0.0 $&$  -0.3 $&$  -1.7 $&$   6.7 $&$   1.5 $&$   0.3 $&$  -0.5 $&$  -0.3 $&$   1.6 $\\
$  5.1 $&$ 0.0056 $&$ 0.0750 $&$  0.2 $&$ 0.0544 $&$  11.6 $&$  10.3 $&$  15.5 $&$   0.0 $&$   0.2 $&$   6.9 $&$  -0.6 $&$   0.2 $&$  -2.6 $&$  -0.3 $&$   6.2 $&$   1.5 $&$   1.3 $&$  -1.1 $&$  -2.5 $&$   1.6 $\\
$  5.1 $&$ 0.0056 $&$ 0.0750 $&$  0.4 $&$ 0.0255 $&$  10.2 $&$  10.3 $&$  14.5 $&$   0.0 $&$  -0.6 $&$   7.7 $&$  -0.8 $&$   0.2 $&$   0.7 $&$   0.0 $&$   5.0 $&$   1.5 $&$   2.3 $&$  -1.3 $&$  -3.1 $&$   1.6 $\\
$  5.1 $&$ 0.0056 $&$ 0.0750 $&$  0.6 $&$ 0.0094 $&$  17.1 $&$  10.3 $&$  20.0 $&$   0.0 $&$  -2.0 $&$   9.4 $&$  -1.2 $&$   0.2 $&$  -0.7 $&$   0.3 $&$   0.6 $&$   1.7 $&$   1.3 $&$  -0.9 $&$  -2.2 $&$   1.6 $\\
$  5.1 $&$ 0.0178 $&$ 0.0085 $&$  0.2 $&$ 0.0345 $&$   4.3 $&$  10.6 $&$  11.4 $&$  -7.0 $&$   1.2 $&$   3.6 $&$  -0.8 $&$  -0.2 $&$  -1.7 $&$   0.5 $&$   5.4 $&$   0.1 $&$   1.7 $&$  -1.6 $&$  -2.6 $&$   2.0 $\\
$  5.1 $&$ 0.0178 $&$ 0.0085 $&$  0.4 $&$ 0.0119 $&$   5.5 $&$  10.6 $&$  11.9 $&$  -6.2 $&$   1.0 $&$   4.6 $&$  -0.8 $&$  -0.2 $&$   1.3 $&$   2.2 $&$   5.1 $&$   0.6 $&$   1.1 $&$  -1.4 $&$  -3.1 $&$   2.0 $\\
$  5.1 $&$ 0.0178 $&$ 0.0085 $&$  0.6 $&$ 0.0039 $&$  11.2 $&$  10.6 $&$  15.4 $&$  -8.6 $&$   1.2 $&$   2.9 $&$  -0.5 $&$  -0.2 $&$   0.6 $&$  -0.2 $&$   4.1 $&$   1.6 $&$   0.7 $&$  -0.6 $&$  -1.8 $&$   2.0 $\\
$  5.1 $&$ 0.0178 $&$ 0.0160 $&$  0.2 $&$ 0.0366 $&$   3.6 $&$   9.2 $&$   9.9 $&$   0.0 $&$   0.5 $&$   4.8 $&$  -0.1 $&$   0.2 $&$  -2.3 $&$   4.9 $&$   4.8 $&$  -0.3 $&$   1.3 $&$  -1.2 $&$  -1.4 $&$   1.6 $\\
$  5.1 $&$ 0.0178 $&$ 0.0160 $&$  0.4 $&$ 0.0110 $&$   5.1 $&$   9.2 $&$  10.5 $&$   0.5 $&$  -0.4 $&$   5.6 $&$   0.3 $&$   0.0 $&$   1.2 $&$   0.9 $&$   6.6 $&$   1.1 $&$   1.1 $&$  -0.7 $&$  -1.0 $&$   1.6 $\\
$  5.1 $&$ 0.0178 $&$ 0.0160 $&$  0.6 $&$ 0.0032 $&$  10.2 $&$   9.2 $&$  13.7 $&$   0.0 $&$  -0.3 $&$   6.7 $&$  -0.2 $&$   0.2 $&$   0.1 $&$   1.3 $&$   2.8 $&$   4.6 $&$   1.9 $&$  -0.7 $&$  -1.2 $&$   1.6 $\\
$  5.1 $&$ 0.0178 $&$ 0.0250 $&$  0.2 $&$ 0.0333 $&$   4.7 $&$   8.9 $&$  10.1 $&$   0.0 $&$   0.3 $&$   6.0 $&$  -0.2 $&$   0.0 $&$  -2.0 $&$   0.7 $&$   5.6 $&$   0.1 $&$   1.6 $&$  -1.5 $&$  -0.6 $&$   1.6 $\\
$  5.1 $&$ 0.0178 $&$ 0.0250 $&$  0.4 $&$ 0.0113 $&$   6.4 $&$   8.9 $&$  10.9 $&$   0.0 $&$   1.7 $&$   5.8 $&$  -0.3 $&$   0.1 $&$   1.4 $&$   0.4 $&$   5.2 $&$   2.4 $&$   1.4 $&$  -1.4 $&$  -0.6 $&$   1.6 $\\
$  5.1 $&$ 0.0178 $&$ 0.0250 $&$  0.6 $&$ 0.0038 $&$  13.3 $&$   8.9 $&$  16.0 $&$   0.0 $&$  -2.4 $&$   7.1 $&$   1.2 $&$   0.0 $&$   0.1 $&$  -0.1 $&$   4.4 $&$   0.0 $&$  -0.2 $&$  -0.6 $&$  -0.4 $&$   1.6 $\\
$  5.1 $&$ 0.0178 $&$ 0.0350 $&$  0.2 $&$ 0.0303 $&$   7.1 $&$   8.7 $&$  11.2 $&$   0.0 $&$   0.6 $&$   6.7 $&$  -0.3 $&$   0.1 $&$  -1.9 $&$   1.4 $&$   4.4 $&$   0.2 $&$   0.9 $&$  -1.4 $&$  -0.4 $&$   1.6 $\\
$  5.1 $&$ 0.0178 $&$ 0.0350 $&$  0.4 $&$ 0.0122 $&$   9.0 $&$   8.7 $&$  12.5 $&$   0.0 $&$  -0.7 $&$   7.8 $&$  -1.0 $&$   0.1 $&$   1.2 $&$   0.7 $&$   1.7 $&$   1.9 $&$   0.5 $&$  -1.2 $&$  -0.4 $&$   1.6 $\\
$  5.1 $&$ 0.0178 $&$ 0.0350 $&$  0.6 $&$ 0.0039 $&$  13.9 $&$   8.7 $&$  16.4 $&$   0.0 $&$   1.6 $&$  -0.8 $&$   0.0 $&$   0.0 $&$   0.5 $&$  -2.3 $&$   7.7 $&$   1.9 $&$   0.2 $&$  -0.8 $&$  -0.3 $&$   1.6 $\\
$  5.1 $&$ 0.0178 $&$ 0.0500 $&$  0.2 $&$ 0.0449 $&$   7.1 $&$   8.8 $&$  11.3 $&$   0.0 $&$   0.4 $&$   6.0 $&$  -1.1 $&$   0.1 $&$  -2.3 $&$   0.8 $&$   5.2 $&$   1.2 $&$   0.2 $&$  -1.5 $&$  -0.3 $&$   1.6 $\\
$  5.1 $&$ 0.0178 $&$ 0.0500 $&$  0.4 $&$ 0.0133 $&$   9.0 $&$   8.8 $&$  12.6 $&$   0.4 $&$   1.2 $&$   3.0 $&$  -0.7 $&$   0.2 $&$   1.2 $&$   1.6 $&$   5.0 $&$   4.9 $&$   0.6 $&$  -3.1 $&$  -0.7 $&$   1.6 $\\
$  5.1 $&$ 0.0178 $&$ 0.0500 $&$  0.6 $&$ 0.0050 $&$  15.2 $&$   8.8 $&$  17.6 $&$   0.0 $&$   0.4 $&$   8.4 $&$   0.2 $&$   0.2 $&$  -0.5 $&$   1.5 $&$  -0.3 $&$  -1.2 $&$   0.2 $&$  -0.7 $&$  -0.2 $&$   1.6 $\\
\hline
\end{tabular}
\end{tiny}
 \caption{
 The  reduced diffractive  cross sections $\xpom \sigma_r^{D(4)}$
as a function of
 $Q^2, \beta$, $\xpom$ and $t$ values (columns $1-5$). 
The statistical ($\delta_{stat}$), systematic ($\delta_{sys}$) and 
total ($\delta_{tot}$) uncertainties are given in columns 6 to 8.
The remaining columns give the changes of the cross sections  
due to a $+ 1 \sigma$ variation of the various systematic error
sources described in section~\ref{systs}:
the hadronic energy scale ($\delta_{had}$); 
the electromagnetic energy scale ($\delta_{ele}$); 
the scattering angle of the electron ($\delta_{\theta}$); 
the reweighting of the simulation in $\beta$ ($\delta_{\beta}$), 
$\xpom$ ($\delta_{\xpom}$) and
$t$ ($\delta_{t}$); 
the leading proton energy $E_p$ ($\delta_{E_p}$), 
the proton transverse momentum components  
$p_x$ 
 ($\delta_{p_x}$) 
and $p_y$
 ($\delta_{p_y}$);
the reweighting of the simulation in $Q^2$ ($\delta_{Q^2}$);
the background from beam halo, 
photoproduction and proton dissociation processes ($\delta_{bgn}$); 
the vertex reconstruction efficiency ($\delta_{vtx}$) and 
 the bin centre corrections ($\delta_{bcc}$). 
All uncertainties are given in per cent. 
 The normalisation uncertainty of $4.3\%$ is not included. The table continues on the next pages.}
\label{table:f2d4}
\end{table}

\newpage
\begin{table}[ht]
\centering
\begin{tiny}
\begin{tabular}{| c| c| c| c| c| c| c| c| c| c| c| c| c| c| c| c| c| c| c| c| c|}
\hline
 $Q^2$ & $\beta$ & $\xpom$ & $-t$ & $\xpom\sigma_r^{D(4)}$ & $\delta_{stat}$ & $\delta_{sys}$ & $\delta_{tot}$ &
 $\delta_{had}$ & $\delta_{ele}$ & $\delta_{\theta}$ & $\delta_{\beta}$ & $\delta_{\xpom}$ & $\delta_{t}$ &
 $\delta_{E_p}$ & $\delta_{p_x}$ & $\delta_{p_y}$ & $\delta_{Q^2}$ & $\delta_{vtx}$ & $\delta_{bgn}$ & $\delta_{bcc}$ \\
 $[\GeV^2]$ & & & $[\GeV^2]$ & $[\GeV^{-2}]$ & $[\%]$ & $[\%]$ & $[\%]$ & $[\%]$ & $[\%]$ & $[\%]$ & $[\%]$ & $[\%]$ & $[\%]$ & $[\%]$ & $[\%]$ &
 $[\%]$ & $[\%]$ & $[\%]$ & $[\%]$ & $[\%]$ \\
\hline
$  5.1 $&$ 0.0178 $&$ 0.0750 $&$  0.2 $&$ 0.0508 $&$  12.7 $&$   8.2 $&$  15.1 $&$   0.0 $&$  -0.5 $&$   3.4 $&$  -0.5 $&$   0.2 $&$  -2.8 $&$   0.7 $&$   6.0 $&$   0.3 $&$  -0.3 $&$  -1.3 $&$  -2.6 $&$   1.6 $\\
$  5.1 $&$ 0.0178 $&$ 0.0750 $&$  0.4 $&$ 0.0208 $&$  15.2 $&$   8.2 $&$  17.3 $&$   0.0 $&$   1.2 $&$   6.4 $&$  -0.8 $&$   0.4 $&$   0.6 $&$   0.8 $&$  -1.2 $&$   0.9 $&$  -0.7 $&$  -1.8 $&$  -3.8 $&$   1.6 $\\
$  5.1 $&$ 0.0178 $&$ 0.0750 $&$  0.6 $&$ 0.0093 $&$  20.3 $&$   8.2 $&$  21.9 $&$   0.0 $&$   0.9 $&$   2.8 $&$  -0.1 $&$   0.0 $&$  -0.3 $&$  -1.2 $&$   7.2 $&$   0.9 $&$   0.2 $&$  -0.6 $&$  -1.3 $&$   1.6 $\\
$  5.1 $&$ 0.0562 $&$ 0.0025 $&$  0.2 $&$ 0.0345 $&$   3.4 $&$  10.1 $&$  10.7 $&$  -4.9 $&$   0.6 $&$   2.8 $&$  -1.5 $&$   0.1 $&$  -1.9 $&$  -5.8 $&$   3.5 $&$  -0.1 $&$   1.3 $&$  -1.5 $&$  -2.9 $&$   2.6 $\\
$  5.1 $&$ 0.0562 $&$ 0.0025 $&$  0.4 $&$ 0.0102 $&$   4.5 $&$  10.1 $&$  11.1 $&$  -4.7 $&$   0.3 $&$   2.5 $&$  -1.2 $&$   0.1 $&$   0.8 $&$  -4.4 $&$   6.0 $&$   1.1 $&$   1.0 $&$  -1.0 $&$  -2.7 $&$   2.6 $\\
$  5.1 $&$ 0.0562 $&$ 0.0025 $&$  0.6 $&$ 0.0030 $&$   9.1 $&$  10.1 $&$  13.7 $&$  -1.8 $&$  -1.2 $&$   5.8 $&$  -1.4 $&$   0.3 $&$   0.2 $&$  -4.9 $&$   3.6 $&$   2.3 $&$   1.6 $&$  -0.9 $&$  -3.2 $&$   2.6 $\\
\hline
\end{tabular}
\end{tiny}
\end{table}

\newpage
\begin{table}[ht]
\centering
\begin{tiny}
  \begin{tabular}{| c| c| c| c| c| c| c| c| c| c| c| c| c| c| c| c| c| c| c| c| c|}
\hline
 $Q^2$ & $\beta$ & $\xpom$ & $-t$ & $\xpom\sigma_r^{D(4)}$ & $\delta_{stat}$ & $\delta_{sys}$ & $\delta_{tot}$ &
 $\delta_{had}$ & $\delta_{ele}$ & $\delta_{\theta}$ & $\delta_{\beta}$ & $\delta_{\xpom}$ & $\delta_{t}$ &
 $\delta_{E_p}$ & $\delta_{p_x}$ & $\delta_{p_y}$ & $\delta_{Q^2}$ & $\delta_{vtx}$ & $\delta_{bgn}$ & $\delta_{bcc}$ \\
 $[\GeV^2]$ & & & $[\GeV^2]$ & $[\GeV^{-2}]$ & $[\%]$ & $[\%]$ & $[\%]$ & $[\%]$ & $[\%]$ & $[\%]$ & $[\%]$ & $[\%]$ & $[\%]$ & $[\%]$ & $[\%]$ &
 $[\%]$ & $[\%]$ & $[\%]$ & $[\%]$ & $[\%]$ \\
\hline
$  8.8 $&$ 0.0178 $&$ 0.0350 $&$  0.2 $&$ 0.0432 $&$   6.3 $&$   7.4 $&$   9.7 $&$   0.1 $&$   0.7 $&$   1.9 $&$  -0.4 $&$   0.1 $&$  -2.8 $&$   1.2 $&$   5.8 $&$  -0.6 $&$   1.3 $&$  -1.7 $&$  -0.6 $&$   1.6 $\\
$  8.8 $&$ 0.0178 $&$ 0.0350 $&$  0.4 $&$ 0.0128 $&$   9.5 $&$   7.4 $&$  12.0 $&$   0.0 $&$  -0.7 $&$   2.8 $&$  -0.4 $&$   0.1 $&$   0.4 $&$   0.0 $&$   5.3 $&$   3.6 $&$   0.7 $&$  -1.2 $&$  -0.5 $&$   1.6 $\\
$  8.8 $&$ 0.0178 $&$ 0.0350 $&$  0.6 $&$ 0.0059 $&$  14.3 $&$   7.4 $&$  16.1 $&$   0.0 $&$   0.1 $&$   2.1 $&$   0.7 $&$  -0.1 $&$   0.9 $&$   0.0 $&$   5.5 $&$   3.9 $&$   0.5 $&$  -0.7 $&$  -0.3 $&$   1.6 $\\
$  8.8 $&$ 0.0178 $&$ 0.0500 $&$  0.2 $&$ 0.0484 $&$   7.0 $&$   6.8 $&$   9.8 $&$   0.5 $&$  -0.3 $&$   2.5 $&$  -1.0 $&$   0.1 $&$  -2.4 $&$   0.2 $&$   5.4 $&$   0.0 $&$   0.7 $&$  -1.2 $&$  -0.2 $&$   1.6 $\\
$  8.8 $&$ 0.0178 $&$ 0.0500 $&$  0.4 $&$ 0.0192 $&$   8.1 $&$   6.8 $&$  10.6 $&$   0.0 $&$   0.7 $&$   0.7 $&$  -0.3 $&$   0.0 $&$   0.5 $&$   0.9 $&$   5.6 $&$   3.1 $&$   0.4 $&$  -0.7 $&$  -0.2 $&$   1.6 $\\
$  8.8 $&$ 0.0178 $&$ 0.0500 $&$  0.6 $&$ 0.0057 $&$  13.9 $&$   6.8 $&$  15.5 $&$   0.0 $&$   2.4 $&$   1.2 $&$  -1.3 $&$  -0.2 $&$   0.5 $&$   1.6 $&$   5.3 $&$   0.5 $&$   1.0 $&$  -1.7 $&$  -0.5 $&$   1.6 $\\
\hline
\end{tabular}
\end{tiny}
\end{table}

\newpage
\begin{table}[ht]
\centering
\begin{tiny}
  \begin{tabular}{| c| c| c| c| c| c| c| c| c| c| c| c| c| c| c| c| c| c| c| c| c|}
\hline
 $Q^2$ & $\beta$ & $\xpom$ & $-t$ & $\xpom\sigma_r^{D(4)}$ & $\delta_{stat}$ & $\delta_{sys}$ & $\delta_{tot}$ &
 $\delta_{had}$ & $\delta_{ele}$ & $\delta_{\theta}$ & $\delta_{\beta}$ & $\delta_{\xpom}$ & $\delta_{t}$ &
 $\delta_{E_p}$ & $\delta_{p_x}$ & $\delta_{p_y}$ & $\delta_{Q^2}$ & $\delta_{vtx}$ & $\delta_{bgn}$ & $\delta_{bcc}$ \\
 $[\GeV^2]$ & & & $[\GeV^2]$ & $[\GeV^{-2}]$ & $[\%]$ & $[\%]$ & $[\%]$ & $[\%]$ & $[\%]$ & $[\%]$ & $[\%]$ & $[\%]$ & $[\%]$ & $[\%]$ & $[\%]$ &
 $[\%]$ & $[\%]$ & $[\%]$ & $[\%]$ & $[\%]$ \\
\hline
$ 15.3 $&$ 0.0178 $&$ 0.0250 $&$  0.2 $&$ 0.0549 $&$   6.6 $&$   7.5 $&$  10.0 $&$   0.0 $&$  -1.0 $&$   3.2 $&$  -0.7 $&$   0.0 $&$  -1.8 $&$   0.3 $&$   5.8 $&$   0.4 $&$   1.9 $&$  -0.9 $&$  -0.7 $&$   1.6 $\\
$ 15.3 $&$ 0.0178 $&$ 0.0250 $&$  0.4 $&$ 0.0164 $&$   9.0 $&$   7.5 $&$  11.7 $&$   0.4 $&$   0.2 $&$  -0.1 $&$  -0.5 $&$   0.0 $&$   2.4 $&$   1.5 $&$   4.3 $&$   4.2 $&$   1.1 $&$  -2.2 $&$  -1.9 $&$   1.6 $\\
$ 15.3 $&$ 0.0178 $&$ 0.0250 $&$  0.6 $&$ 0.0053 $&$  18.3 $&$   7.5 $&$  19.8 $&$  -1.2 $&$   1.0 $&$   0.3 $&$  -0.1 $&$   0.0 $&$  -0.3 $&$   4.3 $&$   5.3 $&$  -1.9 $&$   0.2 $&$  -0.4 $&$  -0.5 $&$   1.6 $\\
$ 15.3 $&$ 0.0178 $&$ 0.0350 $&$  0.2 $&$ 0.0488 $&$   9.2 $&$   7.3 $&$  11.8 $&$   0.0 $&$   0.6 $&$   3.0 $&$  -0.3 $&$   0.0 $&$  -2.4 $&$   0.3 $&$   5.8 $&$   0.6 $&$   0.5 $&$  -1.3 $&$  -0.5 $&$   1.6 $\\
$ 15.3 $&$ 0.0178 $&$ 0.0350 $&$  0.4 $&$ 0.0199 $&$  11.3 $&$   7.3 $&$  13.4 $&$   0.0 $&$  -0.9 $&$   2.7 $&$   0.1 $&$   0.0 $&$   0.5 $&$   2.8 $&$   4.9 $&$   3.1 $&$   0.4 $&$  -1.0 $&$  -0.6 $&$   1.6 $\\
$ 15.3 $&$ 0.0178 $&$ 0.0350 $&$  0.6 $&$ 0.0076 $&$  20.6 $&$   7.3 $&$  21.9 $&$   0.7 $&$  -0.6 $&$  -3.3 $&$   0.1 $&$  -0.2 $&$   0.4 $&$  -5.1 $&$  -1.2 $&$  -3.2 $&$   0.3 $&$  -0.9 $&$  -0.6 $&$   1.6 $\\
\hline
\end{tabular}
\end{tiny}
\end{table}

\newpage
\begin{table}[ht]
\centering
\begin{tiny}
  \begin{tabular}{| c| c| c| c| c| c| c| c| c| c| c| c| c| c| c| c| c| c| c| c| c|}
\hline
 $Q^2$ & $\beta$ & $\xpom$ & $-t$ & $\xpom\sigma_r^{D(4)}$ & $\delta_{stat}$ & $\delta_{sys}$ & $\delta_{tot}$ &
 $\delta_{had}$ & $\delta_{ele}$ & $\delta_{\theta}$ & $\delta_{\beta}$ & $\delta_{\xpom}$ & $\delta_{t}$ &
 $\delta_{E_p}$ & $\delta_{p_x}$ & $\delta_{p_y}$ & $\delta_{Q^2}$ & $\delta_{vtx}$ & $\delta_{bgn}$ & $\delta_{bcc}$ \\
 $[\GeV^2]$ & & & $[\GeV^2]$ & $[\GeV^{-2}]$ & $[\%]$ & $[\%]$ & $[\%]$ & $[\%]$ & $[\%]$ & $[\%]$ & $[\%]$ & $[\%]$ & $[\%]$ & $[\%]$ & $[\%]$ &
 $[\%]$ & $[\%]$ & $[\%]$ & $[\%]$ & $[\%]$ \\
\hline
$ 26.5 $&$ 0.0056 $&$ 0.0750 $&$  0.2 $&$ 0.1363 $&$  29.5 $&$   7.3 $&$  30.4 $&$   0.0 $&$  -0.4 $&$   2.1 $&$  -1.5 $&$   0.3 $&$  -2.5 $&$   0.6 $&$   5.3 $&$  -0.2 $&$   0.5 $&$  -0.6 $&$  -2.6 $&$   2.1 $\\
$ 26.5 $&$ 0.0056 $&$ 0.0750 $&$  0.4 $&$ 0.0307 $&$  33.9 $&$   7.3 $&$  34.7 $&$   0.0 $&$  -2.0 $&$   5.4 $&$  -1.0 $&$   0.1 $&$   0.9 $&$   0.4 $&$   0.9 $&$   1.6 $&$   0.7 $&$  -0.9 $&$  -3.1 $&$   2.1 $\\
$ 26.5 $&$ 0.0056 $&$ 0.0750 $&$  0.6 $&$ 0.0188 $&$  39.8 $&$   7.3 $&$  40.5 $&$   0.0 $&$  -0.1 $&$   0.8 $&$  -0.8 $&$   0.1 $&$   0.0 $&$   1.2 $&$   6.3 $&$   1.8 $&$   0.3 $&$  -0.6 $&$  -1.9 $&$   2.1 $\\
$ 26.5 $&$ 0.0178 $&$ 0.0250 $&$  0.2 $&$ 0.0574 $&$  12.4 $&$   7.6 $&$  14.6 $&$   0.5 $&$   0.1 $&$   2.5 $&$  -1.4 $&$  -0.1 $&$  -2.6 $&$   4.1 $&$   3.9 $&$   0.6 $&$   1.3 $&$  -1.1 $&$  -1.9 $&$   1.9 $\\
$ 26.5 $&$ 0.0178 $&$ 0.0250 $&$  0.4 $&$ 0.0200 $&$  15.1 $&$   7.6 $&$  16.9 $&$   0.2 $&$  -0.9 $&$   2.7 $&$  -1.3 $&$   0.0 $&$   1.3 $&$   1.7 $&$   6.0 $&$   0.9 $&$   0.6 $&$  -0.9 $&$  -1.5 $&$   1.9 $\\
$ 26.5 $&$ 0.0178 $&$ 0.0250 $&$  0.6 $&$ 0.0074 $&$  29.9 $&$   7.6 $&$  30.9 $&$   0.0 $&$   3.1 $&$   0.3 $&$  -1.1 $&$  -0.1 $&$  -0.2 $&$   2.9 $&$   5.8 $&$  -1.1 $&$   0.5 $&$  -0.4 $&$  -0.8 $&$   1.9 $\\
\hline
\end{tabular}
\end{tiny}
\end{table}

\newpage
\begin{table}[ht]
\centering
\begin{tiny}
  \begin{tabular}{| c| c| c| c| c| c| c| c| c| c| c| c| c| c| c| c| c| c| c| c| c|}
\hline
 $Q^2$ & $\beta$ & $\xpom$ & $-t$ & $\xpom\sigma_r^{D(4)}$ & $\delta_{stat}$ & $\delta_{sys}$ & $\delta_{tot}$ &
 $\delta_{had}$ & $\delta_{ele}$ & $\delta_{\theta}$ & $\delta_{\beta}$ & $\delta_{\xpom}$ & $\delta_{t}$ &
 $\delta_{E_p}$ & $\delta_{p_x}$ & $\delta_{p_y}$ & $\delta_{Q^2}$ & $\delta_{vtx}$ & $\delta_{bgn}$ & $\delta_{bcc}$ \\
 $[\GeV^2]$ & & & $[\GeV^2]$ & $[\GeV^{-2}]$ & $[\%]$ & $[\%]$ & $[\%]$ & $[\%]$ & $[\%]$ & $[\%]$ & $[\%]$ & $[\%]$ & $[\%]$ & $[\%]$ & $[\%]$ &
 $[\%]$ & $[\%]$ & $[\%]$ & $[\%]$ & $[\%]$ \\
\hline
$ 26.5 $&$ 0.5620 $&$ 0.0025 $&$  0.2 $&$ 0.0711 $&$   4.6 $&$  10.4 $&$  11.3 $&$   5.4 $&$  -1.5 $&$   2.4 $&$   0.5 $&$  -1.3 $&$  -1.9 $&$  -3.3 $&$   6.0 $&$   0.2 $&$   0.3 $&$  -0.9 $&$  -2.2 $&$   3.4 $\\
$ 26.5 $&$ 0.5620 $&$ 0.0025 $&$  0.4 $&$ 0.0215 $&$   6.2 $&$  10.4 $&$  12.1 $&$   6.0 $&$  -0.8 $&$   1.0 $&$   1.2 $&$  -1.2 $&$   1.0 $&$  -4.7 $&$   4.4 $&$   1.5 $&$   0.4 $&$  -1.0 $&$  -3.0 $&$   3.4 $\\
$ 26.5 $&$ 0.5620 $&$ 0.0025 $&$  0.6 $&$ 0.0059 $&$  13.1 $&$  10.4 $&$  16.7 $&$   4.6 $&$  -1.1 $&$   2.4 $&$   0.6 $&$  -0.6 $&$   0.0 $&$  -1.8 $&$   7.2 $&$   2.2 $&$   0.2 $&$  -0.7 $&$  -2.6 $&$   3.4 $\\
$ 26.5 $&$ 0.5620 $&$ 0.0085 $&$  0.2 $&$ 0.0534 $&$   5.2 $&$   9.5 $&$  10.9 $&$   5.7 $&$  -0.5 $&$   1.4 $&$   0.9 $&$   0.0 $&$  -1.3 $&$   6.1 $&$   2.9 $&$  -0.1 $&$   0.1 $&$  -0.7 $&$  -0.3 $&$   2.8 $\\
$ 26.5 $&$ 0.5620 $&$ 0.0085 $&$  0.4 $&$ 0.0144 $&$   7.4 $&$   9.5 $&$  12.1 $&$   5.6 $&$  -0.6 $&$   1.3 $&$   0.7 $&$   0.1 $&$   0.6 $&$   4.9 $&$   5.0 $&$   0.2 $&$   0.0 $&$  -0.5 $&$  -0.3 $&$   2.8 $\\
$ 26.5 $&$ 0.5620 $&$ 0.0085 $&$  0.6 $&$ 0.0062 $&$  13.3 $&$   9.5 $&$  16.4 $&$   0.2 $&$   0.0 $&$   0.1 $&$   1.0 $&$  -0.4 $&$  -0.3 $&$   5.2 $&$  -5.1 $&$   5.3 $&$   0.1 $&$  -0.9 $&$  -0.5 $&$   2.8 $\\
\hline
\end{tabular}
\end{tiny}
\end{table}

\newpage
\begin{table}[ht]
\centering
\begin{tiny}
  \begin{tabular}{| c| c| c| c| c| c| c| c| c| c| c| c| c| c| c| c| c| c| c| c| c|}
\hline
 $Q^2$ & $\beta$ & $\xpom$ & $-t$ & $\xpom\sigma_r^{D(4)}$ & $\delta_{stat}$ & $\delta_{sys}$ & $\delta_{tot}$ &
 $\delta_{had}$ & $\delta_{ele}$ & $\delta_{\theta}$ & $\delta_{\beta}$ & $\delta_{\xpom}$ & $\delta_{t}$ &
 $\delta_{E_p}$ & $\delta_{p_x}$ & $\delta_{p_y}$ & $\delta_{Q^2}$ & $\delta_{vtx}$ & $\delta_{bgn}$ & $\delta_{bcc}$ \\
 $[\GeV^2]$ & & & $[\GeV^2]$ & $[\GeV^{-2}]$ & $[\%]$ & $[\%]$ & $[\%]$ & $[\%]$ & $[\%]$ & $[\%]$ & $[\%]$ & $[\%]$ & $[\%]$ & $[\%]$ & $[\%]$ &
 $[\%]$ & $[\%]$ & $[\%]$ & $[\%]$ & $[\%]$ \\
\hline
$ 46.0 $&$ 0.5620 $&$ 0.0085 $&$  0.2 $&$ 0.0490 $&$   7.9 $&$   9.8 $&$  12.6 $&$   6.2 $&$  -0.8 $&$   1.3 $&$   0.6 $&$  -0.2 $&$  -1.7 $&$   4.9 $&$   4.0 $&$   0.5 $&$   0.1 $&$  -0.7 $&$  -0.9 $&$   3.3 $\\
$ 46.0 $&$ 0.5620 $&$ 0.0085 $&$  0.4 $&$ 0.0160 $&$  10.3 $&$   9.8 $&$  14.2 $&$   6.7 $&$  -1.1 $&$   2.6 $&$   0.6 $&$  -0.1 $&$   0.8 $&$   5.0 $&$   1.8 $&$  -1.3 $&$   0.1 $&$  -0.6 $&$  -0.9 $&$   3.3 $\\
$ 46.0 $&$ 0.5620 $&$ 0.0085 $&$  0.6 $&$ 0.0066 $&$  17.8 $&$   9.8 $&$  20.3 $&$   1.9 $&$  -1.1 $&$   1.6 $&$   0.8 $&$   0.0 $&$   1.4 $&$   5.9 $&$   4.9 $&$   3.8 $&$   0.0 $&$  -0.4 $&$  -1.0 $&$   3.3 $\\
$ 46.0 $&$ 0.5620 $&$ 0.0160 $&$  0.2 $&$ 0.0384 $&$  12.5 $&$   8.6 $&$  15.2 $&$   0.0 $&$  -1.0 $&$   6.0 $&$   1.1 $&$   0.2 $&$  -1.9 $&$  -2.3 $&$   3.3 $&$   0.1 $&$   0.0 $&$  -1.4 $&$  -1.9 $&$   3.2 $\\
$ 46.0 $&$ 0.5620 $&$ 0.0160 $&$  0.4 $&$ 0.0177 $&$  14.6 $&$   8.6 $&$  17.0 $&$   0.0 $&$  -3.2 $&$   1.9 $&$   2.4 $&$   0.4 $&$   2.3 $&$   0.8 $&$   5.6 $&$  -0.4 $&$   0.2 $&$  -1.1 $&$  -2.3 $&$   3.2 $\\
$ 46.0 $&$ 0.5620 $&$ 0.0160 $&$  0.6 $&$ 0.0036 $&$  39.6 $&$   8.6 $&$  40.5 $&$   0.0 $&$  -4.1 $&$   0.0 $&$   2.2 $&$   0.1 $&$  -0.3 $&$   2.1 $&$   4.7 $&$   2.6 $&$   0.6 $&$  -1.3 $&$  -2.5 $&$   3.2 $\\
\hline
\end{tabular}
\end{tiny}
\end{table}

\newpage
\begin{table}[ht]
\centering
\begin{tiny}
  \begin{tabular}{| c| c| c| c| c| c| c| c| c| c| c| c| c| c| c| c| c| c| c| c|}
\hline
 $Q^2$ & $\beta$ & $\xpom$ & $\xpom\sigma_r^{D(3)}$ & $\delta_{stat}$ & $\delta_{sys}$ & $\delta_{tot}$ &
 $\delta_{had}$ & $\delta_{ele}$ & $\delta_{\theta}$ & $\delta_{\beta}$ & $\delta_{\xpom}$ & $\delta_{t}$ &
 $\delta_{E_p}$ & $\delta_{p_x}$ & $\delta_{p_y}$ & $\delta_{Q^2}$ & $\delta_{vtx}$ & $\delta_{bgn}$ & $\delta_{bcc}$ \\
 $[\GeV^2]$ & & & & $[\%]$ & $[\%]$ & $[\%]$ & $[\%]$ & $[\%]$ & $[\%]$ & $[\%]$ & $[\%]$ & $[\%]$ & $[\%]$ & $[\%]$ &
 $[\%]$ & $[\%]$ & $[\%]$ & $[\%]$ & $[\%]$ \\
\hline
$   5.1 $&$ 0.0018 $&$ 0.0500 $&$ 0.0340 $&$   5.9 $&$   8.5 $&$  10.3 $&$  -0.2 $&$   4.2 $&$   0.7 $&$  -0.9 $&$   0.0 $&$  -2.8 $&$   1.1 $&$   5.0 $&$   2.9 $&$   1.5 $&$  -2.4 $&$  -1.1 $&$   1.0 $\\
$   5.1 $&$ 0.0018 $&$ 0.0750 $&$ 0.0352 $&$   6.7 $&$   7.1 $&$   9.8 $&$  -0.1 $&$   1.1 $&$   2.7 $&$  -0.4 $&$   0.1 $&$  -3.1 $&$   0.8 $&$   4.7 $&$   2.1 $&$   1.0 $&$  -0.9 $&$  -1.7 $&$   0.3 $\\
$   5.1 $&$ 0.0056 $&$ 0.0160 $&$ 0.0217 $&$   4.1 $&$   9.4 $&$  10.2 $&$  -0.5 $&$   2.8 $&$   3.1 $&$  -0.9 $&$  -0.1 $&$   1.1 $&$   3.0 $&$   6.4 $&$   1.3 $&$   1.9 $&$  -3.0 $&$  -1.6 $&$   1.2 $\\
$   5.1 $&$ 0.0056 $&$ 0.0250 $&$ 0.0220 $&$   3.6 $&$   8.2 $&$   9.0 $&$   0.0 $&$   2.2 $&$   3.5 $&$  -0.4 $&$   0.0 $&$   0.2 $&$   1.5 $&$   6.3 $&$   1.5 $&$   1.7 $&$  -1.5 $&$  -0.7 $&$   0.5 $\\
$   5.1 $&$ 0.0056 $&$ 0.0350 $&$ 0.0237 $&$   4.2 $&$   7.2 $&$   8.3 $&$  -0.3 $&$   1.2 $&$   3.7 $&$   0.0 $&$   0.1 $&$   0.1 $&$   2.2 $&$   5.0 $&$   2.2 $&$   1.2 $&$  -0.8 $&$  -0.3 $&$   0.3 $\\
$   5.1 $&$ 0.0056 $&$ 0.0500 $&$ 0.0253 $&$   4.3 $&$   8.8 $&$   9.8 $&$   0.0 $&$  -1.0 $&$   6.9 $&$  -0.2 $&$   0.1 $&$  -1.1 $&$   0.3 $&$   4.1 $&$   2.9 $&$   1.4 $&$  -0.8 $&$  -0.4 $&$   0.2 $\\
\hline
\end{tabular}
\end{tiny}
 \caption{
 The  reduced diffractive  cross sections $\xpom \sigma_r^{D(3)}$
 as a function of  $Q^2, \beta$ and $\xpom$ values (columns $1-4$).
The statistical ($\delta_{stat}$), systematic ($\delta_{sys}$) and 
total ($\delta_{tot}$) uncertainties are given in columns 6 to 8.
The remaining columns give the changes of the cross sections  
due to a $+ 1 \sigma$ variation of the various systematic error
sources described in section~\ref{systs}:
the hadronic energy scale ($\delta_{had}$); 
the electromagnetic energy scale ($\delta_{ele}$); 
the scattering angle of the electron ($\delta_{\theta}$); 
the reweighting of the simulation in $\beta$ ($\delta_{\beta}$), 
$\xpom$ ($\delta_{\xpom}$) and
$t$ ($\delta_{t}$); 
the leading proton energy $E_p$ ($\delta_{E_p}$), 
the proton transverse momentum components  
$p_x$ 
 ($\delta_{p_x}$) 
and $p_y$
 ($\delta_{p_y}$);
the reweighting of the simulation in $Q^2$ ($\delta_{Q^2}$);
the background from beam halo, 
photoproduction and proton dissociation processes ($\delta_{bgn}$); 
the vertex reconstruction efficiency ($\delta_{vtx}$) and 
 the bin centre corrections ($\delta_{bcc}$). 
All uncertainties are given in per cent.
 The normalisation uncertainty of $6\%$ is not included. The table continues on the next pages.}
\label{table:f2d3}
\end{table}

\newpage
\begin{table}[ht]
\centering
\begin{tiny}
  \begin{tabular}{| c| c| c| c| c| c| c| c| c| c| c| c| c| c| c| c| c| c| c| c|}
\hline
 $Q^2$ & $\beta$ & $\xpom$ & $\xpom\sigma_r^{D(3)}$ & $\delta_{stat}$ & $\delta_{sys}$ & $\delta_{tot}$ &
 $\delta_{had}$ & $\delta_{ele}$ & $\delta_{\theta}$ & $\delta_{\beta}$ & $\delta_{\xpom}$ & $\delta_{t}$ &
 $\delta_{E_p}$ & $\delta_{p_x}$ & $\delta_{p_y}$ & $\delta_{Q^2}$ & $\delta_{vtx}$ & $\delta_{bgn}$ & $\delta_{bcc}$ \\
 $[\GeV^2]$ & & & & $[\%]$ & $[\%]$ & $[\%]$ & $[\%]$ & $[\%]$ & $[\%]$ & $[\%]$ & $[\%]$ & $[\%]$ & $[\%]$ & $[\%]$ &
 $[\%]$ & $[\%]$ & $[\%]$ & $[\%]$ & $[\%]$ \\
\hline
$   8.8 $&$ 0.0562 $&$ 0.0500 $&$ 0.0252 $&$   6.0 $&$   8.8 $&$  10.7 $&$   0.0 $&$  -0.3 $&$   6.5 $&$   1.2 $&$   0.2 $&$  -1.3 $&$  -0.3 $&$   5.1 $&$   2.4 $&$   0.0 $&$  -0.1 $&$  -0.5 $&$   0.3 $\\
$   8.8 $&$ 0.1780 $&$ 0.0025 $&$ 0.0204 $&$   2.5 $&$   9.5 $&$   9.9 $&$  -2.8 $&$  -0.2 $&$   3.6 $&$  -0.3 $&$  -0.7 $&$  -0.6 $&$  -4.0 $&$   6.1 $&$   1.6 $&$   0.8 $&$  -1.3 $&$  -2.7 $&$   1.8 $\\
$   8.8 $&$ 0.1780 $&$ 0.0085 $&$ 0.0182 $&$   2.7 $&$   8.8 $&$   9.2 $&$   2.6 $&$  -0.2 $&$   2.0 $&$   1.1 $&$   0.1 $&$  -0.1 $&$   7.5 $&$   2.8 $&$   0.5 $&$   0.2 $&$  -0.3 $&$  -0.2 $&$   0.6 $\\
$   8.8 $&$ 0.1780 $&$ 0.0160 $&$ 0.0164 $&$   4.3 $&$   8.4 $&$   9.5 $&$   0.0 $&$   0.6 $&$   2.8 $&$   3.1 $&$   0.2 $&$  -0.2 $&$   4.5 $&$   5.5 $&$   1.2 $&$   0.0 $&$  -0.1 $&$  -0.9 $&$   0.7 $\\
$   8.8 $&$ 0.5620 $&$ 0.0025 $&$ 0.0328 $&$   1.5 $&$  10.8 $&$  10.9 $&$   6.6 $&$  -1.3 $&$   4.4 $&$   0.9 $&$  -1.2 $&$  -0.6 $&$   0.3 $&$   5.5 $&$   1.5 $&$   0.4 $&$  -0.8 $&$  -1.4 $&$   3.9 $\\
$  15.3 $&$ 0.0056 $&$ 0.0500 $&$ 0.0456 $&$   8.1 $&$   7.0 $&$  10.7 $&$  -0.1 $&$   1.6 $&$   1.7 $&$  -1.6 $&$   0.1 $&$  -1.5 $&$   1.1 $&$   4.8 $&$   3.5 $&$   0.7 $&$  -1.0 $&$  -0.6 $&$   0.9 $\\
$  15.3 $&$ 0.0056 $&$ 0.0750 $&$ 0.0498 $&$  10.2 $&$   6.6 $&$  12.2 $&$   0.0 $&$   1.9 $&$   0.0 $&$  -0.7 $&$   0.1 $&$  -3.7 $&$   0.7 $&$   3.0 $&$   3.3 $&$   0.5 $&$  -0.8 $&$  -2.2 $&$   0.3 $\\
$  15.3 $&$ 0.0178 $&$ 0.0160 $&$ 0.0335 $&$   5.1 $&$   8.0 $&$   9.4 $&$  -0.2 $&$  -0.9 $&$   4.0 $&$  -1.5 $&$  -0.2 $&$   0.8 $&$   2.9 $&$   5.5 $&$   0.8 $&$   0.7 $&$  -1.3 $&$  -1.2 $&$   0.8 $\\
$  15.3 $&$ 0.0178 $&$ 0.0250 $&$ 0.0318 $&$   4.7 $&$   7.3 $&$   8.7 $&$  -0.1 $&$  -1.2 $&$   3.1 $&$  -0.5 $&$   0.0 $&$  -0.1 $&$   1.1 $&$   5.9 $&$   1.6 $&$   1.5 $&$  -0.9 $&$  -0.8 $&$   0.2 $\\
$  15.3 $&$ 0.0178 $&$ 0.0350 $&$ 0.0313 $&$   6.1 $&$   7.1 $&$   9.4 $&$   0.0 $&$   0.4 $&$   3.1 $&$  -0.2 $&$   0.0 $&$   1.1 $&$   0.7 $&$   5.5 $&$   2.6 $&$   0.5 $&$  -1.1 $&$  -0.6 $&$   0.1 $\\
$  15.3 $&$ 0.0178 $&$ 0.0500 $&$ 0.0338 $&$   5.9 $&$   7.2 $&$   9.3 $&$   0.0 $&$  -0.7 $&$   2.5 $&$  -0.6 $&$   0.1 $&$  -1.4 $&$   0.9 $&$   5.5 $&$   3.2 $&$   0.5 $&$  -1.2 $&$  -0.4 $&$   0.2 $\\
$  15.3 $&$ 0.0178 $&$ 0.0750 $&$ 0.0407 $&$   8.9 $&$   6.4 $&$  11.0 $&$   0.0 $&$   0.9 $&$  -0.3 $&$  -0.4 $&$   0.2 $&$  -2.3 $&$   1.2 $&$   4.0 $&$   3.4 $&$   0.1 $&$  -0.8 $&$  -2.1 $&$   0.2 $\\
$  15.3 $&$ 0.0562 $&$ 0.0085 $&$ 0.0214 $&$   4.7 $&$   9.4 $&$  10.5 $&$  -3.8 $&$   0.0 $&$   3.0 $&$  -0.6 $&$  -0.2 $&$   0.2 $&$   2.9 $&$   6.2 $&$   1.4 $&$   0.7 $&$  -1.2 $&$  -3.8 $&$   0.4 $\\
$  15.3 $&$ 0.0562 $&$ 0.0160 $&$ 0.0221 $&$   4.1 $&$   8.2 $&$   9.1 $&$   0.0 $&$   0.4 $&$   2.7 $&$  -0.4 $&$   0.0 $&$  -0.8 $&$   5.1 $&$   5.2 $&$   0.8 $&$   0.5 $&$  -1.3 $&$  -1.6 $&$   0.2 $\\
$  15.3 $&$ 0.0562 $&$ 0.0250 $&$ 0.0203 $&$   4.9 $&$   7.3 $&$   8.8 $&$   0.0 $&$   0.5 $&$   1.8 $&$  -0.3 $&$   0.0 $&$   0.8 $&$   1.8 $&$   6.5 $&$   1.4 $&$   0.4 $&$  -1.0 $&$  -0.9 $&$   0.2 $\\
$  15.3 $&$ 0.0562 $&$ 0.0350 $&$ 0.0221 $&$   6.3 $&$   6.7 $&$   9.2 $&$   0.0 $&$  -1.3 $&$   2.1 $&$  -0.3 $&$   0.0 $&$   0.6 $&$   1.3 $&$   4.8 $&$   3.5 $&$   0.4 $&$  -1.0 $&$  -0.8 $&$   0.3 $\\
$  15.3 $&$ 0.0562 $&$ 0.0500 $&$ 0.0240 $&$   6.2 $&$   6.6 $&$   9.1 $&$   0.0 $&$  -0.3 $&$   1.5 $&$  -0.3 $&$   0.1 $&$  -1.6 $&$   0.3 $&$   5.1 $&$   3.5 $&$   0.1 $&$  -0.7 $&$  -0.3 $&$   0.3 $\\
$  15.3 $&$ 0.0562 $&$ 0.0750 $&$ 0.0259 $&$  11.6 $&$   7.4 $&$  13.8 $&$   0.0 $&$  -1.3 $&$   3.3 $&$   0.5 $&$   0.1 $&$  -1.9 $&$  -0.6 $&$   3.9 $&$   4.4 $&$   0.0 $&$  -0.6 $&$  -1.8 $&$   0.3 $\\
$  15.3 $&$ 0.1780 $&$ 0.0025 $&$ 0.0218 $&$   3.7 $&$   9.3 $&$  10.0 $&$  -2.8 $&$  -0.5 $&$   2.0 $&$  -0.3 $&$  -0.7 $&$  -0.4 $&$  -5.5 $&$   5.2 $&$   1.0 $&$   0.4 $&$  -0.9 $&$  -3.7 $&$   1.2 $\\
$  15.3 $&$ 0.1780 $&$ 0.0085 $&$ 0.0193 $&$   3.6 $&$   8.8 $&$   9.5 $&$   2.1 $&$  -0.2 $&$   1.3 $&$   0.5 $&$   0.0 $&$   0.1 $&$   7.2 $&$   4.1 $&$   0.7 $&$   0.2 $&$  -0.6 $&$  -0.7 $&$   0.5 $\\
$  15.3 $&$ 0.1780 $&$ 0.0160 $&$ 0.0182 $&$   4.4 $&$   8.2 $&$   9.3 $&$   0.0 $&$   0.3 $&$   2.5 $&$   2.2 $&$   0.1 $&$   0.0 $&$   4.5 $&$   5.6 $&$   0.2 $&$   0.2 $&$  -0.9 $&$  -1.6 $&$   0.3 $\\
$  15.3 $&$ 0.1780 $&$ 0.0250 $&$ 0.0182 $&$   6.6 $&$   7.6 $&$  10.0 $&$   0.0 $&$  -0.8 $&$   3.4 $&$   3.0 $&$   0.2 $&$   0.5 $&$   2.0 $&$   5.3 $&$   1.5 $&$   0.0 $&$  -0.9 $&$  -1.0 $&$   0.3 $\\
$  15.3 $&$ 0.5620 $&$ 0.0025 $&$ 0.0335 $&$   2.4 $&$  10.2 $&$  10.4 $&$   6.0 $&$  -2.1 $&$   3.0 $&$   0.8 $&$  -1.3 $&$  -0.2 $&$  -2.6 $&$   5.3 $&$   1.3 $&$   0.3 $&$  -0.8 $&$  -1.7 $&$   3.4 $\\
$  15.3 $&$ 0.5620 $&$ 0.0085 $&$ 0.0294 $&$   3.4 $&$  10.0 $&$  10.6 $&$   4.9 $&$   1.1 $&$   0.9 $&$   1.3 $&$   0.5 $&$   0.0 $&$   7.7 $&$   2.6 $&$   0.7 $&$  -0.1 $&$  -0.5 $&$   0.0 $&$   2.2 $\\
$  26.5 $&$ 0.0056 $&$ 0.0750 $&$ 0.0554 $&$  16.9 $&$   7.1 $&$  18.3 $&$   0.0 $&$  -1.2 $&$   2.9 $&$  -1.0 $&$   0.2 $&$  -3.5 $&$   0.8 $&$   4.1 $&$   1.8 $&$   0.4 $&$  -0.6 $&$  -2.1 $&$   1.3 $\\
$  26.5 $&$ 0.0178 $&$ 0.0250 $&$ 0.0346 $&$   8.7 $&$   7.4 $&$  11.4 $&$   0.3 $&$   0.4 $&$   2.5 $&$  -1.1 $&$  -0.1 $&$  -1.0 $&$   4.3 $&$   4.6 $&$   1.2 $&$   0.9 $&$  -0.9 $&$  -1.5 $&$   1.1 $\\
$  26.5 $&$ 0.0178 $&$ 0.0350 $&$ 0.0341 $&$   8.8 $&$   6.8 $&$  11.1 $&$   0.0 $&$   2.6 $&$   1.1 $&$  -1.0 $&$   0.0 $&$   0.1 $&$   0.9 $&$   5.1 $&$   3.1 $&$   0.5 $&$  -0.9 $&$  -0.5 $&$   0.3 $\\
$  26.5 $&$ 0.0178 $&$ 0.0500 $&$ 0.0489 $&$   7.4 $&$   7.2 $&$  10.3 $&$   0.0 $&$  -0.7 $&$   3.2 $&$  -0.9 $&$   0.1 $&$  -0.1 $&$   0.6 $&$   4.9 $&$   3.7 $&$   0.2 $&$  -0.7 $&$  -0.3 $&$   0.1 $\\
$  26.5 $&$ 0.0178 $&$ 0.0750 $&$ 0.0414 $&$  11.9 $&$   7.9 $&$  14.3 $&$   0.0 $&$  -0.5 $&$   2.7 $&$  -0.7 $&$   0.1 $&$  -3.6 $&$   0.9 $&$   5.3 $&$   2.8 $&$   0.3 $&$  -0.8 $&$  -2.0 $&$   0.2 $\\
$  26.5 $&$ 0.0562 $&$ 0.0085 $&$ 0.0266 $&$   7.1 $&$   8.4 $&$  11.1 $&$  -6.1 $&$  -0.8 $&$   0.8 $&$  -1.1 $&$   0.0 $&$   0.6 $&$  -2.2 $&$   3.5 $&$   0.3 $&$   0.3 $&$  -0.6 $&$  -3.6 $&$   1.0 $\\
$  26.5 $&$ 0.0562 $&$ 0.0160 $&$ 0.0266 $&$   5.3 $&$   8.0 $&$   9.6 $&$   0.2 $&$   0.3 $&$   2.7 $&$  -0.3 $&$   0.0 $&$   0.7 $&$   3.3 $&$   6.4 $&$   0.3 $&$   0.3 $&$  -1.0 $&$  -1.5 $&$   0.2 $\\
$  26.5 $&$ 0.0562 $&$ 0.0250 $&$ 0.0249 $&$   6.4 $&$   7.2 $&$   9.7 $&$   0.2 $&$  -0.3 $&$   2.8 $&$  -0.6 $&$   0.0 $&$   0.3 $&$   1.8 $&$   6.1 $&$   0.8 $&$   0.4 $&$  -1.1 $&$  -1.2 $&$   0.2 $\\
$  26.5 $&$ 0.0562 $&$ 0.0350 $&$ 0.0290 $&$   7.6 $&$   6.9 $&$  10.3 $&$   0.4 $&$  -0.1 $&$   3.1 $&$  -0.3 $&$   0.0 $&$   0.1 $&$   0.6 $&$   5.4 $&$   2.9 $&$   0.2 $&$  -1.0 $&$  -0.2 $&$   0.2 $\\
$  26.5 $&$ 0.0562 $&$ 0.0500 $&$ 0.0293 $&$   7.5 $&$   7.2 $&$  10.4 $&$   0.0 $&$   1.9 $&$   0.8 $&$  -0.4 $&$   0.1 $&$  -1.3 $&$   1.8 $&$   4.9 $&$   4.2 $&$   0.2 $&$  -0.8 $&$  -0.5 $&$   0.3 $\\
$  26.5 $&$ 0.0562 $&$ 0.0750 $&$ 0.0325 $&$  12.3 $&$   6.5 $&$  14.0 $&$   0.0 $&$  -1.2 $&$  -0.2 $&$  -0.6 $&$   0.1 $&$  -4.6 $&$  -0.5 $&$   3.0 $&$   2.3 $&$   0.2 $&$  -0.7 $&$  -2.2 $&$   0.3 $\\
$  26.5 $&$ 0.1780 $&$ 0.0025 $&$ 0.0235 $&$   5.5 $&$   9.2 $&$  10.7 $&$  -1.7 $&$  -0.5 $&$   1.4 $&$  -0.4 $&$  -0.4 $&$  -0.4 $&$  -6.6 $&$   3.9 $&$   0.8 $&$   0.2 $&$  -0.6 $&$  -4.0 $&$   1.5 $\\
$  26.5 $&$ 0.1780 $&$ 0.0085 $&$ 0.0217 $&$   5.1 $&$   8.3 $&$   9.7 $&$   1.2 $&$  -0.2 $&$   1.5 $&$   0.1 $&$  -0.1 $&$   0.0 $&$   6.8 $&$   4.1 $&$   0.5 $&$   0.3 $&$  -0.7 $&$  -1.3 $&$   0.5 $\\
$  26.5 $&$ 0.1780 $&$ 0.0160 $&$ 0.0215 $&$   5.6 $&$   7.9 $&$   9.7 $&$   0.0 $&$   0.0 $&$   1.2 $&$   0.5 $&$   0.0 $&$   0.1 $&$   4.4 $&$   6.1 $&$   1.1 $&$   0.3 $&$  -1.1 $&$  -1.4 $&$   0.3 $\\
$  26.5 $&$ 0.1780 $&$ 0.0250 $&$ 0.0171 $&$   7.0 $&$   6.8 $&$   9.8 $&$   0.0 $&$  -0.3 $&$   0.9 $&$   1.2 $&$   0.0 $&$   0.9 $&$   1.7 $&$   6.0 $&$   1.8 $&$   0.1 $&$  -0.7 $&$  -1.0 $&$   0.2 $\\
$  26.5 $&$ 0.1780 $&$ 0.0350 $&$ 0.0181 $&$   9.9 $&$   7.6 $&$  12.5 $&$   0.0 $&$  -3.4 $&$   2.7 $&$   2.5 $&$   0.1 $&$   0.5 $&$   2.4 $&$   4.9 $&$   1.3 $&$   0.2 $&$  -1.1 $&$  -0.7 $&$   0.2 $\\
$  26.5 $&$ 0.5620 $&$ 0.0025 $&$ 0.0339 $&$   3.5 $&$  10.2 $&$  10.8 $&$   5.1 $&$  -2.2 $&$   2.2 $&$   0.6 $&$  -1.1 $&$  -0.6 $&$  -4.3 $&$   5.6 $&$   1.4 $&$   0.3 $&$  -0.8 $&$  -2.3 $&$   3.0 $\\
$  26.5 $&$ 0.5620 $&$ 0.0085 $&$ 0.0292 $&$   4.1 $&$   9.4 $&$  10.2 $&$   4.9 $&$  -0.7 $&$   1.4 $&$   0.7 $&$   0.0 $&$  -0.2 $&$   6.8 $&$   3.0 $&$   0.4 $&$   0.0 $&$  -0.5 $&$  -0.3 $&$   2.3 $\\
$  46.0 $&$ 0.0178 $&$ 0.0500 $&$ 0.0687 $&$  12.2 $&$   7.3 $&$  14.2 $&$  -0.3 $&$  -0.1 $&$   1.8 $&$  -1.2 $&$   0.1 $&$  -0.6 $&$   1.5 $&$   5.8 $&$   3.3 $&$   0.5 $&$  -0.7 $&$  -0.2 $&$   0.6 $\\
$  46.0 $&$ 0.0178 $&$ 0.0750 $&$ 0.0401 $&$  19.0 $&$   7.0 $&$  20.3 $&$   0.3 $&$   0.6 $&$   2.2 $&$  -1.1 $&$   0.0 $&$  -2.9 $&$  -1.5 $&$   4.1 $&$   3.2 $&$   0.4 $&$  -0.6 $&$  -1.9 $&$   0.1 $\\
$  46.0 $&$ 0.0562 $&$ 0.0160 $&$ 0.0360 $&$   8.9 $&$   8.0 $&$  11.9 $&$   0.1 $&$  -0.2 $&$   2.0 $&$  -1.1 $&$   0.0 $&$  -0.7 $&$   4.1 $&$   6.1 $&$   1.5 $&$   0.5 $&$  -0.7 $&$  -0.8 $&$   0.6 $\\
$  46.0 $&$ 0.0562 $&$ 0.0250 $&$ 0.0310 $&$   8.7 $&$   7.5 $&$  11.5 $&$   0.1 $&$   0.6 $&$   1.7 $&$  -1.2 $&$   0.0 $&$  -0.4 $&$  -1.2 $&$   6.9 $&$   1.0 $&$   0.2 $&$  -0.8 $&$  -0.6 $&$   0.1 $\\
$  46.0 $&$ 0.0562 $&$ 0.0350 $&$ 0.0301 $&$  10.6 $&$   7.0 $&$  12.8 $&$   0.3 $&$  -0.8 $&$   1.5 $&$  -0.1 $&$   0.1 $&$  -1.0 $&$   3.4 $&$   5.3 $&$   2.1 $&$   0.2 $&$  -0.7 $&$  -1.0 $&$   0.1 $\\
$  46.0 $&$ 0.0562 $&$ 0.0500 $&$ 0.0400 $&$   9.5 $&$   6.9 $&$  11.7 $&$   0.0 $&$  -1.0 $&$   2.0 $&$  -0.5 $&$   0.1 $&$  -1.9 $&$  -0.6 $&$   5.8 $&$   1.9 $&$   0.2 $&$  -1.1 $&$  -0.3 $&$   0.2 $\\
\hline
\end{tabular}
\end{tiny}
\end{table}

\newpage
\begin{table}[ht]
\centering
\begin{tiny}
  \begin{tabular}{| c| c| c| c| c| c| c| c| c| c| c| c| c| c| c| c| c| c| c| c|}
\hline
 $Q^2$ & $\beta$ & $\xpom$ & $\xpom\sigma_r^{D(3)}$ & $\delta_{stat}$ & $\delta_{sys}$ & $\delta_{tot}$ &
 $\delta_{had}$ & $\delta_{ele}$ & $\delta_{\theta}$ & $\delta_{\beta}$ & $\delta_{\xpom}$ & $\delta_{t}$ &
 $\delta_{E_p}$ & $\delta_{p_x}$ & $\delta_{p_y}$ & $\delta_{Q^2}$ & $\delta_{vtx}$ & $\delta_{bgn}$ & $\delta_{bcc}$ \\
 $[\GeV^2]$ & & & & $[\%]$ & $[\%]$ & $[\%]$ & $[\%]$ & $[\%]$ & $[\%]$ & $[\%]$ & $[\%]$ & $[\%]$ & $[\%]$ & $[\%]$ &
 $[\%]$ & $[\%]$ & $[\%]$ & $[\%]$ & $[\%]$ \\
\hline
$  46.0 $&$ 0.0562 $&$ 0.0750 $&$ 0.0420 $&$  16.1 $&$   6.7 $&$  17.4 $&$   0.0 $&$  -1.2 $&$   2.1 $&$  -0.7 $&$   0.1 $&$  -3.7 $&$   0.1 $&$   3.9 $&$   2.5 $&$   0.2 $&$  -0.6 $&$  -1.7 $&$   0.2 $\\
$  46.0 $&$ 0.1780 $&$ 0.0085 $&$ 0.0219 $&$   8.2 $&$   9.2 $&$  12.3 $&$   0.2 $&$  -1.7 $&$   2.7 $&$  -0.6 $&$   0.0 $&$   0.0 $&$   6.0 $&$   5.3 $&$   1.0 $&$   0.3 $&$  -0.9 $&$  -3.0 $&$   0.5 $\\
$  46.0 $&$ 0.1780 $&$ 0.0160 $&$ 0.0249 $&$   7.2 $&$   7.7 $&$  10.5 $&$   0.4 $&$  -0.5 $&$   1.1 $&$  -0.2 $&$   0.0 $&$  -0.8 $&$   4.3 $&$   5.8 $&$   0.9 $&$   0.2 $&$  -0.9 $&$  -1.5 $&$   0.2 $\\
$  46.0 $&$ 0.1780 $&$ 0.0250 $&$ 0.0244 $&$   8.6 $&$   7.4 $&$  11.3 $&$   0.0 $&$  -0.4 $&$   1.5 $&$  -0.4 $&$   0.0 $&$  -0.1 $&$   3.8 $&$   5.8 $&$   1.0 $&$   0.1 $&$  -1.0 $&$  -1.3 $&$   0.2 $\\
$  46.0 $&$ 0.1780 $&$ 0.0350 $&$ 0.0238 $&$  10.9 $&$   7.1 $&$  13.0 $&$   0.0 $&$  -0.2 $&$   2.6 $&$   0.2 $&$   0.2 $&$   0.0 $&$   1.7 $&$   5.8 $&$   2.6 $&$   0.1 $&$  -0.9 $&$  -0.4 $&$   0.2 $\\
$  46.0 $&$ 0.1780 $&$ 0.0500 $&$ 0.0250 $&$  11.2 $&$   8.1 $&$  13.8 $&$   0.0 $&$   1.3 $&$   5.1 $&$   0.9 $&$   0.3 $&$  -3.2 $&$   1.1 $&$   3.6 $&$   3.1 $&$   0.1 $&$  -1.2 $&$  -0.4 $&$   0.3 $\\
$  46.0 $&$ 0.1780 $&$ 0.0750 $&$ 0.0341 $&$  20.8 $&$   7.1 $&$  22.0 $&$   0.0 $&$  -2.4 $&$   0.2 $&$   1.6 $&$   0.1 $&$  -3.7 $&$  -1.2 $&$   3.6 $&$   2.7 $&$  -0.2 $&$  -1.0 $&$  -2.2 $&$   0.3 $\\
$  46.0 $&$ 0.5620 $&$ 0.0025 $&$ 0.0400 $&$   5.5 $&$  10.8 $&$  12.1 $&$   5.0 $&$  -2.8 $&$   2.2 $&$   0.3 $&$  -0.9 $&$   0.2 $&$  -6.1 $&$   4.3 $&$   1.1 $&$   0.2 $&$  -0.7 $&$  -2.7 $&$   3.6 $\\
$  46.0 $&$ 0.5620 $&$ 0.0085 $&$ 0.0301 $&$   5.6 $&$   9.6 $&$  11.2 $&$   5.6 $&$  -1.3 $&$   1.7 $&$   0.5 $&$  -0.1 $&$  -0.4 $&$   6.0 $&$   3.4 $&$   0.7 $&$   0.1 $&$  -0.5 $&$  -0.9 $&$   2.9 $\\
$  46.0 $&$ 0.5620 $&$ 0.0160 $&$ 0.0242 $&$   8.7 $&$   8.4 $&$  12.1 $&$   0.0 $&$  -2.6 $&$   5.5 $&$   1.3 $&$   0.2 $&$   0.7 $&$  -1.9 $&$   3.9 $&$   0.4 $&$   0.1 $&$  -1.2 $&$  -2.1 $&$   2.8 $\\
$  46.0 $&$ 0.5620 $&$ 0.0250 $&$ 0.0167 $&$  13.2 $&$   7.5 $&$  15.2 $&$   0.0 $&$   1.3 $&$   3.0 $&$   3.0 $&$   0.2 $&$  -1.3 $&$  -0.9 $&$   4.0 $&$   2.7 $&$  -0.9 $&$  -1.5 $&$  -0.7 $&$   2.7 $\\
$  80.0 $&$ 0.0562 $&$ 0.0350 $&$ 0.0375 $&$  19.2 $&$   7.9 $&$  20.8 $&$   0.3 $&$  -2.1 $&$   2.8 $&$  -2.8 $&$  -0.1 $&$   0.6 $&$   1.8 $&$   5.2 $&$   2.0 $&$   1.7 $&$  -0.8 $&$  -2.2 $&$   0.1 $\\
$  80.0 $&$ 0.0562 $&$ 0.0500 $&$ 0.0385 $&$  15.6 $&$   6.9 $&$  17.0 $&$   0.0 $&$  -1.1 $&$   2.7 $&$  -1.2 $&$   0.0 $&$   0.2 $&$   0.3 $&$   5.0 $&$   3.3 $&$   0.9 $&$  -0.8 $&$   0.0 $&$   0.1 $\\
$  80.0 $&$ 0.0562 $&$ 0.0750 $&$ 0.0354 $&$  24.2 $&$   7.4 $&$  25.3 $&$   0.0 $&$   0.5 $&$   1.8 $&$  -0.7 $&$   0.1 $&$  -3.0 $&$   0.5 $&$   3.0 $&$   5.3 $&$   0.2 $&$  -0.8 $&$  -1.8 $&$   0.2 $\\
$  80.0 $&$ 0.1780 $&$ 0.0085 $&$ 0.0396 $&$  14.9 $&$   8.6 $&$  17.2 $&$  -2.0 $&$  -1.5 $&$   2.2 $&$  -1.6 $&$  -0.1 $&$  -0.9 $&$  -0.8 $&$   5.8 $&$   1.1 $&$   0.4 $&$  -0.8 $&$  -4.8 $&$   1.0 $\\
$  80.0 $&$ 0.1780 $&$ 0.0160 $&$ 0.0227 $&$  12.8 $&$   7.9 $&$  15.0 $&$   0.0 $&$   0.5 $&$   2.5 $&$  -0.4 $&$   0.0 $&$   0.2 $&$   5.2 $&$   5.1 $&$   0.9 $&$   0.3 $&$  -0.8 $&$  -1.3 $&$   0.2 $\\
$  80.0 $&$ 0.1780 $&$ 0.0250 $&$ 0.0246 $&$  11.8 $&$   7.8 $&$  14.1 $&$   0.0 $&$   0.5 $&$   1.7 $&$  -0.9 $&$   0.1 $&$   0.2 $&$   0.7 $&$   6.8 $&$   2.3 $&$   0.5 $&$  -1.5 $&$  -1.4 $&$   0.2 $\\
$  80.0 $&$ 0.1780 $&$ 0.0350 $&$ 0.0290 $&$  14.1 $&$   6.9 $&$  15.8 $&$   0.0 $&$   0.8 $&$  -0.3 $&$  -3.5 $&$   0.1 $&$  -0.2 $&$   4.1 $&$   4.2 $&$   0.3 $&$   0.2 $&$  -0.7 $&$  -0.7 $&$   0.2 $\\
$  80.0 $&$ 0.1780 $&$ 0.0500 $&$ 0.0251 $&$  15.1 $&$   6.8 $&$  16.6 $&$   0.0 $&$  -1.5 $&$   0.9 $&$  -0.5 $&$  -0.1 $&$  -1.4 $&$  -0.2 $&$   3.9 $&$   4.5 $&$   0.4 $&$  -1.7 $&$  -1.2 $&$   0.3 $\\
$  80.0 $&$ 0.1780 $&$ 0.0750 $&$ 0.0226 $&$  25.8 $&$   6.8 $&$  26.7 $&$   0.0 $&$  -1.5 $&$   1.3 $&$   0.8 $&$   0.1 $&$  -2.8 $&$   0.8 $&$   5.4 $&$   1.2 $&$   0.0 $&$  -0.6 $&$  -1.8 $&$   0.3 $\\
$  80.0 $&$ 0.5620 $&$ 0.0085 $&$ 0.0266 $&$   9.8 $&$  10.4 $&$  14.3 $&$   4.7 $&$  -1.4 $&$   1.1 $&$   0.3 $&$  -0.2 $&$   0.3 $&$   5.0 $&$   6.1 $&$   1.1 $&$   0.3 $&$  -1.0 $&$  -3.1 $&$   2.9 $\\
$  80.0 $&$ 0.5620 $&$ 0.0160 $&$ 0.0192 $&$  13.0 $&$   9.3 $&$  16.0 $&$   0.0 $&$  -5.2 $&$   0.3 $&$   0.5 $&$   0.2 $&$  -0.7 $&$  -2.9 $&$   4.8 $&$   0.0 $&$   0.2 $&$  -0.4 $&$  -4.5 $&$   2.8 $\\
$  80.0 $&$ 0.5620 $&$ 0.0250 $&$ 0.0180 $&$  16.1 $&$   8.2 $&$  18.1 $&$   0.0 $&$  -5.5 $&$   2.8 $&$   1.2 $&$   0.2 $&$  -2.1 $&$  -0.8 $&$   2.3 $&$   2.5 $&$   0.4 $&$  -1.2 $&$  -1.5 $&$   2.7 $\\
$  80.0 $&$ 0.5620 $&$ 0.0350 $&$ 0.0199 $&$  24.6 $&$  10.0 $&$  26.6 $&$   0.0 $&$  -4.7 $&$   0.6 $&$   1.8 $&$   0.1 $&$   0.4 $&$  -3.1 $&$   7.5 $&$   0.9 $&$   0.0 $&$  -0.9 $&$  -0.2 $&$   2.5 $\\
$ 200.0 $&$ 0.0562 $&$ 0.0500 $&$ 0.0285 $&$  27.4 $&$   6.2 $&$  28.1 $&$   0.0 $&$   1.4 $&$   1.6 $&$  -0.6 $&$   0.0 $&$  -1.0 $&$   0.8 $&$   3.7 $&$   2.9 $&$   0.3 $&$  -0.9 $&$  -2.6 $&$   1.6 $\\
$ 200.0 $&$ 0.0562 $&$ 0.0750 $&$ 0.0490 $&$  36.4 $&$   7.3 $&$  37.2 $&$   0.0 $&$  -0.2 $&$   0.6 $&$   0.1 $&$   0.1 $&$  -3.2 $&$  -1.3 $&$   5.4 $&$   3.0 $&$  -0.1 $&$  -0.7 $&$  -1.5 $&$   0.7 $\\
$ 200.0 $&$ 0.1780 $&$ 0.0160 $&$ 0.0260 $&$  19.6 $&$   8.4 $&$  21.3 $&$  -1.0 $&$   1.9 $&$   0.1 $&$  -0.4 $&$   0.1 $&$   0.5 $&$   5.3 $&$   4.7 $&$   2.2 $&$   0.5 $&$  -0.8 $&$  -2.3 $&$   1.7 $\\
$ 200.0 $&$ 0.1780 $&$ 0.0250 $&$ 0.0353 $&$  16.4 $&$   7.5 $&$  18.1 $&$   0.0 $&$   1.9 $&$   0.3 $&$  -0.5 $&$   0.1 $&$   0.1 $&$   6.0 $&$   3.8 $&$   1.3 $&$  -0.1 $&$  -0.6 $&$  -0.5 $&$   0.7 $\\
$ 200.0 $&$ 0.1780 $&$ 0.0350 $&$ 0.0290 $&$  22.1 $&$   6.9 $&$  23.1 $&$   0.2 $&$  -0.1 $&$   0.0 $&$  -0.2 $&$   0.0 $&$   1.0 $&$  -3.9 $&$   5.4 $&$   1.4 $&$  -0.1 $&$  -0.6 $&$  -0.6 $&$   0.2 $\\
$ 200.0 $&$ 0.1780 $&$ 0.0500 $&$ 0.0405 $&$  20.2 $&$   6.8 $&$  21.4 $&$   0.0 $&$  -2.0 $&$   0.7 $&$  -0.6 $&$   0.1 $&$  -0.6 $&$   1.1 $&$   4.6 $&$   4.2 $&$  -0.3 $&$  -0.7 $&$  -0.5 $&$   0.3 $\\
$ 200.0 $&$ 0.1780 $&$ 0.0750 $&$ 0.0359 $&$  34.5 $&$   7.0 $&$  35.2 $&$   0.0 $&$  -1.3 $&$   0.8 $&$  -1.3 $&$   0.0 $&$  -2.5 $&$  -1.0 $&$   4.0 $&$   4.3 $&$  -0.4 $&$  -0.8 $&$  -1.9 $&$   0.4 $\\
$ 200.0 $&$ 0.5620 $&$ 0.0085 $&$ 0.0210 $&$  18.6 $&$  10.1 $&$  21.2 $&$   3.3 $&$  -1.6 $&$   1.0 $&$   0.3 $&$   0.0 $&$  -0.5 $&$   3.1 $&$   7.7 $&$   1.2 $&$   0.1 $&$  -1.0 $&$  -2.6 $&$   3.1 $\\
$ 200.0 $&$ 0.5620 $&$ 0.0160 $&$ 0.0174 $&$  22.4 $&$   9.0 $&$  24.1 $&$   0.0 $&$  -2.2 $&$  -0.3 $&$  -0.2 $&$   0.0 $&$   0.6 $&$  -2.3 $&$   6.6 $&$   0.8 $&$  -0.6 $&$  -0.7 $&$  -4.2 $&$   2.7 $\\
$ 200.0 $&$ 0.5620 $&$ 0.0250 $&$ 0.0139 $&$  26.6 $&$   8.1 $&$  27.8 $&$   0.0 $&$  -6.0 $&$   2.1 $&$   0.7 $&$   0.2 $&$  -0.3 $&$  -0.7 $&$   3.6 $&$   0.7 $&$  -1.1 $&$  -0.8 $&$  -1.3 $&$   2.5 $\\
$ 200.0 $&$ 0.5620 $&$ 0.0350 $&$ 0.0275 $&$  33.1 $&$   9.9 $&$  34.6 $&$   0.0 $&$  -2.0 $&$   1.4 $&$   1.5 $&$  -0.1 $&$  -0.5 $&$  -3.5 $&$   7.8 $&$   3.1 $&$  -0.9 $&$  -0.9 $&$  -0.4 $&$   2.3 $\\
$ 200.0 $&$ 0.5620 $&$ 0.0500 $&$ 0.0261 $&$  28.7 $&$   6.9 $&$  29.5 $&$   0.0 $&$  -1.2 $&$   0.7 $&$  -0.3 $&$   0.2 $&$   0.7 $&$   0.3 $&$   6.0 $&$   1.9 $&$  -0.8 $&$  -0.4 $&$   0.0 $&$   2.1 $\\
$ 200.0 $&$ 0.5620 $&$ 0.0750 $&$ 0.0407 $&$  48.3 $&$   6.7 $&$  48.8 $&$   0.0 $&$  -0.9 $&$   0.4 $&$   1.7 $&$   0.0 $&$  -4.7 $&$   0.4 $&$   3.0 $&$   0.9 $&$  -1.3 $&$  -0.7 $&$  -1.8 $&$   1.7 $\\
\hline
\end{tabular}
\end{tiny}
\end{table}

\end{landscape}

\renewcommand{\arraystretch}{1.}

\clearpage
\newpage

\end{document}